\documentclass[aps,prd,floatfix,nofootinbib,showpacs,twocolumn,preprintnumbers,superscriptaddress]{revtex4-1} \usepackage{dcolumn}
\usepackage{bm}
\usepackage{braket}
\usepackage{color}
\usepackage{amssymb}
\usepackage{amsmath}
\usepackage{graphicx}
\usepackage{amsfonts}
\usepackage{slashed}
\usepackage{float}
\usepackage{multirow}
\usepackage{array}
\allowdisplaybreaks
\usepackage{dcolumn}
\usepackage{float}
\usepackage[nice]{nicefrac}

\graphicspath{{./figures/}}

\renewcommand{\vec}{\bm}

\newcommand{\dt}{\delta\tau}
\newcommand{\ham}{H}
\newcommand{\PsiTrial}{\Phi}
\newcommand{\PsiI}{\Phi_I}

\newcommand{\PsiGround}{\Psi_0}

\newcommand{\bigR}{\vec{R}} 

\usepackage[hypertexnames=true]{hyperref}
\usepackage[capitalize]{cleveref}
\hypersetup{
    colorlinks=true,       linkcolor=blue,          citecolor=blue,        filecolor=blue,      urlcolor=blue           }

\begin{document}

\preprint{FERMILAB-PUB-23-128-T}

\title{Mitigating Green's function Monte Carlo signal-to-noise problems \texorpdfstring{\\}{} using contour deformations}

\author{Gurtej Kanwar}
\affiliation{Albert Einstein Center, Institute for Theoretical Physics, University of Bern, 3012 Bern, Switzerland}

\author{Alessandro Lovato}
\affiliation{Physics Division, Argonne National Laboratory, Argonne, IL 60439}
\affiliation{Computational Science Division, Argonne National Laboratory, Argonne, IL 60439}
\affiliation{INFN-TIFPA Trento Institute for Fundamental Physics and Applications, 38123 Trento, Italy}

\author{Noemi Rocco}
\affiliation{Theoretical Physics Department, Fermi National Accelerator Laboratory, P.O. Box 500, Batavia, Illinois 60510, USA}

\author{Michael Wagman}
\affiliation{Theoretical Physics Department, Fermi National Accelerator Laboratory, P.O. Box 500, Batavia, Illinois 60510, USA}

\date{\today}

\begin{abstract}
The Green's function Monte Carlo (GFMC) method provides accurate solutions to the nuclear many-body problem and predicts properties of light nuclei starting from realistic two- and three-body interactions. Controlling the GFMC fermion-sign problem is crucial, as the signal-to-noise ratio decreases exponentially with Euclidean time, requiring significant computing resources. Inspired by similar scenarios in lattice quantum field theory and spin systems, in this work, we employ integration contour deformations to improve the GFMC signal-to-noise ratio. Machine learning techniques are used to select optimal contours with minimal variance from parameterized families of deformations. As a proof of principle, we consider the deuteron binding energies and Euclidean density response functions. We only observe mild signal-to-noise improvement for the binding energy case. On the other hand, we achieve an order of magnitude reduction of the variance for Euclidean density response functions, paving the way for computing electron- and neutrino-nucleus cross-sections of larger nuclei.
\end{abstract}

\maketitle

\section{Introduction}

The overarching goal of nuclear many-body theory is the description of atomic nuclei starting from the interactions among the constituent protons and neutrons. An important step of this endeavor consists in solving the Schr\"odinger equation associated with the nuclear Hamiltonian, a formidable computational task because of the non-perturbative nature and strong spin-isospin dependence of realistic nuclear forces~\cite{Hergert:2020bxy}. 

Continuum quantum Monte Carlo approaches, such as Green's function Monte Carlo (GFMC)~\cite{Carlson:2014vla,Gandolfi:2020pbj}, tackle this challenge with high accuracy using imaginary-time projection techniques. The GFMC method is applicable to nuclei with up to $A \lesssim 12$ nucleons, but encounters difficulties for larger nuclei due to the exponentially large number of spin-isospin degrees of freedom and the reliance on potentially noisy Monte Carlo sampling of particle spatial coordinates. The auxiliary field diffusion Monte Carlo (AFDMC) method~\cite{Schmidt:1999lik} can reach larger systems by sampling the spin and isospin degrees of freedom, at the cost of additional noise. 

Both the GFMC and the AFDMC methods suffer from the so-called fermion sign problem. This problem originates from the imaginary-time projection converging to the boson ground state of the Hamiltonian, making the overlap with the fermion ground state exponentially small. In condensed-matter and quantum chemistry applications, where the ground-state wave function is real, the ``fixed-node'' approximation is often employed to control the sign problem and provides rigorous upper bounds of the true ground-state energy of the system~\cite{Anderson:1976JChPh,Ceperley:1984JChPh}. On the other hand, nuclear ground-state wave functions are typically complex-valued. The ``constrained-path'' approximation employed in GFMC and AFDMC brings about a bias in the energy expectation value~\cite{Pudliner:1997ck,Piarulli:2019pfq}, leading to possible violations of the variational principle. Performing the required ``unconstrained'' propagations is computationally demanding, since a large number of Monte Carlo samples are required to reduce the statistical noise of the calculation. 

In addition to ground-state properties, GFMC has been employed to compute electromagnetic, neutral-current, and charged-current response functions of $^4$He and $^{12}$C in the quasi-elastic region, up to moderate values of the momentum transfer~\cite{Carlson:2001mp,Lovato:2016gkq,Lovato:2017cux,Lovato:2020kba}, and the muon capture rates of $^{4}$He and $^{3}$H~\cite{Lovato:2019fiw}. Within this approach, the electroweak response functions are inferred from their Laplace transforms, or Euclidean responses, that are estimated during unconstrained imaginary-time propagations. Retrieving the energy dependence of the response functions is nontrivial and requires accurate estimates of the Euclidean responses even when inversion methods based on artificial neural networks are employed~\cite{Raghavan:2020bze}.
In particular, the importance of precise cross-section predictions for next-generation neutrino experiments~\cite{Alvarez-Ruso:2014bla,DUNE:2015lol,NuSTEC:2017hzk,Meyer:2022mix,Ruso:2022qes,Simons:2022ltq} and the challenges in these inverse Laplace transform approaches motivate the development of noise reduction methods to improve the applicability and precision of GFMC cross-section predictions.

This work introduces a method of noise reduction based on contour deformations of the integration over particle coordinates. Such contour deformations have the ability to modify the noise properties within a Monte Carlo integration scheme while guaranteeing exactness based on analyticity. To be effective, contour deformations should be selected based on the observable under study. In this proof-of-principle work, we demonstrate contour deformations that yield modest improvements in the precision of ground-state energy estimations and significant improvements in the precision of Euclidean density response functions.

Contour deformations have previously been applied to path integrals arising in lattice gauge theory~\cite{Cristoforetti:2012su,Aarts:2013fpa,Mukherjee:2013aga,Aarts:2014nxa,Schmidt:2017gvu,DiRenzo:2017igr,Kashiwa:2018vxr,Alexandru:2018ngw,Kashiwa:2019lkv,Detmold:2020ncp,Pawlowski:2021bbu,Detmold:2021ulb,Kanwar:2021tkd}, low-dimensional lattice field theories of interacting non-relativistic fermions and scalars~\cite{Cristoforetti:2012su,Cristoforetti:2013wha,Aarts:2013fpa,Fujii:2013sra,Mukherjee:2013aga,Aarts:2014nxa,Cristoforetti:2014gsa,Alexandru:2015xva,Alexandru:2015sua,Fujii:2015vha,Alexandru:2016gsd,Alexandru:2016ejd,Alexandru:2017czx,Alexandru:2017lqr,Mori:2017nwj,Tanizaki:2017yow,Alexandru:2018brw,Alexandru:2018fqp,Alexandru:2018ddf,Mou:2019gyl,Lawrence:2021izu,DiRenzo:2021kcw,Lawrence:2022dba}, spin models relevant for condensed matter systems such as the Hubbard model~\cite{Mukherjee:2014hsa,Tanizaki:2015rda,Fukuma:2019uot,Fukuma:2019wbv,Ulybyshev:2019fte,Mishchenko:2021vnx,Rodekamp:2022xpf}, and electronic structure calculations~\cite{Rom:1997,Rom:1998,Baer:1998,Baer:2000,Baer:2000b}; for a recent review see Ref.~\cite{Alexandru:2020wrj}.
Many applications have focused on path integrals where the action is complex due to the inclusion of a non-zero chemical potential~\cite{Cristoforetti:2012su,Cristoforetti:2013wha,Aarts:2013fpa,Mukherjee:2013aga,Aarts:2014nxa,Cristoforetti:2014gsa,Alexandru:2015xva,Alexandru:2015sua,Fujii:2015vha,Tanizaki:2015rda,Alexandru:2016ejd,Schmidt:2017gvu,DiRenzo:2017igr,Alexandru:2017czx,Mori:2017nwj,Tanizaki:2017yow,Kashiwa:2018vxr,Alexandru:2018ngw,Alexandru:2018brw,Alexandru:2018fqp,Alexandru:2018ddf,Kashiwa:2019lkv,Fukuma:2019uot,Fukuma:2019wbv,Ulybyshev:2019fte,Pawlowski:2021bbu,Mishchenko:2021vnx,DiRenzo:2021kcw,Rodekamp:2022xpf,Lawrence:2022dba} or the use of real-time evolution~\cite{Alexandru:2016gsd,Alexandru:2017lqr,Mou:2019gyl,Lawrence:2021izu,Kanwar:2021tkd}.
While some studies work towards the construction of contours with appealing formal properties such as Lefschetz thimbles~\cite{Witten:2010cx,Witten:2010zr}, several more recent works have pioneered the application of machine learning techniques towards the parameterization and numerical determination of optimal contour deformations~\cite{Alexandru:2017czx,Mori:2017nwj,Alexandru:2018fqp,Alexandru:2018ddf,Detmold:2020ncp,Detmold:2021ulb,Lawrence:2021izu,Rodekamp:2022xpf,Lawrence:2022dba}.

Recently, contour deformations have also been applied to lattice field theory path integrals where the action is real but severe complex phase fluctuations arise from the inclusion of a noisy observable in expectation values~\cite{Detmold:2020ncp,Detmold:2021ulb}.
In these works, variances of observables are minimized by applying machine learning techniques to find optimal contours within parameterized families of deformations with a given observable's variance treated as the loss function to be minimized.

This work applies a similar strategy to the construction of variance-reducing integration contours for the GFMC evaluation of nuclear observables.
As a starting point, we first analytically continue the Argonne $v_{18}$ (AV18) potentials and deuteron ground-state variational wavefunctions to complex values of the particle coordinates by fitting to appropriate Chebyshev polynomial series.
This defines a holomorphic integrand which can be evaluating on complexified contours of integration without modifying the expectation values of observables.
Working with simple parameterizations of contour deformation families, we then optimize choices of contour deformations of the integration over particle coordinates specifically for the cases of
the deuteron binding energy and the deuteron Euclidean density response functions.
In the latter case, the phase fluctuations leading to signal-to-noise problems can be easily identified, facilitating significant reductions in the variance.
Future work will explore whether generalizations of these deformations or more sophisticated families of contour deformations based for example on neural networks will be sufficient to improve signal-to-noise problems in GFMC calculations of larger nuclei.

The remainder of this work is structured as follows. The GFMC methods and Hamiltonian used in this work are briefly reviewed in Sec.~\ref{sec:gfmc} with emphasis on the integral representation of GFMC observables suitable for applying contour deformations.
Contour deformation methods for GFMC, including analytic continuation of the potential and wavefunctions, parameterization of the integration contour, and numerical optimization techniques used to search for variance-reducing deformations within these families are discussed in Sec.~\ref{sec:contour}.
Finally, applications of these methods to calculations of the deuteron binding energy and density response function are discussed in Sec.~\ref{sec:results}.

\section{Green's function Monte Carlo \texorpdfstring{\\}{} for nuclei}\label{sec:gfmc}

\subsection{Nuclear Hamiltonian}

The nonrelativistic nuclear Hamiltonian adopted in this work reads
\begin{align}
\ham= \sum_{i=1}^A\frac{{\bf p}_i^2}{2M_N}+\sum_{j>i=1}^A v_{ij}\ .
\end{align}
In the above equation, $A$ is the number of nucleons, $\vec{p}$ and $M_N$ are the nucleon momentum and mass, respectively, and $v_{ij}$ is the two-nucleon potential.
There are many ways to parameterize nucleon-nucleon (NN) interactions using potentials, which can then be constrained using NN scattering data and other properties of nuclei, including potentials based on chiral effective field theory (EFT) as reviewed in Refs.~\cite{Epelbaum:2008ga,Navratil:2016ycn,Tews:2020hgp,vanKolck:2020llt,Epelbaum:2022cyo}.
One such parameterization is the AV18 potential~\cite{Wiringa:1994wb}, a state-of-the-art phenomenological model of NN interaction given by the sum of 18 operators commonly used in quantum Monte Carlo calculations of light nuclei and infinite nuclear matter~\cite{Carlson:2014vla,Piarulli:2019pfq,Lovato:2022apd}.

The AV18 potential can be expressed as a
sum of products of radial functions and spin-isospin operators
\begin{align}
v_{ij}(\vec{r}_{ij})=\sum_p v_p(|\vec{r}_{ij}|) \, O^p_{ij}(\vec{r}_{ij})
\end{align}
where $\vec{r}_{ij} = \vec{r}_i-\vec{r}_j$ is the relative displacement between the coordinates $\vec{r}_i$ and $\vec{r}_j$ of the $i$-th and $j$-th nucleons, respectively. The radial functions $v_p$ only depend on this relative distance, while the operators $O^p_{ij}$ encompass the strong spin–isospin dependence of nuclear forces and the occurrence of noncentral interactions. The first six operators are the most important ones and read 
\begin{equation}
\begin{aligned}
O^{1}_{ij} &= 1, \quad & & O^{2}_{ij} = \bm{\tau}_i \cdot \bm{\tau}_j \\
O^{3}_{ij} &= \bm{\sigma}_i \cdot \bm{\sigma}_j, \quad & & O^{4}_{ij} = (\bm{\tau}_i \cdot \bm{\tau}_j)(\bm{\sigma}_i \cdot \bm{\sigma}_j) \\
O^{5}_{ij} &= S_{ij}(\vec{r}_{ij}), \quad & & O^{6}_{ij} = (\bm{\tau}_i \cdot \bm{\tau}_j) \, S_{ij}(\vec{r}_{ij})
\end{aligned} \label{eq:o6}
\end{equation}
where ${\bm \sigma}_i$ and ${\bm \tau}_i$ are the Pauli matrices operating over the nucleon spin and isospin degrees of freedom, respectively, of the $i$-th particle, and $S_{ij}(\vec{r}_{ij})=3({\bm \sigma}_i\cdot \hat{r}_{ij}) ({\bm \sigma}_j\cdot \hat{r}_{ij})-{\bm \sigma}_i\cdot{\bm \sigma}_j$ is the tensor operator. The AV18 radial functions are determined by fitting the  
full Nijmegen NN scattering data and deuteron properties~\cite{Wiringa:1994wb}.

As the detailed form of the potential is not relevant for applications of contour deformations using the methods described below, in this work we adopt
the simplified Argonne $v_6^\prime$ (AV6P) potential, constructed by projecting the
full AV18 on the basis of the six operators of Eq.~\eqref{eq:o6}, as well as the Argonne $v_4^\prime$ (AV4P) potential, which only retains $O^1_{ij},\ldots,O^4_{ij}$~\cite{Wiringa:2002ja}.

\subsection{Green's function Monte Carlo methods}

The GFMC method begins with a variational wave function $\PsiTrial$ that approximates the nuclear state under study. This is typically defined through the application of correlation operators on a Slater determinant of single-particle orbitals. The optimal set of variational parameters are found by applying Variational Monte Carlo (VMC) to minimize the energy expectation value on the variational wave function. 

The GFMC method then finds the ground-state of the system by evolving the variational wave function by imaginary time $\tau$ as  
\begin{equation}
    \ket{\Psi(\tau)} = e^{-H\tau} \ket{\Psi(0)}, \quad \text{where} \; \ket{\Psi(0)} = \ket{\PsiTrial}.
\end{equation}
For sufficiently large values of $\tau$, all excited-state contamination present in $\PsiTrial$ is suppressed and the true ground state $\PsiGround$ is recovered, i.e.,
\begin{align} \label{eq:ground-state-convergence}
\lim_{\tau \to \infty} \ket{\Psi(\tau)} = \lim_{\tau \to \infty} e^{-H\tau} \ket{\PsiTrial} \propto \ket{\PsiGround} \, .
\end{align}
In practice, the limit in Eq.~\eqref{eq:ground-state-convergence} is approximated by studying the properties of $\ket{\Psi(\tau)}$ as a function of $\tau$ and extrapolating to the infinite limit.

The direct computation of the propagator $\exp(-H\tau)$ for arbitrary values of $\tau$ is typically not possible. However, the calculation becomes tractable if the time evolution is carried out as a series of $N$ small steps of size $\dt = \tau/N$.
For small $\dt$ a Trotter expansion can be performed,
\begin{equation}
e^{-H\dt } = e^{-V \dt / 2} e^{-T \dt} e^{-V\dt / 2}  + \mathcal{O}(\dt^2)
\end{equation}
in terms of the kinetic operator $T = \sum_{i=1}^{A} \bm{p}_i^2 / 2M_N$ and the interaction potential $V = \sum_{j > i=1}^{A} v_{ij}$.

Any wave function $\Psi$ can be written as a function of the particle coordinates,
\begin{equation}
\Psi(\vec{r}_1, \vec{r}_2, \dots, \vec{r}_A) \equiv \Psi(\bigR) = \braket{\bigR | \Psi},
\end{equation}
where the notation $\bigR$ indicates the collection of position vectors $\{ \vec{r}_1, \dots, \vec{r}_A \}$.
The wavefunction $\Psi(\bigR)$ is itself a complex vector encompassing spin and isospin degrees of freedom whose dimension is $2^A \times \binom{A}{Z}$, where $Z$ is the number of protons and $\binom{A}{Z}$ counts the number of ways to label the nucleons as neutrons and protons. To be concise, these spin-isospin indices are suppressed in the following.

Maintaining full spin-isospin wave functions while inserting complete sets of position states using the notation above, the propagation can be written in terms of the path integral 
\begin{equation}
\begin{aligned}
    \Psi(\tau,\bigR^N)=&\int \prod_{n=0}^{N-1} d\bigR^n \braket{ \bigR^N|e^{-H\delta\tau}|\bigR^{N-1}} \\
    &\times \ldots \times \braket{ \bigR^1|e^{-H\delta\tau}|\bigR^{0} } \PsiTrial(\bigR^0) \, .
    \label{eq:psidef}
\end{aligned}
\end{equation}
The short-time propagator (or Green's function) can be approximated using the Trotter expansion as
\begin{equation}
\begin{aligned}
G_{\dt}(\bigR^\prime, \bigR)   &\equiv \langle \bigR^\prime| e^{- H \delta\tau} |\bigR\rangle \\
&\approx S_{\frac{1}{2}\dt}(\bigR') G_{\dt}^0(\bigR^\prime, \bigR) S_{\frac{1}{2}\dt}(\bigR)
\end{aligned} \label{eq:prop}
\end{equation}
where
\begin{equation}
\begin{aligned}
S_{\frac{1}{2}\dt}(\bigR) &= e^{-\frac{1}{2}V_{SI}(\bigR)\dt} \\
&\hspace{10pt} \times \left[ 1 - \frac{\dt}{2} V_{SD}(\bigR) + \frac{\dt^2}{8} V_{SD}^2(\bigR) \right]\, .
\end{aligned}
\label{eq:prop-M2}
\end{equation}
In the above equation, we have separated the spin-isospin-independent (SI) term $v_1 O^1_{ij}$ and spin-isospin-dependent (SD) terms of the interaction and have taken the quadratic expansion of the exponential term containing $V_{SD}$, which is a matrix in spin-isospin space. The kinetic term is the free propagator, which can be expressed as a simple Gaussian in the space of particle coordinates,
\begin{equation} 
\begin{aligned}
G_{\dt}^0(\bigR^\prime, \bigR) &= \langle \bigR^\prime |e^{-T\dt} |\bigR \rangle \\ 
&= \Big[ \sqrt{\frac{M_N}{2\pi \dt}} \Big]^{3A} \exp\Big[ -\frac{(\bigR^\prime -\bigR)^2}{2\dt/M_N} \Big]. \label{eq:kinetic}
\end{aligned}
\end{equation}

Substituting the Trotter expansion from Eqs.~\eqref{eq:prop}--\eqref{eq:kinetic} into the path integral definition of $\ket{\Psi(\tau)}$, the wavefunction can be evaluated to $O(\dt^2)$ as
\begin{equation}
\begin{aligned}
    \Psi(\tau, \bigR^{N}) &= \int \prod_{n=0}^{N-1} \left[ d\bigR^{n} G^0_{\dt}(\bigR^{n+1}, \bigR^{n}) \right] \\
    &\hspace{10pt} \times \prod_{n=0}^{N-1} \left[ S_{\frac{1}{2}\dt}(\bigR^{n+1}) S_{\frac{1}{2}\dt}(\bigR^{n}) \right] \PsiTrial(\bigR^0).
    \hspace{-20pt} \end{aligned} \label{eq:trotter-path-integral}
\end{equation}
For future convenience, the spin-isospin matrix consisting of products of $S_{\frac{1}{2}\dt}$ will be denoted
\begin{equation}
    S(\bigR^{N}, \dots, \bigR^{0}) \equiv \prod_{n=0}^{N-1} \left[ S_{\frac{1}{2}\dt}(\bigR^{n+1}) S_{\frac{1}{2}\dt}(\bigR^{n}) \right].
\end{equation}

The integrals in Eq.~\eqref{eq:trotter-path-integral} typically cannot be evaluated directly and one has to resort to Monte Carlo sampling. To this aim, we define the (complex) scalar density
\begin{equation}
\begin{aligned}
    I(\bigR^N, \dots, \bigR^0) &= \prod_{n=0}^{N-1}  G^0_{\dt}(\bigR^{n+1}, \bigR^{n}) \\
    &\times \PsiTrial^\dag(\bigR^N) S(\bigR^N, \dots, \bigR^0) \PsiTrial(\bigR^0),
\end{aligned}
\end{equation}
where the importance sampling wave function is here chosen as customary in GFMC calculations to be the trial wavefunction, $\PsiTrial^\dag$, though other choices $\PsiI^\dag$ are possible. Note that the integral of $I(\dots)$ gives the Trotter approximation to the normalization $\braket{\PsiTrial | e^{-H \tau} | \PsiTrial}$. If $I(\dots)$ is positive, it can therefore be interpreted as the appropriate probability distribution to estimate observables,
\begin{equation} \label{eq:observable}
    \braket{O} = \frac{\braket{\PsiTrial | O e^{-H \tau} | \PsiTrial}}{\braket{\PsiTrial | e^{-H \tau} | \PsiTrial}} \, .
\end{equation}
However, the positivity of this function is not guaranteed, since the complex phase structure of $\PsiTrial^\dagger(\bigR)$ is in general different the one of the ground state, leading to the notorious fermion sign problem. Further, for spin-isospin-dependent interactions it is not straightforward to sample the coordinates at all imaginary times simultaneously according to such a distribution, even when $I$ is positive\footnote{Simultaneously sampling the degrees of freedom in a path integral is often used in lattice quantum field theory calculations, where Markov Chain Monte Carlo is used, with challenges arising due to thermalization and autocorrelation times in such a framework.}, although a promising development in this direction has recently been put forward in Ref.~\cite{Chen:2022ndh}.

In diffusion Monte Carlo methods, a recursive approach is instead adopted to sample each $\bigR^n$, commonly denoted as a {\it walker}, based on the previous coordinates $\bigR^{n-1}, \dots, \bigR^0$ with the density $I(...)$ serving to guide the sampling.
We can relate the
density of walkers for $N+1$ steps to the density for $N$ steps as
\begin{equation}
\begin{aligned}
    &I(\bigR^{N+1}, \dots, \bigR^0) = G^0_{\dt}(\bigR^{N+1}, \bigR^N) \\
    &\hspace{20pt} \times w^{N+1}(\bigR^{N+1}, \dots, \bigR^0) \, I(\bigR^N, \dots, \bigR^0) \, ,
\end{aligned} \label{eq:importance-density-recursion}
\end{equation}
where
\begin{equation}
\begin{aligned}
&w^{N+1}(\bigR^{N+1}, \dots, \bigR^0) \equiv \\
&\hspace{10pt} \left[ \frac{\PsiTrial^\dag(\bigR^{N+1}) S(\bigR^{N+1}, \dots, \bigR^0) \PsiTrial(\bigR^0) }{\PsiTrial^\dag(\bigR^N) S(\bigR^N, \dots, \bigR^0) \PsiTrial(\bigR^0)} \right] \, .
\end{aligned}
\end{equation}
The free propagator $G^0_{\dt}(\bigR^{N+1}, \bigR^{N})$ of Eq.~\eqref{eq:kinetic} is easily sampled to obtain $\bigR^{N+1}$ given $\bigR^N$ and $w^{N+1}$ in Eq.~\eqref{eq:importance-density-recursion} is utilized as a re-weighting factor. The initial coordinates $\bigR^0$ are drawn from the distribution proportional to
\begin{equation}
    I(\bigR^0) = \PsiTrial^\dag(\bigR^0) \PsiTrial(\bigR^0) \, .
\end{equation}
by performing a VMC calculation. Note that if $\PsiI \neq \PsiTrial$ is chosen for the importance sampling wave function, then an additional phase factor $I(\bigR^0) / |I(\bigR^0)|$ should be included as a weight. The total re-weighting factor
\begin{align}
    & W(\bigR^{N}, \dots, \bigR^{0}) \equiv \nonumber\\
    & \qquad w^{N}(\bigR^{N}, \dots, \bigR^{0}) \times \dots \times w^{1}(\bigR^{1}, \bigR^{0})
    \label{eq:weight_full}
\end{align}
has been chosen so as to satisfy
\begin{equation} \label{eq:weight-invariant}
    I(\bigR^N, \dots, \bigR^0) = W(\bigR^N, \dots, \bigR^0) \, P(\bigR^N, \dots, \bigR^0)
\end{equation}
at any stage of the sampling, where $P(\bigR^N, \dots, \bigR^0)$ is the probability density associated with the coordinates sampled so far using the free propagator. This choice allows observables as defined in Eq.~\eqref{eq:observable} to be evaluated by a re-weighted estimate,
\begin{equation} \label{eq:observable-mcmc}
    \braket{O} = \frac{\left< W(\bigR^N, \dots, \bigR^0) 
    \frac{\PsiTrial^\dag(\bigR^N) O S(\bigR^N, \dots, \bigR^0) \PsiTrial(\bigR^0)}
    {\PsiTrial^\dag(\bigR^N) S(\bigR^N, \dots, \bigR^0) \PsiTrial(\bigR^0)}
     \right>_P}{ \left< W(\bigR^N, \dots, \bigR^0) \right>_P},
\end{equation}
where $\braket{ \cdot }_P$ denotes the statistical expectation value with respect to the distribution $P$. In practice, this is estimated by averaging according to an ensemble of coordinates sampled by the procedure described above.

To remove the linear terms coming from the exponential of Eq.~\eqref{eq:kinetic} and enforce the symmetry of the Gaussian $G^0_{\dt}$, we apply a \emph{forward-backward heatbath} to simultaneously propose the mirror points $\bigR^{N+1}_\pm = \bigR^N \pm \delta\bigR$,
which have the same weight under $G^0_{\dt}$. The effect of the full importance sampling distribution is then captured by applying a heat bath according to the relative weights
\begin{equation} \label{eq:fwd-bwd-weights}
\begin{aligned}
&w_{\pm} \equiv W(\bigR^{N+1}_{\pm}, \bigR^{N}, \dots, \bigR^0)
\end{aligned}
\end{equation}
to select between the proposals $\bigR^{N+1}_{\pm}$. Since these weights may be non-positive, the selected walker is taken with probability $|w_{\pm}| / (|w_+| + |w_-|)$ and the reweighting factor is multiplied by $\frac{1}{2} \frac{w_{\pm}}{|w_{\pm}|}(|w_+| + |w_-|)$ to maintain Eq.~\eqref{eq:weight-invariant}.
Note that including this resampling step causes the probability distribution $P$ appearing in Eq.~\eqref{eq:observable-mcmc} to differ from a convolution of free propagators $G_{\delta \tau}^0$.

After several imaginary-time steps, many of the walkers diffuse into regions where their weight becomes very small, making practically no contribution to the expectation value of Eq.~\eqref{eq:observable-mcmc}. At this point, only a few walkers make most of the contribution, resulting in increased statistical noise. A ``branching'' algorithm is commonly applied to remedy this problem: walkers with small weights are more likely to be discarded, while those with large weights are replicated~\cite{Pudliner:1997ck,Foulkes:2001}. In this work, we do not apply the branching algorithm, as discarding a walker makes it difficult to apply gradient-based optimization. However, there would in principle be no obstacle to incorporating this step after determining an optimal contour by the gradient-based methods described below.

\section{Contour deformation}\label{sec:contour}

The path integral definition of $\Psi(\tau,\bigR^N)$ in Eq.~\eqref{eq:psidef} can be used as a starting point for contour deformation techniques previously applied to lattice quantum field theory
~\cite{Cristoforetti:2012su,Aarts:2013fpa,Mukherjee:2013aga,Aarts:2014nxa,Schmidt:2017gvu,DiRenzo:2017igr,Kashiwa:2018vxr,Alexandru:2018ngw,Kashiwa:2019lkv,Detmold:2020ncp,Pawlowski:2021bbu,Detmold:2021ulb,Kanwar:2021tkd,Cristoforetti:2013wha,Fujii:2013sra,Cristoforetti:2014gsa,Alexandru:2015xva,Alexandru:2015sua,Fujii:2015vha,Alexandru:2016gsd,Alexandru:2016ejd,Alexandru:2017czx,Alexandru:2017lqr,Mori:2017nwj,Tanizaki:2017yow,Alexandru:2018brw,Alexandru:2018fqp,Alexandru:2018ddf,Mou:2019gyl,Lawrence:2021izu,DiRenzo:2021kcw,Lawrence:2022dba,Mukherjee:2014hsa,Tanizaki:2015rda,Fukuma:2019uot,Fukuma:2019wbv,Ulybyshev:2019fte,Mishchenko:2021vnx,Rodekamp:2022xpf,Alexandru:2020wrj}. 
Under the assumption that $G_{\dt}(\bigR',\bigR)$ and $\PsiTrial(\bigR)$ are holomorphic function of $\bigR$ and $\bigR'$, Eqs.~\eqref{eq:psidef}--\eqref{eq:prop} show that GFMC coordinate integration contours can be deformed with the value of $\Psi(\tau,\bigR^N)$ guaranteed to be conserved through a multi-dimensional generalization of Cauchy's theorem~\cite{Alexandru:2020wrj}.
The kinetic operator
\begin{equation}
    \sum_{i=1}^{A} \frac{\bm{p}_i^2}{2M_N} \Psi(\tau,\bigR^N) = -\sum_{i=1}^{A} \frac{\bm{\nabla}_i^2}{2M_N} \Psi(\tau,\bigR^N)
\end{equation}
is holomorphic if the path integrand is holomorphic, meaning the kinetic term in $H$ does not introduce any complications. Contour deformations are therefore applicable as long as the potential $V(\bigR)$ and trial wavefunction $\PsiTrial(\bigR)$ are holomorphic functions of $\bigR$.
Standard parameterizations of nucleon-nucleon potentials and trial wavefunctions commonly used in GFMC calculations are given in terms of real coordinates $\bigR$, meaning analytic continuation is needed to yield a well-defined and holomorphic integrand.

Although the value of $\Psi(\tau,\bigR^N)$ and therefore GFMC results for observables are independent of the choice of integration contour under these holomorphicity assumptions, the distributions of the coordinates $\bigR^1,\ldots,\bigR^N$ generated during GFMC evolution can be modified by contour deformation, causing the variance of GFMC results for observables to be modified by this procedure.
In particular, sign problems and associated signal-to-noise problems arising from fluctuations in the complex phase or sign of $G_{\dt}(\bigR^n,\bigR^{n-1})$ can be improved (or worsened) by deforming the $\bigR^1,\ldots,\bigR^N$ integration contours.
Contour deformation methods can also be applied to reduce phase fluctuations of noisy GFMC observables in analogy to the methods applied to lattice quantum field theory in Refs.~\cite{Detmold:2020ncp,Detmold:2021ulb}.

The remainder of this section discusses analytic continuation of the GFMC potential and trial wavefunction, parameterization of families of contour deformations of the coordinate integrals, and a method for numerically optimizing the parameters to minimize the variances of GFMC observables of interest.

\subsection{Analytic continuation of GFMC path integrals}\label{sec:analytic}

The AV18 potentials discussed above, as well as chiral EFT potentials that are minimally nonlocal in coordinate space~\cite{Gezerlis:2013ipa,Lynn:2015jua,Piarulli:2014bda,Piarulli:2017dwd},
are standardly written in terms of the magnitude of coordinate differences $|\bm{r}_{ij}| = \sqrt{\bm{r}_{ij}\cdot\bm{r}_{ij}}$, where $\vec{r}_{ij} = \bm{r}_i - \bm{r}_j$.
The presence of the square root would lead to nonholomorphic dependence on $\bigR$ if this is used to define the potential on complexified coordinates.
To instead obtain a holomorphic path integrand over all coordinates, we begin by defining the holomorphic variable $\varrho_{ij} = \bm{r}_{ij} \cdot \bm{r}_{ij}$, which on the base manifold is equal to  $|\bm{r}_{ij}|^2$. This allows us to then define a polynomial function $V_H(\varrho_{ij})$ which approximates the potential on the original real coordinates.

For a finite interval $\varrho_{ij} \in [0, \varrho_{\rm max}]$, an arbitrary function can be described by an infinite series of Chebyshev polynomials.
Our holomorphic approximation to the potential is defined by truncating this series to obtain a polynomial of degree $n_{\rm max}$,
\begin{equation}
  V_H(\varrho_{ij} ) = \sum_{n=0}^{n_{\rm max}} C_n\, T_n^*\left( \frac{ \varrho_{ij} }{ \varrho_{\rm max} } \right) \label{eq:VCheb}
\end{equation}
where the $T_n^*(x) = T_n(2x - 1)$ are shifted versions of the Chebyshev polynomials $T_n(u)$, which can be defined recursively by
\begin{equation}
\begin{aligned}
    T_0(u) &= 1 \\
    T_1(u) &= u \\
    T_{n+1}(u) &= 2uT_n(u) - T_{n-1}(u).
\end{aligned}
\end{equation}
The $C_n$ in Eq.~\eqref{eq:VCheb} are coefficients chosen such that $V_H(\varrho_{ij} ) \approx V(|\bm{r}_{ij}|)$ over the interval $\varrho_{ij} \in [0, \varrho_{\rm max}]$.
In particular, the $C_n$ can be chosen to minimize
\begin{equation}
  \chi^2(C_n) = \sum_k \left[ V_H\left(\varrho_{ij}^k\right) - V\left(\sqrt{\varrho_{ij}^k}\right) \right]^2,
\end{equation}
where the $\varrho_{ij}^k$ are a one-dimensional grid of points inside the interval $[0, \varrho_{\rm max}]$.
For the calculations below, we obtain $C_n$ coefficients using a uniform grid of $\varrho_{ij}^k$ with spacing $0.001\,\mathrm{fm}^2$, $\varrho_{\rm max} = (20\text{ fm})^2$, and $n_{\rm max} = 250$.

By increasing $n_{\rm max}$, the holomorphic approximation $V_H(\varrho_{ij} )$ can be made arbitrarily close to $V(|\bm{r}_{ij}|)$ for real-valued $\bigR$ with $\varrho_{ij} < \varrho_{\rm max}$.
Since only real-valued $\bigR$ are used in performing global fits to extract $V$, for sufficiently large $n_{\rm max}$ and $\varrho_{\rm max}$ calculations using $V_H$ provide equally good fits to nucleon-nucleon phase shifts and other experimental data used to constrain $V$.
However, away from the real axis $V$ and $V_H$ are not constrained to be similar, and in particular $V_H$ does not contain the branch cut singularities related to the appearance of $|\bm{r}_{ij}|$ in $V$. 

For a variational wavefunction $\Psi(|\bm{r}_{ij}|)$ defined in terms of coordinate differences, an identical procedure can be applied to analytically continue to the holomorphic function $\Psi(\varrho_{ij})$. In the numerical results presented below, a Chebyshev fit with identical choices of $\varrho^k_{ij}$, $\varrho_{\mathrm{max}}$, and $n_{\mathrm{max}}$ is used to achieve this analytic continuation for the variational deuteron wavefunction.

The Laplacian required to calculate the kinetic term appearing in $\langle \PsiTrial| \ham | \Psi(\tau)\rangle$ is typically calculated in spherical coordinates involving $|\bm{r}_i|$. This term must therefore also be analytically continued to allow contour deformations to be applied.
This is easily achieved when the spherical-coordinates Laplacian $\nabla^2_i \Psi_H$ is written in terms of $\rho_i \equiv \bm{r}_i \cdot \bm{r}_i$ and the usual spherical angles $\theta_i$ and $\phi_i$ as
\begin{align}
\nabla^2_i \Psi_H = & \ 6 \frac{\partial \Psi_H}{\partial \rho_i} + 4 \rho_i \frac{\partial^2 \Psi_H}{\partial \rho_i^2} \\\nonumber
  &+ \frac{1}{\rho_i \sin\theta_i} \left[  \frac{\partial}{\partial\theta_i}\left( \sin\theta_i \frac{\partial \Psi_H}{\partial\theta_i} \right) + \frac{\partial^2 \Psi_H}{\partial \phi_{i}^2} \right],
\end{align}
which is a holomorphic function of $\rho_i$, $\theta_i$, and $\phi_i$  after multiplying by the spherical coordinate Jacobian $\rho_i \sin\theta_i$.
With this coordinate parameterization and the holomorphic approximations to the potential and wavefunction discussed above, the path integral construction of $\langle \PsiTrial| \ham | \Psi(\tau)\rangle$ discussed in Sec.~\ref{sec:gfmc} has a holomorphic integrand.
Matrix elements of arbitrary local operators $\langle \PsiTrial| \mathcal{O} | \Psi(\tau)\rangle$ and derivative operators have holomorphic descriptions by identical arguments provided that derivatives are taken using $\{\rho_i, \theta_i,\phi_i \}$ coordinates.
Contour deformations can therefore be applied without introducing bias as long as certain conditions regarding the endpoints of the integration contours are met; these are discussed in the next section below.

\subsection{Contour parameterization}

Many different parameterizations of the particle coordinates $\bm{r}_i$ are possible. Within each parameterization, the relevant degrees of freedom may then be complexified and the contour of integration deformed.
Although contours defined using one coordinate parameterization are related to those defined using another by a smooth change of basis, a contour that is simple in one coordinate parameterization could be complicated in another. Different coordinate parameterizations can therefore be practically advantageous for improving signal-to-noise problems of different observables.

One simple choice of coordinate parameterization is to use spherical coordinates for each particle $\{r_i, \theta_i,\phi_i \}$ as base coordinates; the squared coordinates $\rho_i$ and coordinate differences $\varrho_{ij}$ discussed in Sec.~\ref{sec:analytic} are holomorphic function of these coordinates and therefore so are $\nabla_a^2 \Psi_H$, $V_H$, and other GFMC observables.
A contour deformation then corresponds to integrating on a complexified manifold written in terms of complex coordinates $\widetilde{r}_i, \widetilde{\theta}_i, \widetilde{\phi}_i \in \mathbb{C}$.
It is convenient to define a deformed contour of integration by a map from the base coordinates to the deformed coordinates
$\{r_i, \theta_i,\phi_i \} \rightarrow \{\widetilde{r}_i, \widetilde{\theta}_i,\widetilde{\phi}_i \} $. Though such a deformation could in principle be a function of the coordinates of all particles, we restrict to writing the transformed coordinates of particle $i$ as a function of the original coordinates of only particle $i$  itself, such as $\widetilde{r}_i(r_i, \theta_i, \phi_i)$. The measure on the deformed contour of integration for the coordinates of particle $i$ can then be written in terms of the usual spherical coordinates measure as
\begin{equation}
d\widetilde{\bigR} = d\bigR \ \prod_{i=1}^{A} J(r_i, \theta_i, \phi_i) \ M(r_i, \theta_i, \phi_i),
\end{equation}
where $J(r_i, \theta_i, \phi_i)$ is the Jacobian determinant associated with the coordinate map into the manifold,
\begin{equation}
J(r_i, \theta_i, \phi_i) =
\det\begin{bmatrix}
\partial\widetilde{r}_i / \partial r_i & \partial\widetilde{r}_i / \partial \theta_i & \partial\widetilde{r}_i / \partial \phi_i \\
\partial\widetilde{\theta}_i / \partial r_i & \partial\widetilde{\theta}_i / \partial \theta_i & \partial\widetilde{\theta}_i / \partial \phi_i \\
\partial\widetilde{\phi}_i / \partial r_i & \partial\widetilde{\phi}_i / \partial \theta_i & \partial\widetilde{\phi}_i / \partial \phi_i \\
\end{bmatrix}
\end{equation}
and $M(r_i, \theta_i, \phi_i)$ is a measure factor that is given for spherical coordinates by
\begin{equation}
M(r_i, \theta_i, \phi_i) = \frac{\widetilde{r}_i^2 \sin\widetilde{\theta}_i}{r_i^2 \sin\theta_i} = \frac{\widetilde{\rho}_i \sin\widetilde{\theta}_i}{\rho_i \sin\theta_i}.
\end{equation}
The effect of such a contour deformation on GFMC evolution can therefore be described in explicit coordinates by
\begin{equation}
\begin{split}
  d\bigR &= \prod_{i=1}^A dr_i d\theta_i d\phi_i \ r_i^2 \sin\theta_i \\
    \rightarrow d\widetilde{\bigR} &= \prod_{i=1}^A dr_i d\theta_i d\phi_i \ r_i^2 \sin\theta_i \ \mathcal{J}(r_i, \theta_i, \phi_i),
    \end{split} \label{eq:curlyJ}
\end{equation}
for each step of GFMC evolution in Eq.~\eqref{eq:psidef}, where $\mathcal{J}(r_i, \theta_i, \phi_i) \equiv J(r_i, \theta_i, \phi_i) M(r_i, \theta_i, \phi_i)$.

In this parameterization, there are two varieties of angles, the $\theta_i$ which are integrated initially over $[0, \pi/2]$ with distinguished endpoints, and the $\phi_i$ which are integrated initially over $[0, 2\pi]$ with identified endpoints. To guarantee that deformed path integrals exactly agree with their undeformed counterparts by Cauchy's theorem, the complexified integration contours of the $\theta_i$ must preserve the location of the endpoints of the integration interval and the deformed contours of the $\phi_i$ must preserve $2\pi$-periodicity in the real component. The radial coordinates $r_i$ or $\rho_i$ are initially integrated over the non-compact domain $[0, \infty)$. 
Care is therefore required to define the asymptotic behavior of the complexified integration contour at infinity in order to ensure that the value of the integral is not changed. This issue is related to the homology classes of the integration contour~\cite{Alexandru:2020wrj}. Here, we restrict our deformations to contours that asymptotically point in the same direction as the real line at infinity, which is sufficient to guarantee that the integration contour belongs to the same homology class and that the values of all observables are unchanged by contour deformation.

An illustrative example of a simple contour deformation using this spherical coordinate parameterization is the constant shift 
\begin{equation}
\phi_i \rightarrow \widetilde{\phi}_i \equiv \phi_i + i \lambda . \label{eq:phi_shift_def}
\end{equation}
In Cartesian coordinates, this corresponds to 
\begin{equation}
\begin{split}
    (r_i^1,r_i^2,r_i^3) \rightarrow (\widetilde{r}_i^1,\widetilde{r}_i^2,\widetilde{r}_i^3)= & \, (r_i^1\cosh\lambda - i r_i^2\sinh\lambda,\\
    & \; r_i^2 \cosh\lambda + i r_i^1\sinh\lambda, r_i^3). 
    \end{split}
\end{equation}
The resulting modification of the kinetic energy can be included in GFMC calculations by sampling new coordinates $\bigR'$ using the undeformed probability distribution proportional to $G^0_{\delta \tau} (\bigR', \bigR)$ and then reweighting the wavefunction at $\bigR'$ by multiplying by the ratio
\begin{equation}
\begin{split}
    &\frac{G^0_{\delta \tau}(\widetilde{\bigR}', \widetilde{\bigR})}{G^0_{\delta \tau} (\bigR', \bigR)} = \exp\left[ - \frac{(\widetilde{\bigR}^\prime - \widetilde{\bigR})^2 - (\bigR' - \bigR)^2}{2 \delta \tau / M_N} \right] \\
   &\hspace{10pt}= \exp\left[ \frac{ 2 (R^1 R'^1 + R^2 R'^2)(\cosh(\lambda - \lambda') - 1)}{2 \delta \tau / M_N} \right] \\
   &\hspace{20pt} \times \exp\left[-\frac{2 i (R'^1 R^2 - R^1 R'^2) \sinh(\lambda - \lambda')}{2 \delta \tau / M_N} \right]. 
\end{split}
\end{equation}
When $\lambda \neq \lambda'$, the magnitude of this ratio is not equal to one and there is also a non-zero phase.
If  $\lambda - \lambda'$ is sufficiently large, then the fluctuations of the magnitude of this ratio could introduce an overlap problem that spoils the precision of calculations using reweighting. However, for contour deformations in which $\lambda$ is a not too rapidly varying function of $\tau$, reweighting is not problematic even when $\lambda$ is large.
Large magnitude fluctuations arising from products of these ratios over many steps of GFMC evolution can be reduced by standard resampling or branching techniques if necessary~\cite{Pudliner:1997ck}.
The non-zero phase arising from contour deformations must be included as a reweighting factor.
If the wavefunction and potential are real, then this phase will introduce a sign problem and associated signal-to-noise problem to GFMC path integrals. On the other hand, if there are complex phases arising elsewhere in the path integral then it is possible for there to be destructive interference with the phase introduced by contour deformation.
In this case, it is possible for contour deformations to significantly improve sign and signal-to-noise problems arising in GFMC calculations.

The same strategy of reweighting by $G^0_{\delta \tau}(\widetilde{\bigR}', \widetilde{\bigR})/G^0_{\delta \tau} (\bigR', \bigR)$ can be applied for generic contour deformations.
Several qualitative features discussed for the constant shift deformation above apply to the more general case: contour deformations generically give such ratios both non-unit magnitude and non-zero phases that can interfere constructively or destructively with other path integral phases and therefore lead to signal-to-noise degradation or improvement, respectively.    
The construction of contour deformations that have been optimized to achieve destructive phase interference and improve GFMC signal-to-noise using this and other contour deform parameterizations is discussed in Sec.~\ref{sec:optimization}.

A more general class of contour deformations using spherical coordinates can be defined as 
\begin{widetext}
\begin{equation}
    \begin{split}
    r_i & \rightarrow \widetilde{r}_i \equiv r_i + i e^{-r_i^2/(2\sigma^2)} \sum_{m=1}^{\Lambda}  \kappa_{im}^{rr} r_i^m      \left\lbrace 1 + \sum_{n=1}^{\Lambda} \left[ \lambda_{imn}^{r\theta} \sin(2n \theta_i) +  \lambda_{imn}^{r\phi}  \sin(n\phi_i   + \chi_{imn}^{r\phi} ) \right]  \right\rbrace \\
    \theta_i &\rightarrow \widetilde{\theta}_i \equiv  \theta_i + i \sum_{m=1}^{\Lambda}  \kappa_{im}^{\theta \theta} \sin\left( m \theta_i  \right)      \left\lbrace 1 + \sum_{n=1}^{\Lambda} \left[ \lambda_{imn}^{\theta r} e^{-r_i^2 / (2\sigma^2)} r_i^n + \lambda_{imn}^{\theta\phi}  \sin(n\phi_i   + \chi_{imn}^{\theta\phi} ) \right]  \right\rbrace \\
        \phi_i &\rightarrow \widetilde{\phi}_i \equiv  \phi_i + i \kappa_{i0}^{\phi} + i \sum_{m=1}^{\Lambda} \kappa_{im}^{\phi\phi} \sin(m\phi_i + \zeta_{im}^{\phi\phi} ) \left\lbrace 1 + \sum_{n=1}^{\Lambda} \left[ \lambda_{imn}^{\phi r} e^{-r_i^2 /( 2 \sigma^2)}  r_i^n  + \lambda_{imn}^{\phi \theta} \sin \left(n \theta_i   \right)   \right] \right\rbrace
    \end{split}
    \label{eq:spherical_def}
\end{equation}
\end{widetext}
where $\Lambda$ is a hyperparameter controlling the number of tunable parameters in the deformation and $\sigma$, $\kappa_{im}^{rr}$, $\kappa_{im}^{\theta\theta}$, $\kappa_{im}^{\phi\phi}$, $\lambda_{imn}^{r\theta}$, $\lambda_{imn}^{r\phi}$, $\lambda_{imn}^{\theta r}$, $\lambda_{imn}^{\theta \phi}$, $\lambda_{imn}^{\phi r}$, $\lambda_{imn}^{\phi \theta}$, $\zeta_{im}^{\phi \phi}$, $\chi_{imn}^{r \phi}$, and $\chi_{imn}^{\theta \phi}$ are the tunable parameters that can be chosen to specify different contour deformations.
In the limit $\Lambda \rightarrow \infty$ this deformation is expressive enough to describe an arbitrary continuous ``vertical deformation'' in which the integration contour for each variable is shifted by $i$ times a real function. This class of contour deformations has been found to successfully reduce phase fluctuations in several lattice quantum field theory applications~\cite{Alexandru:2017czx,Alexandru:2018fqp,Alexandru:2018ddf,Detmold:2020ncp,Detmold:2021ulb}.
The number of parameters appearing in the definition of this contour  deformation increases with $\Lambda$, and therefore both the expressivity and the practical challenges of finding optimal contour deformations discussed in Sec.~\ref{sec:optimization} increase with $\Lambda$.

The Jacobian $J(r_i,\theta_i,\phi_i)$ for this transformation is straightforward to compute analytically, and the transformation of the measure can then be computed using Eq.~\eqref{eq:curlyJ}.
The form of the deformation resembles a Fourier series for $\theta_i$ and $\phi_i$ that is chosen to preserve periodicity of integrals of $\widetilde{\phi}_i$ and ensure that $\widetilde{\theta}_i$ coincides with $\theta_i$ at the endpoints of the integration. The use of a polynomial series in $r_i$ times a Gaussian ensures that the $r_i=0$ integration endpoint is preserved and that $\widetilde{r}_i$ smoothly approaches $r_i$ as $r_i \rightarrow \infty$.
Other functions satisfying these constraints could also be used to parameterize contour deformations, for example appropriately constrained neural nets, but detailed studies of the practical advantages of different parameterizations are deferred to future work.

We define a second class of deformed contours that start from Cartesian rather than spherical coordinate descriptions of integrals of $\vec{r}_i$.
This parameterization is found below to be practically advantageous for improving the signal-to-noise of the deuteron Euclidean density response.
A simple yet effective deformation in Cartesian coordinates is given by the constant shift,
\begin{equation}
  (r_i^1 , r_i^2 , r_i^3) \rightarrow (\widetilde{r}_i^1,\widetilde{r}_i^2,\widetilde{r}_i^3)= (r_i^1  + i \lambda_1, r_i^2  + i\lambda_2, r_i^3 + i\lambda_3), \label{eq:cartesian_def}
\end{equation}
in terms of a vector of tunable parameters $\vec{\lambda}$. Despite the simplicity of this contour deformation in Cartesian coordinates, it cannot be straightforwardly constructed using finite truncations of the spherical coordinate deformations discussed above, making this a useful complementary parameterization to study.

More general contour deformations in which the deformed variables at a given step of GFMC evolution depend on other particle coordinates at that same step or on the values of the coordinates at other steps are also possible but are not explored in this work for simplicity.

\subsection{Signal-to-noise optimization}\label{sec:optimization}
Contour deformations as defined above allow modifying the GFMC signal-to-noise without compromising exactness. Our aim is to then optimize the choice of contour deformation, in particular the choice of parameters defining the contour, to maximize the signal-to-noise ratio. In the following, a stochastic gradient descent procedure is defined analogously to prior applications of contour deformations for path integrals~\cite{Alexandru:2017czx,Mori:2017nwj,Alexandru:2018fqp,Alexandru:2018ddf,Detmold:2020ncp,Detmold:2021ulb,Lawrence:2021izu,Rodekamp:2022xpf,Lawrence:2022dba}.

Deforming the path integral definition of a GFMC observable has two effects:
\begin{enumerate}
    \item The importance sampling weight is modified to
    \begin{equation} \label{eq:deformed-I}
    \begin{aligned}
        \widetilde{I}(\bigR^N, \dots, \bigR^0) &= \mathcal{J}(\bigR^N, \dots, \bigR^0; \Theta) \\
        &\times \prod_{n=0}^{N-1}  G^0_{\dt}(\widetilde{\bigR}^{n+1}, \widetilde{\bigR}^{n}) \\
        &\times \PsiTrial^\dag(\widetilde{\bigR}^N) S(\widetilde{\bigR}^N, \dots, \widetilde{\bigR}^0) \PsiTrial(\widetilde{\bigR}^0),
        \hspace{-35pt} \end{aligned}
    \end{equation}
    where $\widetilde{\bigR}^k = \widetilde{\bigR}^k(\bigR^k, \dots, \bigR^0; \Theta)$ is the deformed coordinate at the $k$-th GFMC step and $\mathcal{J}(\bigR^N, \dots, \bigR^0; \Theta)$ is the collective Jacobian of the integration contour, both given as functions the parameters $\Theta$ defining the contour and all prior coordinates.
    \item The observable under study is also evaluated using the deformed coordinates,
    \begin{equation}
        \PsiTrial^\dag(\widetilde{\bigR}^N) O S(\widetilde{\bigR}^N, \dots, \widetilde{\bigR}^0) \PsiTrial(\widetilde{\bigR}^0).
    \end{equation}
    When $O$ depends on the coordinates of the wavefunctions themselves, these coordinates must also be replaced with the deformed versions.
\end{enumerate}
The sampling procedure described in Sec.~\ref{sec:gfmc} can be applied to the importance weight $\widetilde{I}$ defined in Eq.~\eqref{eq:deformed-I} by additionally incorporating the ratio
\begin{equation}
    \widetilde{I}(\bigR^N, \dots, \bigR^0) / I(\bigR^N, \dots, \bigR^0)
\end{equation}
into the weights carried forward with each walker in the Monte Carlo evaluation. Weights including this ratio will be denoted by $\widetilde{W}(\bigR^N, \dots, \bigR^0)$. Using resampling between the forward and backward steps described in that section implicitly incorporates the deformation into the sampling distribution due to the dependence of the weights in Eq.~\eqref{eq:fwd-bwd-weights} on $\widetilde{I}$, resulting in sampling according to a new distribution which we denote by $\widetilde{P}$.
The evaluation of observable $\braket{O}$ is then given as a statistical expectation value in the deformed case by (cf.~Eq.~\eqref{eq:observable-mcmc})
\begin{equation} \label{eq:deformed-observable-mcmc}
    \braket{O} = \frac{\left< \widetilde{W}(\bigR^N, \dots, \bigR^0) 
    \frac{\PsiTrial^\dag(\widetilde{\bigR}^N) O S(\widetilde{\bigR}^N, \dots, \widetilde{\bigR}^0) \PsiTrial(\widetilde{\bigR}^0)}
    {\PsiTrial^\dag(\widetilde{\bigR}^N) S(\widetilde{\bigR}^N, \dots, \widetilde{\bigR}^0) \PsiTrial(\widetilde{\bigR}^0)}
     \right>_{\widetilde{P}}}{ \left< \widetilde{W}(\bigR^N, \dots, \bigR^0) \right>_{\widetilde{P}}}.
\end{equation}

The variance of a deformed GFMC observable is in general a complicated function of the distribution of the numerator and denominator of Eq.~\eqref{eq:deformed-observable-mcmc}. 
The numerator depends on the observable at hand, while the denominator is universal.
Formally, a stochastic estimate of the ratio may result in an infinite variance when noncentral potentials are considered, due to rarely sampled configurations of walkers yielding exactly zero in the denominator; see for example Ref.~\cite{Shi:2015lyu} for a discussion of this issue in the context of fermionic Quantum Monte Carlo in several systems. This formal issue, as well as the need to define a differentiable function measuring the statistical noise of an observable, motivates constructing one of several possible alternative ``loss functions'' to be minimized as a proxy for the full variance of a specific observable.

For example, given an observable $O$, a linear combination of the logs of the second moments of the numerator and denominator (including both real and imaginary pieces) provides one useful loss function,

\begin{widetext}

\begin{equation} \label{eq:loss-one-O}
\begin{aligned}
    \mathcal{L}_{O}(\alpha) &\equiv
    (1-\alpha) \log \Big\langle
    \left| \widetilde{W}(\bigR^N, \dots, \bigR^0) \right|^2
    \Big\rangle_{\widetilde{P}}  \\
    &\quad + \alpha \log \Big\langle
    \left| \widetilde{W}(\bigR^N, \dots, \bigR^0) 
    \frac{\PsiTrial^\dag(\widetilde{\bigR}^N) O S(\widetilde{\bigR}^N, \dots, \widetilde{\bigR}^0) \PsiTrial(\widetilde{\bigR}^0)}
    {\PsiTrial^\dag(\widetilde{\bigR}^N) S(\widetilde{\bigR}^N, \dots, \widetilde{\bigR}^0) \PsiTrial(\widetilde{\bigR}^0)} \right|^2
    \Big\rangle_{\widetilde{P}}.
\end{aligned}
\end{equation}
Here the second moments correspond to the non-holomorphic parts of the corresponding variances, and the logarithm ensures that the loss function has relatively uniform gradients even if the variance is potentially modified by several orders of magnitude during optimization. A natural choice of linear combination is given by $\alpha = 1/2$.

The loss function above is specific to one choice of observable, including implicitly one choice of $\tau$, but one is often interested in measuring several choices of observables given a single GFMC evaluation. In particular, it is frequently useful to measure an operator across many or all values of $\tau$ accessed from an evaluation of the path integral. To reduce variance of a given observable for multiple choices of $\tau$, it is straightforward to generalize the loss function in Eq.~\eqref{eq:loss-one-O} to average the loss across a range of values of $\tau$, giving the more general loss function
\begin{equation} \label{eq:general-loss-function}
\begin{aligned}
    \mathcal{L}_{O}(\alpha, \tau_{\mathrm{min}}, \tau_{\mathrm{max}}) &\equiv \frac{\dt}{\tau_{\mathrm{max}} - \tau_{\mathrm{min}}} \sum_{N = \tau_{\mathrm{min}}/\dt}^{\tau_{\mathrm{max}}/\dt - 1} \Big[  (1-\alpha) \log \Big\langle
    \left| \widetilde{W}(\bigR^N, \dots, \bigR^0) \right|^2
    \Big\rangle_{\widetilde{P}} \\
    &\hspace{115pt} + \alpha \log \Big\langle
    \left| \widetilde{W}(\bigR^N, \dots, \bigR^0) 
    \frac{\PsiTrial^\dag(\widetilde{\bigR}^N) O S(\widetilde{\bigR}^N, \dots, \widetilde{\bigR}^0) \PsiTrial(\widetilde{\bigR}^0)}
    {\PsiTrial^\dag(\widetilde{\bigR}^N) S(\widetilde{\bigR}^N, \dots, \widetilde{\bigR}^0) \PsiTrial(\widetilde{\bigR}^0)} \right|^2
    \Big\rangle_{\widetilde{P}} \Big].
\end{aligned}
\end{equation}

\end{widetext}

Both families of loss functions described above consist of a linear combination of various expectation values with respect to the distribution $\widetilde{P}$ sampled by the GFMC procedure. A stochastic estimate of the gradient of the loss function with respect to the parameters $\Theta$ defining the contour can thus be estimated by performing GFMC sampling of a set of walkers under the distribution $\widetilde{P}$, then using autodifferentiation techniques over the stochastic evaluation of each expectation value. The loss function may then be minimized as a function of the contour deformation parameters by stochastic gradient descent using these estimates.

For the numerical results presented below, this minimization is performed using the Adam optimizer~\cite{Kingma:2014} with hyperparameters except for the step size set to their default values. The step size is initially set to $10^{-3}$ for optimization of the energy signal-to-noise ratio and to $10^{-2}$ for the response functions. During optimization, a $50$-step-averaged measurement of the loss function is tracked and the step size is multiplied by a factor of $0.3$ whenever no improvement is seen for $250$ consecutive steps. In all cases, optimization is terminated after the step size decreases twice.

The gradients as defined above are evaluated with respect to holding $\widetilde{P}$ as a fixed importance sampling distribution independent of the deformation. In principle, this distribution will also be modified by changing the contour deformation parameters at each step of a gradient descent procedure. However, the resampling steps used in GFMC are designed to reduce variance with respect to the importance sampling weight, meaning each implicit update to $\widetilde{P}$ can be expected to also reduce the observable-independent variance appearing as the first term in the definition of $\mathcal{L}_{O}$ given in Eq.~\eqref{eq:loss-one-O}. The second term of Eq.~\eqref{eq:loss-one-O} may in principle be increased by such a choice of importance sampling scheme, but the two terms are correlated due to the common factor of $\widetilde{W}(\bigR^N, \dots, \bigR^0)$, suggesting this should rarely be the case if the observable factor does not include significant magnitude fluctuations. Investigating definitions of the loss function that exploit the reparameterization trick~\cite{Kingma:2013} to take gradients with respect to the importance sampling distribution is deferred to future work.

In Ref.~\cite{Detmold:2021ulb}, transfer learning was also found to improve optimization time and yield better final values of the signal-to-noise ratio. A similar approach can be applied in the present context of GFMC calculations by first optimizing deformation parameters for GFMC evaluations over a smaller range of imaginary time $\tau$, then using the resulting parameters as the initialization for subsequent optimization on evaluations with an increasingly large range of $\tau$. For the results in this work, this approach was not found to significantly accelerate optimization or yield improved signal-to-noise ratios and in some cases even resulted in slower training.
As such the results given here are based only on the simpler scheme of directly optimizing parameters for the target number of GFMC steps.

\section{Results}\label{sec:results}

In this work, numerical results are restricted to the simplest multi-nucleon bound state, the deuteron, in order to most straightforwardly give a proof-of-principle demonstration of the method. The results of signal-to-noise optimization are shown below for calculations of the binding energy and Euclidean density response. The former is a relatively noise-free observable, but small improvements are demonstrated using deformations in a spherical coordinate parameterization. In the latter, a significant sign problem is mitigated by application of these methods using a Cartesian coordinate parameterization.

\subsection{Energies}\label{sec:energy}

\begin{figure*}
    \includegraphics[width=0.47\textwidth]{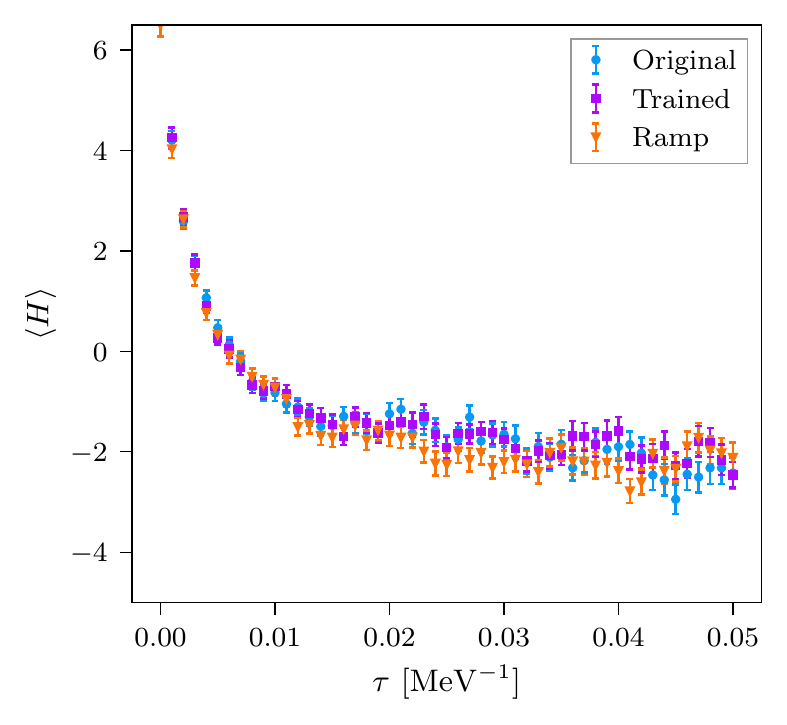}
    \includegraphics[width=0.47\textwidth]{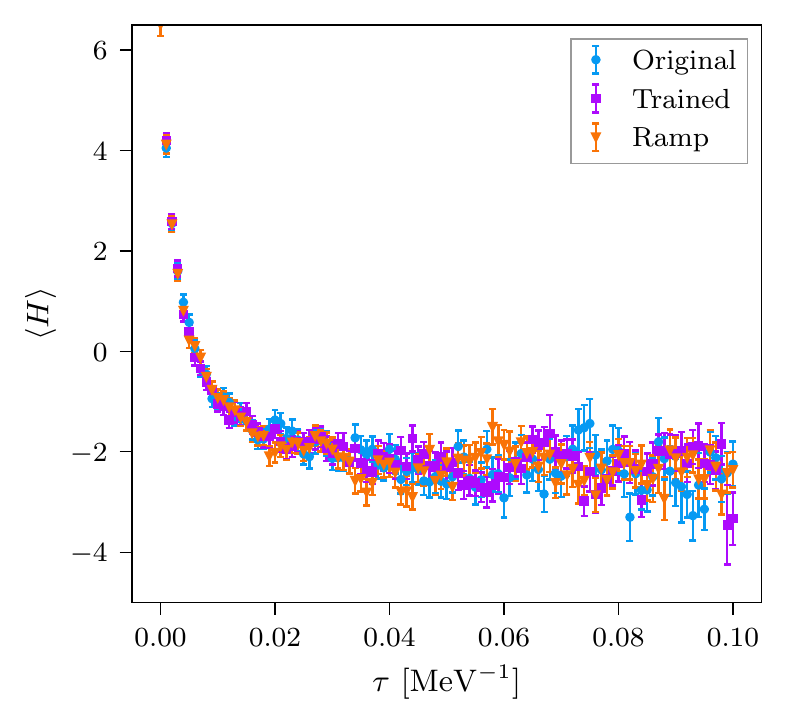}
    \caption{Comparison of the deuteron binding energy evaluated without contour deformation (``original'') versus evaluations using the numerically optimized spherical contour deformation described in the main text (``trained'') as well as a manually selected linear ramp contour (``ramp''), shown for two choices of $N \dt$. Expectation values are evaluated using $N=10,000$ walkers in all cases.}
    \label{fig:H_comparison}
\end{figure*}

The binding energy of the deuteron can be determined using GFMC methods by evaluating the phenomenological nuclear Hamiltonian on an imaginary-time-evolved deuteron state. Variational optimization can yield numerically exact trial states $\ket{\PsiTrial}$ for the deuteron, making the imaginary-time evolution of marginal benefit in typical applications.
To mimic the noise problems seen when computing the binding energy of larger nuclei---whose exact wave functions are not known with the same accuracy as for the deuteron---we here choose to construct an intentionally poor trial wavefunction.
For simplicity, we use a variationally optimized trial wavefunction for the AV4P potential while evolving the system with the more physical AV6P potential. As the AV4P potential does not incorporate noncentral interactions, the structure of the trial wavefunction is quite distinct from the ground state of the AV6P Hamiltonian, in particular taking an unphysical factorized form
\begin{equation} \label{eq:AV4P-trial}
    \PsiTrial(\vec{r}_1, \vec{r}_2) = \frac{1}{\sqrt{2}} \left( \ket{p\uparrow n \uparrow} - \ket{n \uparrow p \uparrow} \right) \times f(|\vec{r}_{12}|),
\end{equation}
where the prefactor indicates the spin-isospin wavefunction.
As shown in Fig.~\ref{fig:H_comparison}, a significant amount of imaginary-time evolution, $\tau \gtrsim 0.03\,\mathrm{MeV}^{-1}$, is then required to converge towards the ground state of the AV6P potential and acquire an estimate of the binding energy.

In the limit of large $\tau$, steadily growing statistical noise sets in, motivating the application of contour deformations for more precise measurements. This noise can be attributed to a combination of fluctuations in the reweighting factors $W(\bigR^N, \dots, \bigR^0)$ and fluctuations in the observable $H$ itself. In particular, a sign problem is expected in the reweighting factors at large $\tau$ because of the presence of the tensor operators
\begin{equation}
    O^5_{12} =  S_{12}(\vec{r}_{12})
    \quad \text{and} \quad
    O^6_{12} = S_{12}(\vec{r}_{12}) (\tau_1 \cdot \tau_2)
\end{equation}
which result in coordinate-dependent spin flips, whereas the spin-isospin factor of the trial wavefunction in Eq.~\eqref{eq:AV4P-trial} is an eigenvector of the operators $O^1_{12}$, \dots, $O^4_{12}$.

\begin{figure*}
\includegraphics[width=0.47\textwidth]{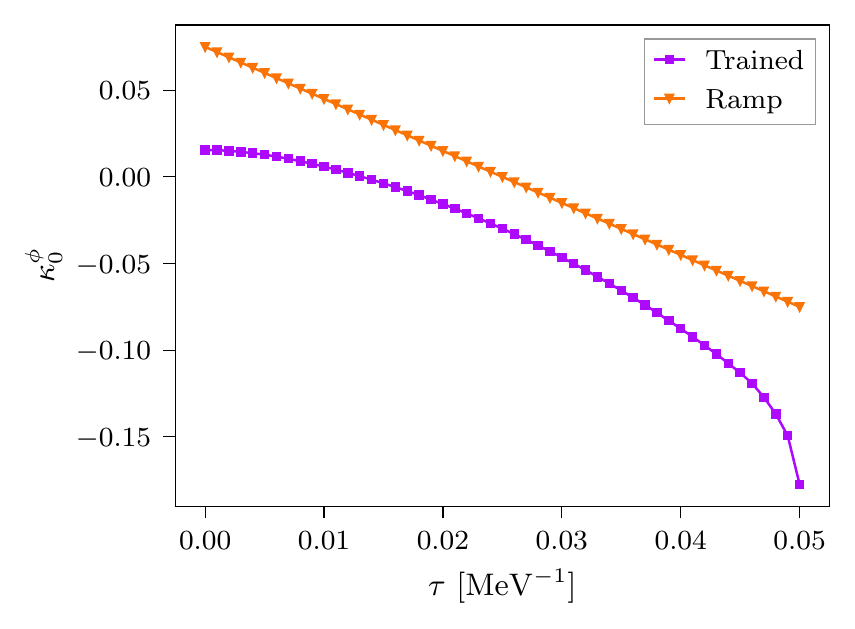}
\includegraphics[width=0.47\textwidth]{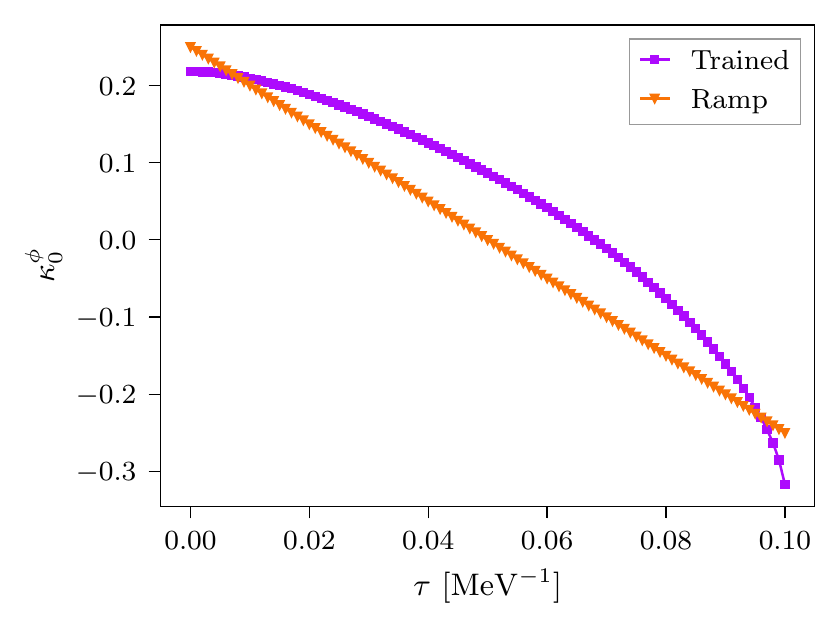}
\caption{Comparison of the numerically optimized shift parameter $\kappa_0^{\phi}$ (``trained'') as a function of the imaginary time $\tau$ versus the manually selected linear ramp contour $\kappa_0^{\phi} = (\tau - N \dt /2) \ell$ (``ramp''). Results are shown for two choices of $N \dt$.}
\label{fig:H_comparison_ramps}
\end{figure*}

\begin{figure*}
    \includegraphics[width=0.47\textwidth]{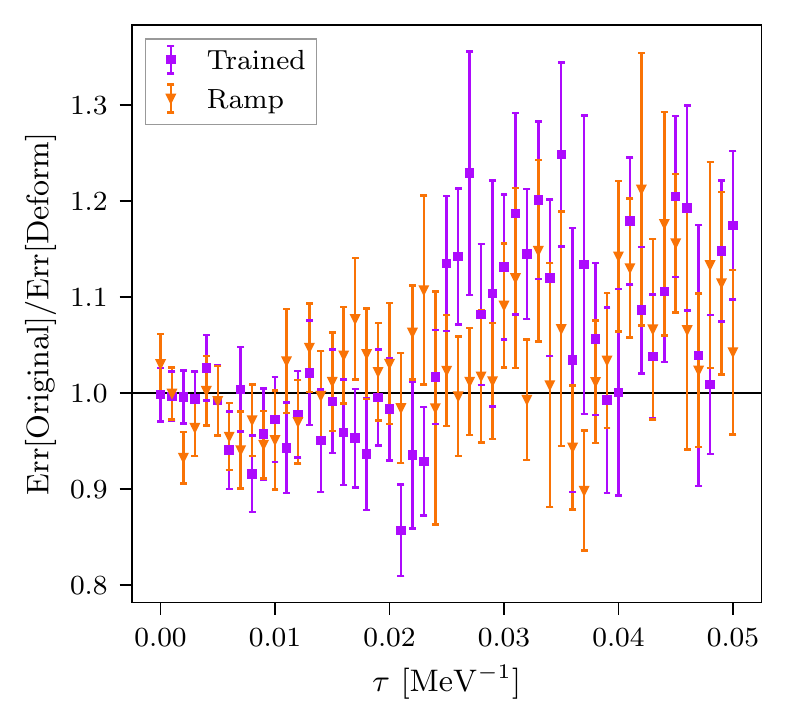}
    \includegraphics[width=0.47\textwidth]{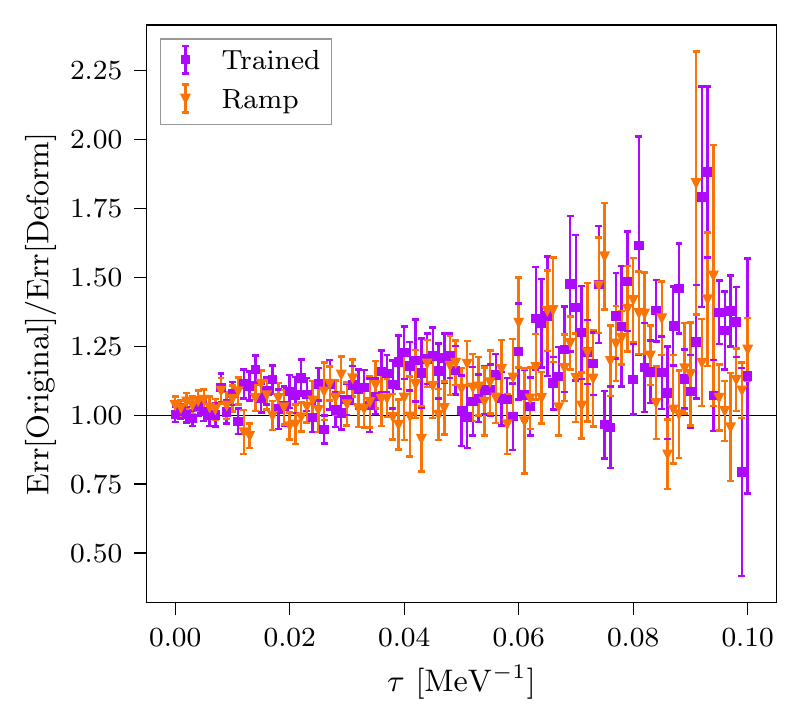}
    \caption{Comparison of the standard error of the deuteron binding energy measurements, given as a ratio between the original measurements and the measurements using contour deformation. The results are compared for the numerically optimized spherical contour deformation (``trained'') and the manually selected linear ramp contour (``ramp''), shown for two choices of $N \dt$. The ratios are estimated by computing the ratio of uncertainties determined by a bootstrap resampling procedure applied to an ensemble of $N=10,000$ walkers, with the uncertainty on the ratio estimated by resampling the ensemble itself in an outer bootstrap resampling step.}
    \label{fig:H_comparison_noise}
\end{figure*}

To understand the origin of this possible sign problem better, we can explicitly evaluate the spin-flipping part of the matrix $S_{12}(\vec{r}_{12})$,
\begin{equation}
    \hat{S}_{12}(\vec{r}_{12}) \equiv ({\bm \sigma}_1\cdot \hat{r}_{12}) \otimes ({\bm \sigma}_2\cdot \hat{r}_{12}).
\end{equation}
It is informative to work in the center-of-mass frame and evaluate in terms of the coordinate $\vec{r}_{12} = 2 \vec{r}_1 \equiv (r, \theta, \phi)$ in spherical coordinates, resulting in
\begin{equation}
\begin{aligned}
    \hat{S}_{12}(r, \theta, \phi) &= \begin{pmatrix}
    \cos{\theta} & \sin{\theta} e^{-i \phi} \\
    \sin{\theta} e^{i \phi} & -\cos{\theta}
    \end{pmatrix} \\
    &\qquad \otimes
    \begin{pmatrix}
    \cos{\theta} & \sin{\theta} e^{-i \phi} \\
    \sin{\theta} e^{i \phi} & -\cos{\theta}
    \end{pmatrix}.
\end{aligned}
\end{equation}
Spin flips are induced by the off-diagonal terms $\sin{\theta} e^{i\phi}$ and $\sin{\theta} e^{-i \phi}$, and in both cases pick up a coordinate-dependent phase.

This structure motivates the parameterization of a contour deformation in terms of spherical coordinates. For example, the simple choice of deformation by an imaginary shift $\widetilde{\phi} = \phi + i \lambda$ modifies this term to
\begin{equation}
\begin{aligned}
\hat{S}_{12}(r, \theta, \widetilde{\phi}) &= \begin{pmatrix}
    \cos{\theta} & \sin{\theta} e^{-i \phi}e^{\lambda} \\
    \sin{\theta} e^{i \phi}e^{-\lambda} & -\cos{\theta}
    \end{pmatrix} \\
    &\qquad \otimes
    \begin{pmatrix}
    \cos{\theta} & \sin{\theta} e^{-i \phi}e^{\lambda} \\
    \sin{\theta} e^{i \phi}e^{-\lambda} & -\cos{\theta}
    \end{pmatrix} \\
    &= X^{-1}(\lambda) \; \hat{S}_{12}(r, \theta, \phi) \; X(\lambda)
\end{aligned}
\end{equation}
where
\begin{equation}
    X(\lambda) \equiv \begin{pmatrix}
        e^{-\lambda / 2} & 0 \\ 0 & e^{\lambda / 2}
    \end{pmatrix} \otimes
    \begin{pmatrix}
        e^{-\lambda / 2} & 0 \\ 0 & e^{\lambda / 2}
    \end{pmatrix}.
\end{equation}
The effect on our trial wavefunction starting in the $\ket{\uparrow \uparrow}$ spin state is therefore to suppress spin flips, while correspondingly affecting the magnitude of the average reweighting factors. Together, the resulting expectation values must remain correct by Cauchy's theorem.

When considering the composition of these operators over multiple imaginary-time steps, an enhanced effect is possible if the deformation $\lambda$ is allowed to depend on $\tau$ as well. For example, we can consider the effect of the simple linear ``ramp'' structure,
\begin{equation}
    \lambda(\tau) = (\tau - N \dt /2) \ell.
\end{equation}
Such a deformation results in insertions of $X(\ell \dt)$ between each appearance of $\hat{S}_{12}$ in the matrix $S(\bigR^{N}, \dots, \bigR^0)$ involved in the GFMC estimation of the importance sampling weight and observable. This suppresses spin flips throughout the imaginary-time evolution, suggesting it may be useful in improving the signal-to-noise ratio for GFMC evaluation of observables at large imaginary time.

The more general spherical-coordinates parameterization of the contour deformation given in Eq.~\eqref{eq:spherical_def} includes this $\tau$-dependent shift of $\phi$ as a special case when the cutoff on Fourier modes is fixed to $\Lambda = 0$, leaving only the $\kappa^\phi_0$ term. We thus proceed by studying the effect of numerically optimizing the parameters in this definition, comparing against a manually selected choice of ramp contour with good signal-to-noise properties. Results for the spherical parameterization with $\Lambda = 1$ do not indicate an improvement over results with $\Lambda = 0$, and as such the following results are restricted to the simpler parameterization with $\Lambda = 0$.

The numerically optimized (``trained'') contour and ramp contour with best signal-to-noise properties are shown in Fig.~\ref{fig:H_comparison_ramps} for two choices of $N \dt$. The average slope as a function of $\tau$ of the imaginary shifts in the trained contour can be seen to be quite similar to the best-performing ramp. Measurements of the binding energy using these contours are compared to the original measurement in Fig.~\ref{fig:H_comparison}. Though the effect is small, the trained and ramp contours result in a more precise estimate of $\left< H \right>$ at large values of $\tau$ in both cases. Though guaranteed by Cauchy's theorem, this figure also confirms the unbiased nature of the deformed measurements. Finally, a quantitative comparison of the standard error of the binding energy measurements is shown in Fig.~\ref{fig:H_comparison_noise}. A trend of increasing improvements can be seen, though the measurement of the errors themselves are quite noisy. At the largest values of $\tau$, we estimate a reduction of the error by $20$--$30\%$.

\subsection{Response functions}\label{sec:response}

Euclidean response functions are important GFMC observables in many contexts, in particular calculations of lepton-nucleus scattering cross sections where they are used to calculate real-time scattering cross sections after applying inverse Laplace transformation methods~\cite{Lovato:2016gkq,Lovato:2017cux,Raghavan:2020bze}.
The response function for a generic pair of momentum-space current operators $J(\vec{q})$ and $J'(\vec{q})$,
\begin{equation}
  \mathcal{R}_{J'J}(\tau,\vec{q}) =  \frac{ \bra{\PsiTrial}    J^{'\dagger}(\vec{q}) e^{-H\tau} J(\vec{q}) \ket{\PsiTrial} }
  {\bra{\PsiTrial}    e^{-H\tau} \ket{\PsiTrial}},
\end{equation}
will have an exponentially severe signal-to-noise problem because of the imaginary-time-evolution operator appearing between the current insertions.
Since $J \ket{\PsiTrial}$ will generically have much worse ground-state overlap than the variationally optimized trial wavefunction $\ket{\PsiTrial}$,
this signal-to-noise problem will be more severe than Hamiltonian matrix elements involving the same imaginary time.
Along with the need for high-precision determinations of response functions for applications to neutrino-nucleus cross section predictions~\cite{Alvarez-Ruso:2014bla,DUNE:2015lol,NuSTEC:2017hzk,Meyer:2022mix,Ruso:2022qes,Simons:2022ltq} and other processes such as superallowed $\beta$-decay rates~\cite{Seng:2018qru,Hardy:2020qwl}, this makes response functions particularly interesting observables for studying the performance of contour deformation techniques. 

As a proof of principle, this section studies the application of contour deformations to the deuteron density response function
\begin{equation}
  \rho(\vec{q},\tau) = \frac{\bra{\PsiTrial} N^\dagger(\vec{q}) e^{-H\tau} N(\vec{q}) \ket{\PsiTrial}}{\bra{\PsiTrial} e^{-H\tau} \ket{\PsiTrial}},
\end{equation}
where $N(\vec{q}) = \int d^3r \ e^{-i\vec{q}\cdot \vec{r}}\, \psi^\dagger(\vec{r}) \psi(\vec{r})$ is the nucleon number operator.
This has the GFMC path integral representation
\begin{align}
    \rho(\vec{q},\tau) &= \sum_{i,j} \rho_{ij}(\vec{q},\tau), \\
    \label{eq:r_gfmc} \rho_{ij}(\vec{q},\tau) &= \frac{1}{Z}\int \prod_{n=0}^{N-1} d\bigR^n \left[ e^{i \vec{q}\cdot (\vec{r}^N_i - \vec{r}^0_j)} \right] \\\nonumber
    &\hspace{20pt} \times I(\bigR^N, \dots, \bigR^0) \\\nonumber
    &= \left< e^{i \vec{q}\cdot (\vec{r}^N_i - \vec{r}^0_j)}\,I(\bigR^N, \dots, \bigR^0) \right>,
\end{align}
where
\begin{equation}
  Z = \int \prod_{n=0}^{N-1} d\bigR^n \  I(\bigR^N, \dots, \bigR^0).
\end{equation}

The phase factors shown explicitly in Eq.~\eqref{eq:r_gfmc} give rise to a sign problem for GFMC response functions with $\vec{q} \neq \vec{0}$.
The expectation value of $\rho(\vec{q},\tau)$ is expected to scale with $\vec{q}^2$ and $\tau$ roughly as $e^{-\tau \vec{q}^2 / 4 M_N}$, where the form of the exponential is specific to the case of the deuteron without subtraction of the elastic contributions~\cite{Carlson:2001mp}. An exponentially decaying path integral mean must therefore arise from the precise cancellation of these fluctuating phase factors and one expects a signal-to-noise problem that is exponentially severe in $\vec{q}^2$; this expectation is confirmed in the numerical results below.
This sign and signal-to-noise problem is a generic consequence of the $\vec{q}$-dependent phase factor in the definition of the response function and its presence does not depend on the type of currents considered, although its practical severity might.

\begin{figure*}[!ht]
	\includegraphics[width=0.47\textwidth]{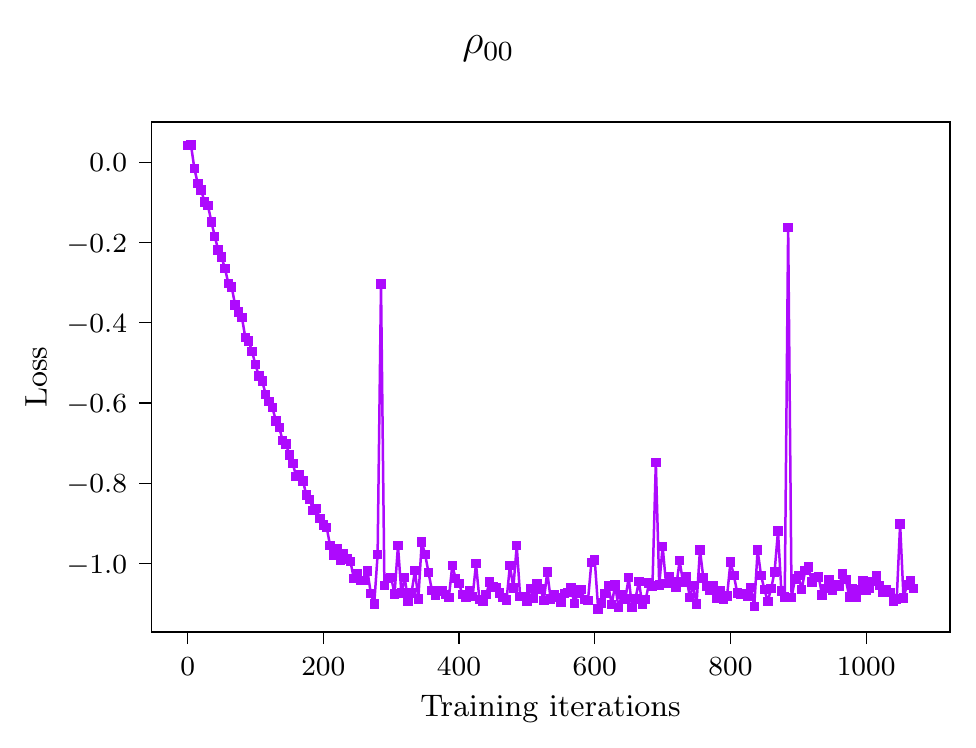}
	\includegraphics[width=0.47\textwidth]{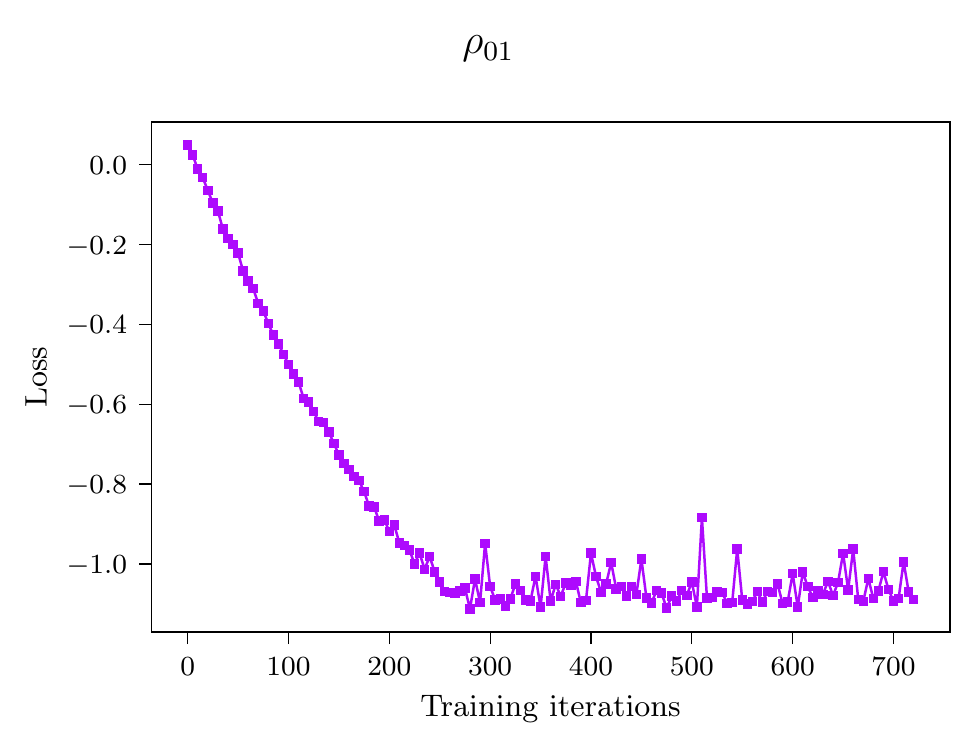}
	\includegraphics[width=0.47\textwidth]{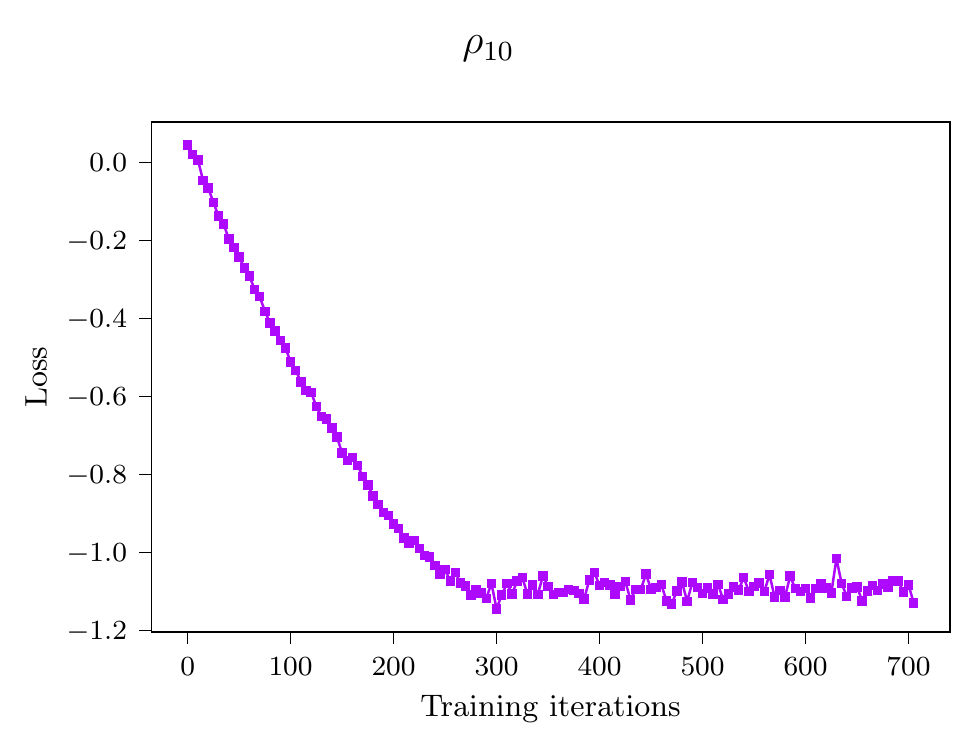}
	\includegraphics[width=0.47\textwidth]{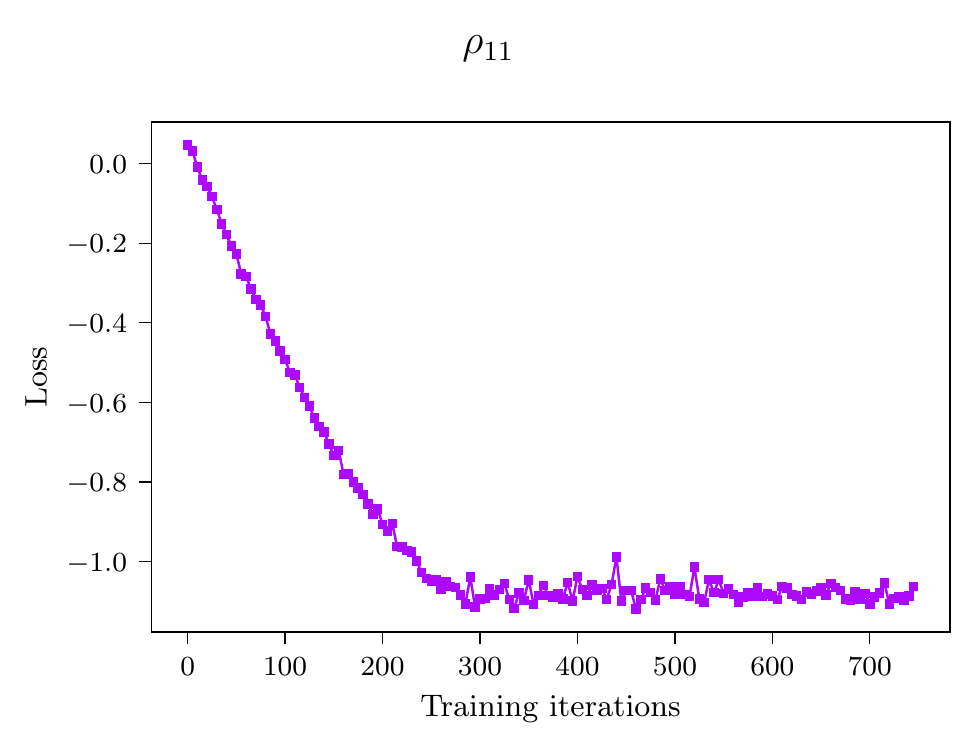}
   \caption{Loss curves from training the numerically optimized contours for each component of $\rho_{ij}$ shown in Fig.~\ref{fig:response_comparison_ramps}. \label{fig:response_loss} }
\end{figure*}

This analysis of the source of sign and signal-to-noise problems in Euclidean response functions suggests that a simple constant deformation in Cartesian coordinates could exponentially decrease their severity.
We define constant Cartesian shifts $\vec{\lambda}_{ij}(\vec{q},\tau)$ for each component of $\rho_{ij}(\vec{q},\tau)$ separately. Their action on GFMC coordinates $\vec{r}_1^n$ and $\vec{r}_2^n$ can be defined for the deuteron as
\begin{equation}
  \begin{split}
    \vec{r}_1^n &\rightarrow \widetilde{\vec{r}}_1^n \equiv \vec{r}_1^n + i \vec{\lambda}_{ij}(\vec{q},\tau), \\
    \vec{r}_2^n &\rightarrow \widetilde{\vec{r}}_2^n \equiv \vec{r}_2^n - i \vec{\lambda}_{ij}(\vec{q},\tau), \\
  \end{split}
\end{equation}
where the opposite-sign deformations of the $i$-th and $j$-th particle coordinates are chosen to preserve the center-of-mass condition $\sum_i \bm{r}_i^n = \vec{0}$ after contour deformation.
Choosing the deformation parameters to scale as $\vec{\lambda}_{00}(\vec{q},\tau) \sim \vec{q} \, \tau / 4M_N$ and $\vec{\lambda}_{00}(\vec{q},0) \sim - \vec{q} \, \tau / 4M_N$ would lead to a decrease in the magnitude of $\rho_{00}(\vec{q},\tau)$ scaling as $e^{i \vec{q} \cdot i \vec{\lambda}_{00}(\vec{q},\tau)} = e^{- \tau \vec{q}^2 / 4 M_N}$ for each sample of the Monte Carlo evaluation.
This matches the expected scaling of the central value of $\rho_{00}(\vec{q}, \tau)$ without the need for any strong cancellation between Monte Carlo samples from phase fluctuations, suggesting that the resulting estimate would be nearly free of the original sign problem.
A similar decrease in the magnitude of $\rho_{11}(\vec{q},\tau)$ can be achieved with an opposite sign shift with $\vec{\lambda}_{11}(\vec{q},\tau) \sim -\vec{q} \, \tau / 4M_N$ and $\vec{\lambda}_{11}(\vec{q},0) \sim \vec{q} \, \tau / 4M_N$. On the other hand, for $\rho_{12}(\vec{q}, \tau)$ and $\rho_{21}(\vec{q}, \tau)$, a decrease in the magnitude can be achieved by shifts
\begin{equation}
\begin{aligned}
    \vec{\lambda}_{10}(\vec{q}, \tau) &\sim \vec{\lambda}_{10}(\vec{q}, 0) \sim \vec{q} \, \tau / 4M_N \\
    \vec{\lambda}_{01}(\vec{q}, \tau) &\sim \vec{\lambda}_{01}(\vec{q}, 0) \sim -\vec{q} \, \tau / 4M_N.
\end{aligned}
\end{equation}
The presence of other sources of variance arising for example from phases introduced to the kinetic-energy evolution factors mean that these precise values of $\vec{\lambda}$ may not be the optimal value for minimizing the variance of the density response, but they suggest that simple one-parameter deformations of the form $\vec{\lambda}_{00}(\vec{q},\tau) \propto (\tau - N \delta \tau / 2) \vec{q}$ and $\vec{\lambda}_{11}(\vec{q},\tau) \propto -(\tau - N \delta \tau / 2) \vec{q}$ which are antisymmetric about the midpoint of GFMC evolution might be a useful family of contour deformations for minimizing signal-to-noise problems arising from phase fluctuations in the diagonal elements of the response function. It is less clear how to treat the off-diagonal elements, but a symmetric ansatz inspired by numerical optimization is adopted below.

\begin{figure*}
	\includegraphics[width=0.47\textwidth]{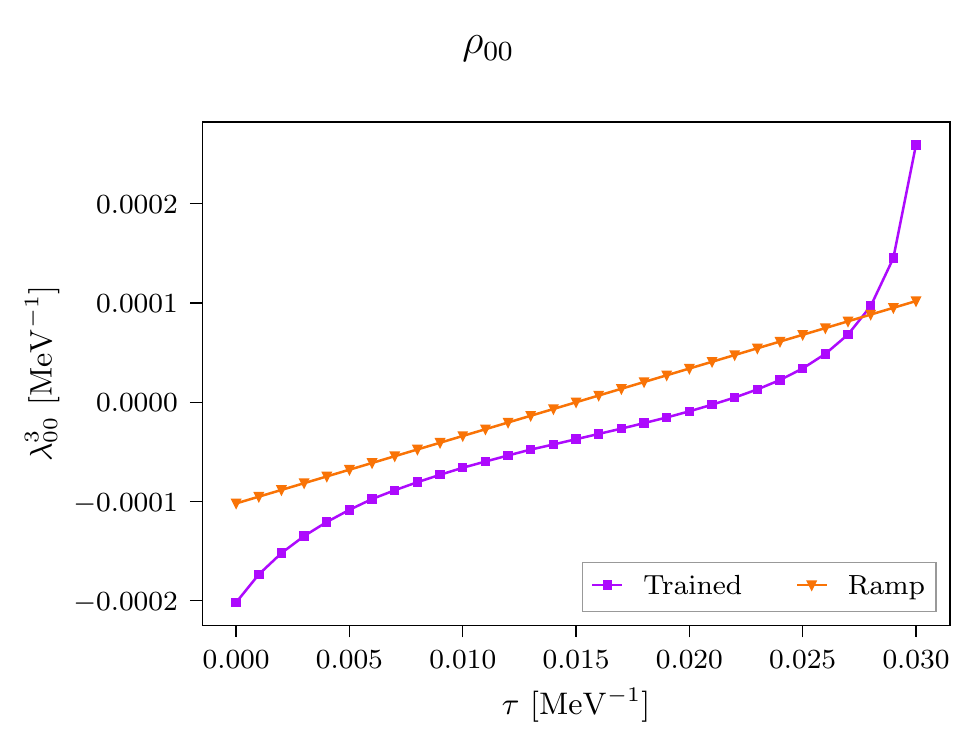}
	\includegraphics[width=0.47\textwidth]{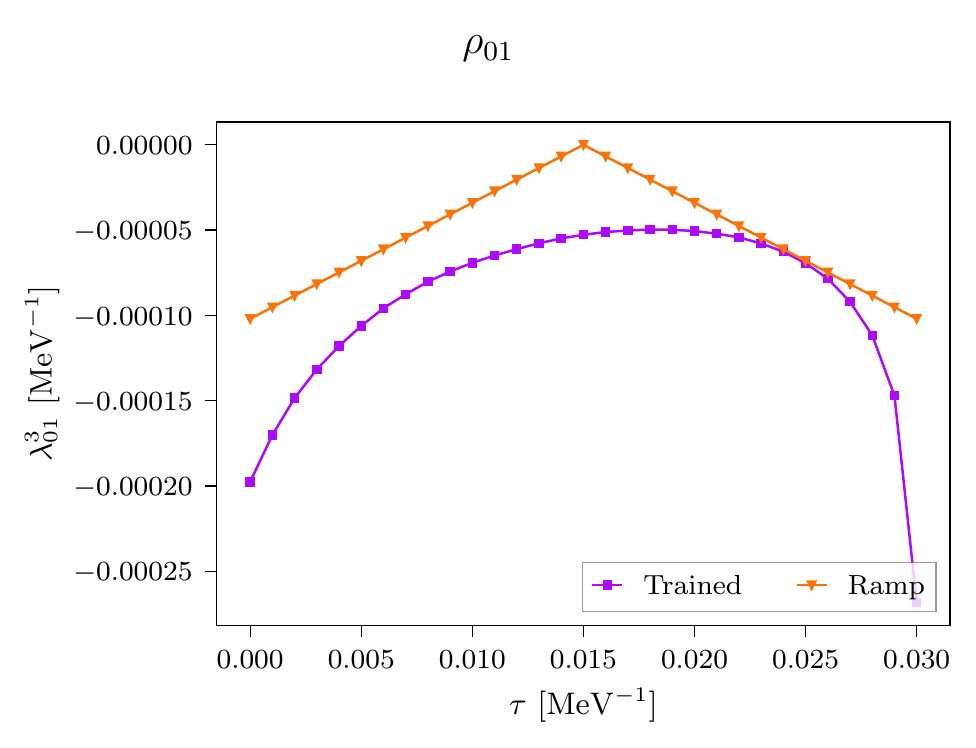}
	\includegraphics[width=0.47\textwidth]{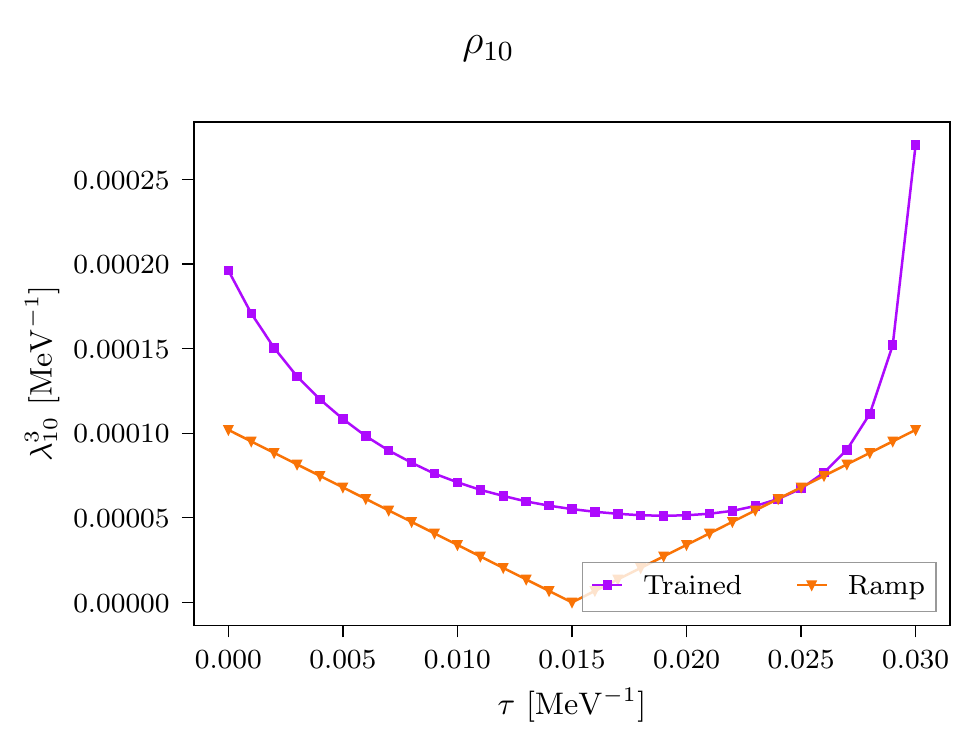}
	\includegraphics[width=0.47\textwidth]{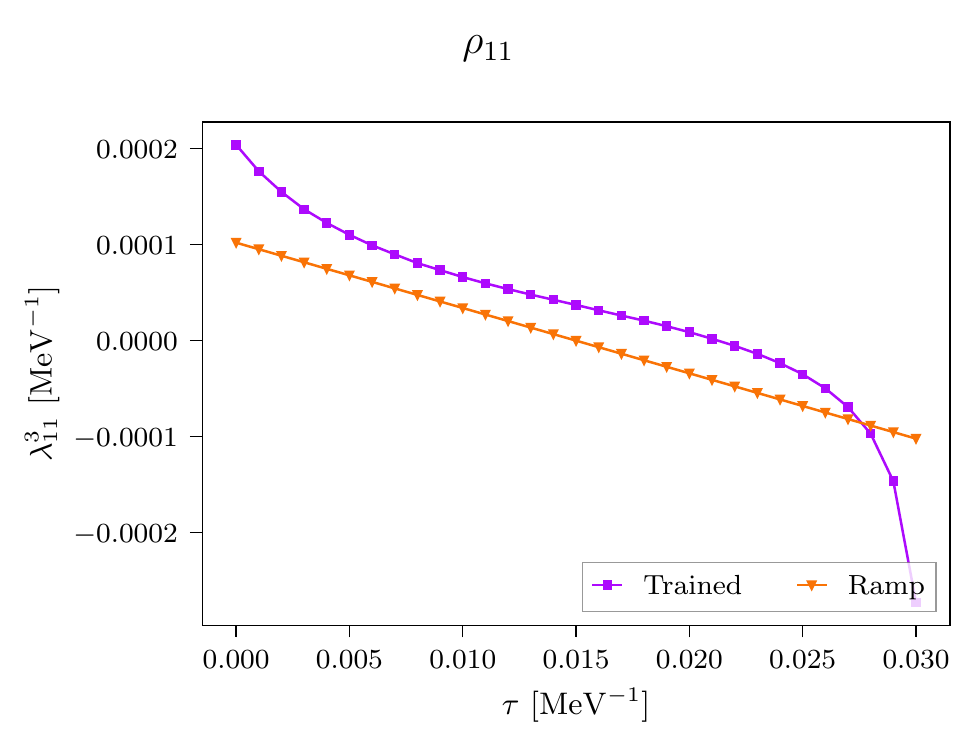}
   \caption{Comparison of the optimal $\lambda$ appearing in the ramp deform and the optimal $\lambda(\vec{q},\tau)$ appearing in the trained deform shown in Fig.~\ref{fig:response_ML_comparison} for $\vec{q} = (0,0,600)$ MeV. \label{fig:response_comparison_ramps} }
\end{figure*}

We begin our numerical studies of Cartesian constant shift deformations by considering generic deformation functions $\vec{\lambda}_{ij}(\vec{q},\tau)$ including free parameters for every $\tau$ in a GFMC calculation at fixed $\vec{q}$.
The optimal $\vec{\lambda}_{ij}(\vec{q},\tau)$ are determined through numerical optimization as described in Sec.~\ref{sec:optimization}. The loss function is here chosen to be $\mathcal{L}_{O}(1/2, 0, N\dt)$, as given in Eq.~\eqref{eq:general-loss-function}, with
\begin{equation}
O \equiv e^{i \vec{q} \cdot (\vec{r}^N_i - \vec{r}^0_j)}.
\end{equation}
This loss function is an estimate of the log of the non-holomorphic piece of the variance of $\rho_{ij}(\vec{q},\tau)$ averaged across all values of $\tau$.
Numerical optimization of $\vec{\lambda}_{ij}(\vec{q},\tau)$ leads to significant reduction of this loss function for all choices of $i,j\in \{1,2\}$.
Interestingly, attempts to numerically optimize the variance of $\rho(\vec{q},\tau)$ using a common choice of contour for all components $\rho_{ij}(\vec{q},\tau)$ by including a sum over $i$ and $j$ in the definition of the loss function do not lead to any significant reduction of the loss function compared to its value on the original contour.
This behavior can be explained using the simple arguments motivating the Cartesian constant shift deformation: any deformation that decreases the magnitude of $e^{i\vec{q}\cdot(\vec{r}^N_1 - \vec{r}^0_2)}$ and therefore $\rho_{12}(\vec{q},\tau)$ necessarily increases the magnitude of $e^{i\vec{q}\cdot(\vec{r}^N_2 - \vec{r}^1_2)}$ and therefore $\rho_{21}(\vec{q},\tau)$ because of the center-of-mass constraint $\sum_i \vec{r}^n_i = \vec{0}$ that is enforced at each step of GFMC evolution.
Similar obstacles of constrained sums over phases whose magnitude could not be decreased through vertical deformations were seen for the case of $SU(N)$ Wilson loops in lattice gauge theory~\cite{Detmold:2021ulb}, and the solution used in that work of deforming individual components of the sum inspired the deformation of the individual $\rho_{ij}(\vec{q},\tau)$ pursued here.

\begin{figure*}
    \centering
	\includegraphics[width=0.47\textwidth]{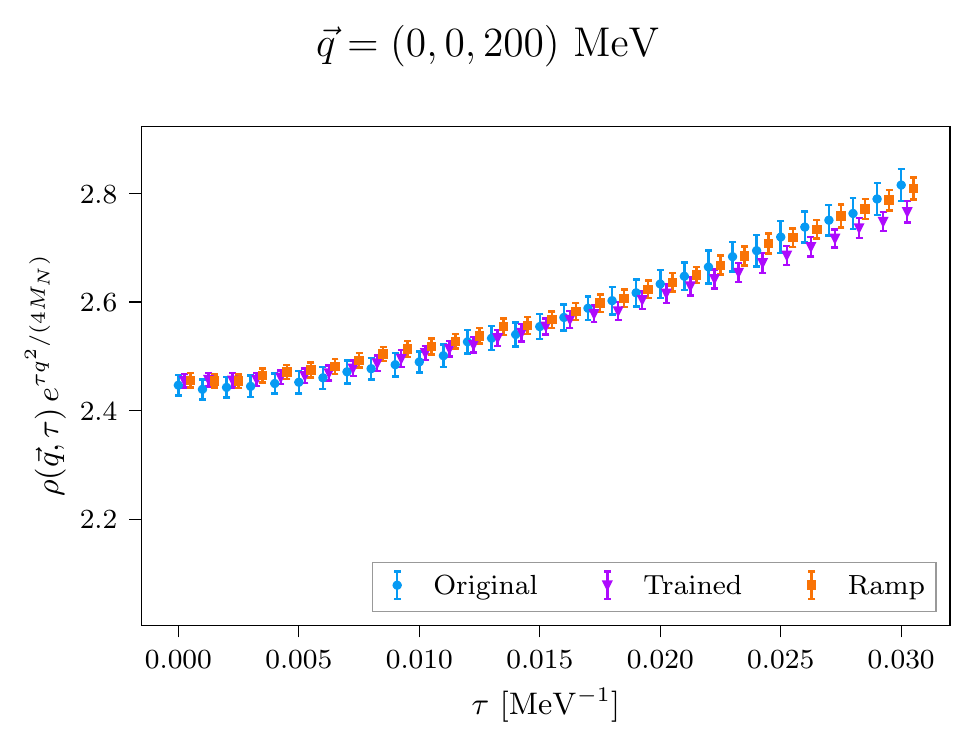}
	\includegraphics[width=0.47\textwidth]{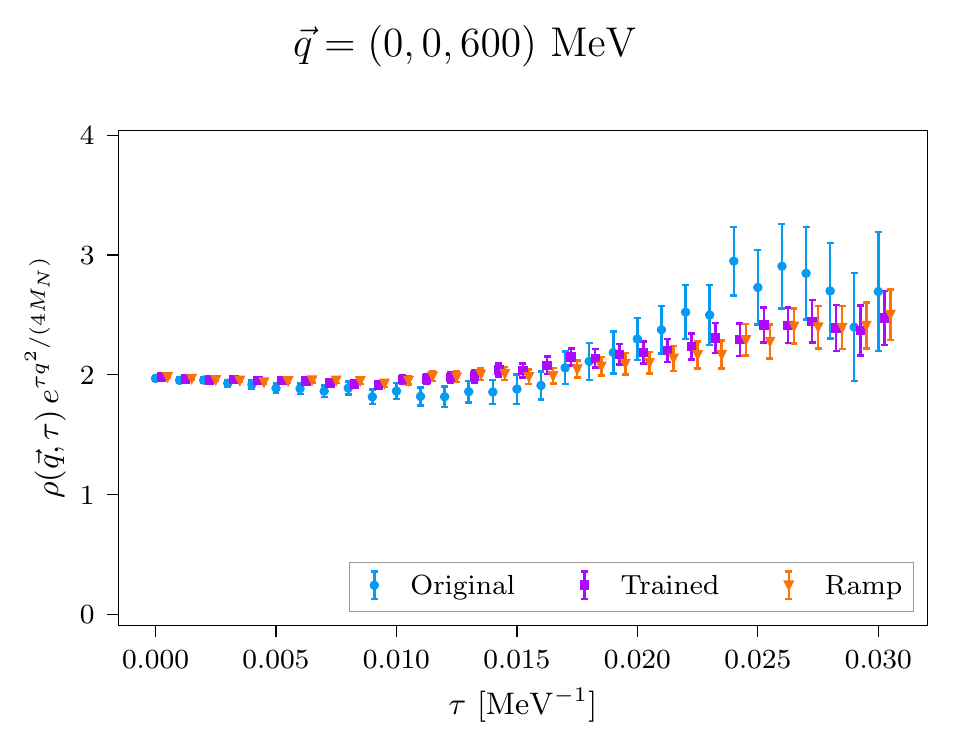}
   \caption{Comparison of GFMC results for Euclidean response functions for the two values of $\vec{q}$ indicated and $\tau \leq 0.03$ MeV${}^{-1}$ using $N=5,000$ walkers without contour deformation (``original'', blue), with the one-parameter ramp contour deformation shown in Eq.~\eqref{eq:ramp} with $\lambda = \pm 0.64\text{ MeV}^{-1}$ (``ramp,'' orange), and with a multi-parameter contour deformation that has been numerically optimized as described in the main text (``trained,'' purple).\label{fig:response_ML_comparison} }
\end{figure*}

Based on the analytical arguments above, the magnitude of $\rho_{ij}(\vec{q},\tau)$ is appropriately decreased by contour deformations which are antisymmetric about the midpoint of GFMC evolution for $i=j$. A simple antisymmetric deformation is the ``ramp'' proportional to $\tau - N \delta \tau / 2$ discussed above. For $i \neq j$, the previous arguments suggest that deformations should shift in the same direction across the range of imaginary time $\tau$. Inspection of the numerically optimized generic contour deformations shown in Fig.~\ref{fig:response_comparison_ramps} confirms this intuition and suggests that deformations proportional to $|\tau - N \delta \tau / 2|$ could be effective in this case.
Generalizing from these examples leads to the simple one-parameter ramp ansatz,
\begin{equation}
  \begin{split}
    \vec{\lambda}_{ij}(\vec{q},\tau) &= \frac{\lambda}{4 M_N} \, \vec{q} \, \eta_{ij} \left[ \delta_{ij} \left( \frac{\tau}{N \delta \tau} - \frac{1}{2} \right) \right. \\
    &\hspace{20pt} \left. + (1 - \delta_{ij}) \left| \frac{\tau}{N \delta \tau} - \frac{1}{2} \right| \right], \label{eq:ramp}
  \end{split}
\end{equation}
where $\eta_{00} = \eta_{10} = 1$, $\eta_{11} = \eta_{01} = -1$,  and $\lambda$ is a free parameter.

Response function results using numerically optimized generic Cartesian constant shifts $\vec{\lambda}_{ij}(\vec{q},\tau)$ are compared with optimized ramp deformations of the form Eq.~\eqref{eq:ramp} in Fig.~\ref{fig:response_ML_comparison} for two values of momentum transfer, $\vec{q} = (0,0,200\text{ MeV})$ and $\vec{q} = (0,0,600\text{ MeV})$.
Only small differences in the response function variance are visible for $\vec{q} = (0,0,200\text{ MeV})$, but for $\vec{q} = (0,0,600\text{ MeV})$ the response function variance is a factor of 4 smaller using optimized contour deformations with either generic $\tau$ dependence or with the specific one-parameter ramp defined in Eq.~\eqref{eq:ramp}.
The relative improvement in variance reduction at larger $\vec{q}^2$ and the relatively similar performance of one-parameter ramp contours as well as generic $\tau$-dependent contours are both consistent with the picture of signal-to-noise problems arising from phase fluctuations used to motivate the Cartesian constant shift above.
The differences between the precise shapes of the numerically optimized generic and one-parameter ramp deformations visible in Fig.~\ref{fig:response_comparison_ramps} do lead to slightly smaller values of the loss function
$\mathcal{L}_{O}(1/2, 0, N \dt)$
for the numerically optimized generic contours, but
they do not lead to more than few percent differences of the bootstrap errors on $\rho(\vec{q},\tau)$ shown in Fig.~\ref{fig:response_ML_comparison}.

\begin{figure*}[p]
	\includegraphics[width=0.47\textwidth]{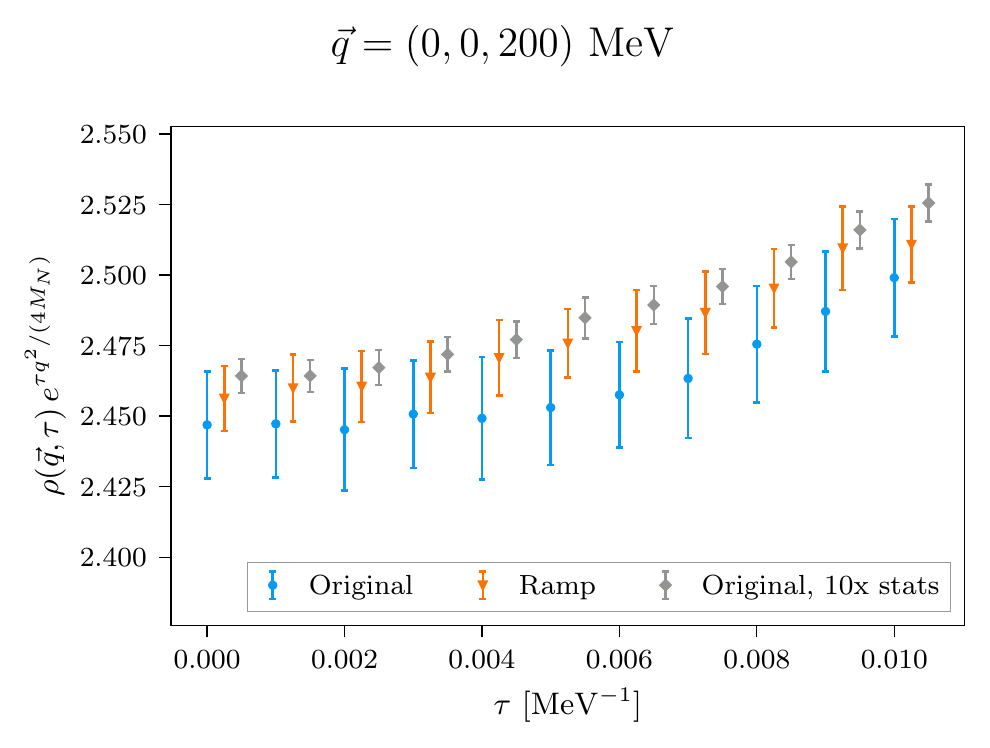}
	\includegraphics[width=0.47\textwidth]{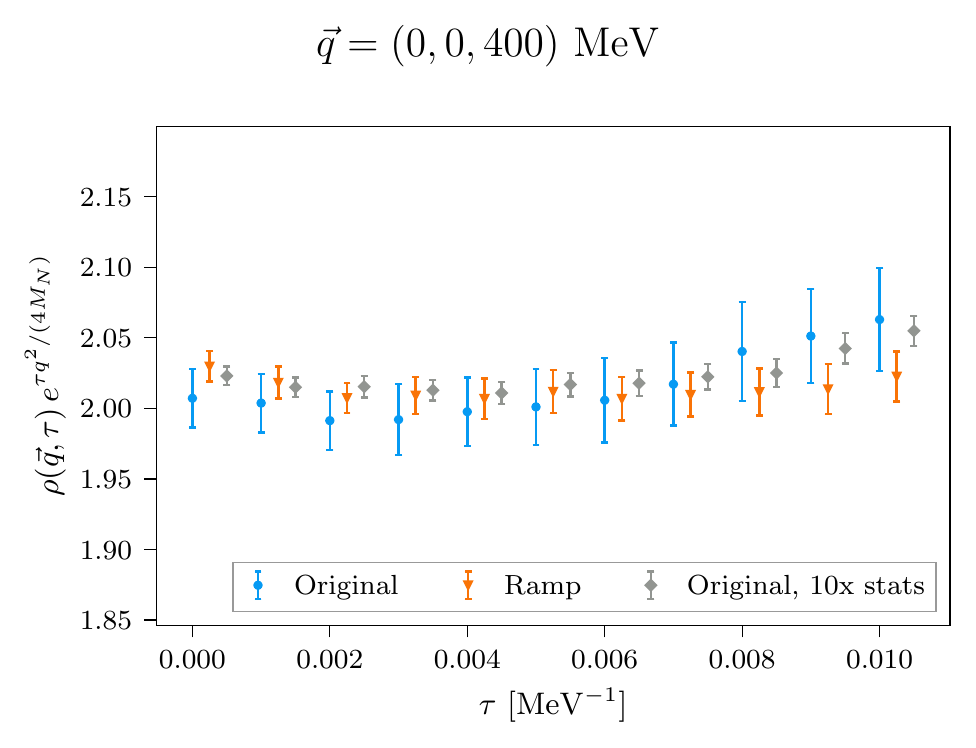}
	\includegraphics[width=0.47\textwidth]{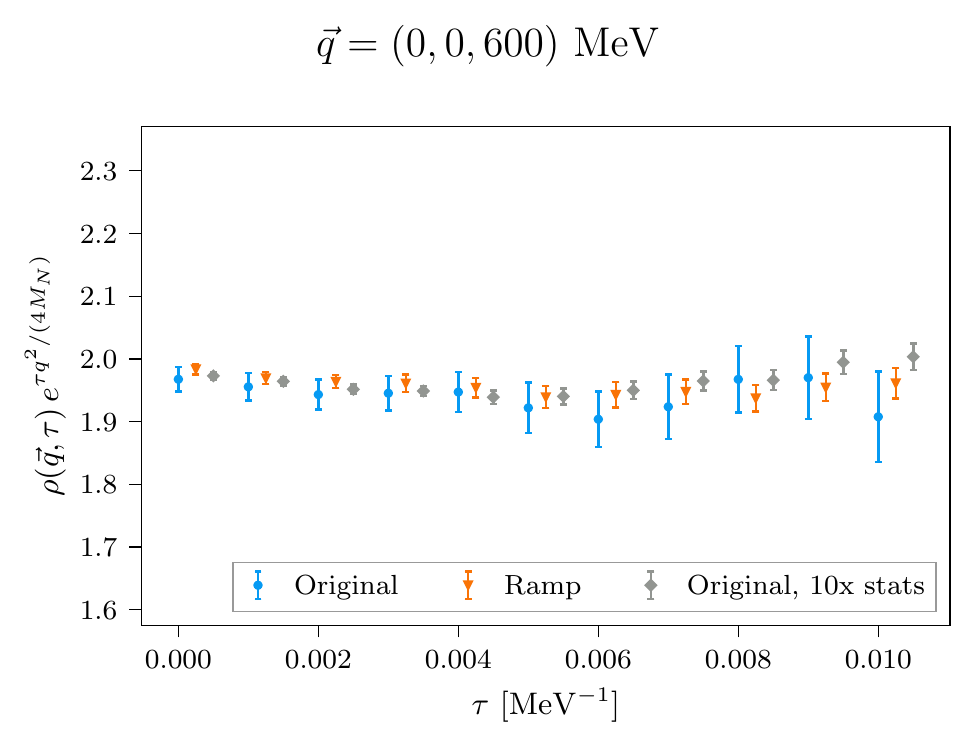}
	\includegraphics[width=0.47\textwidth]{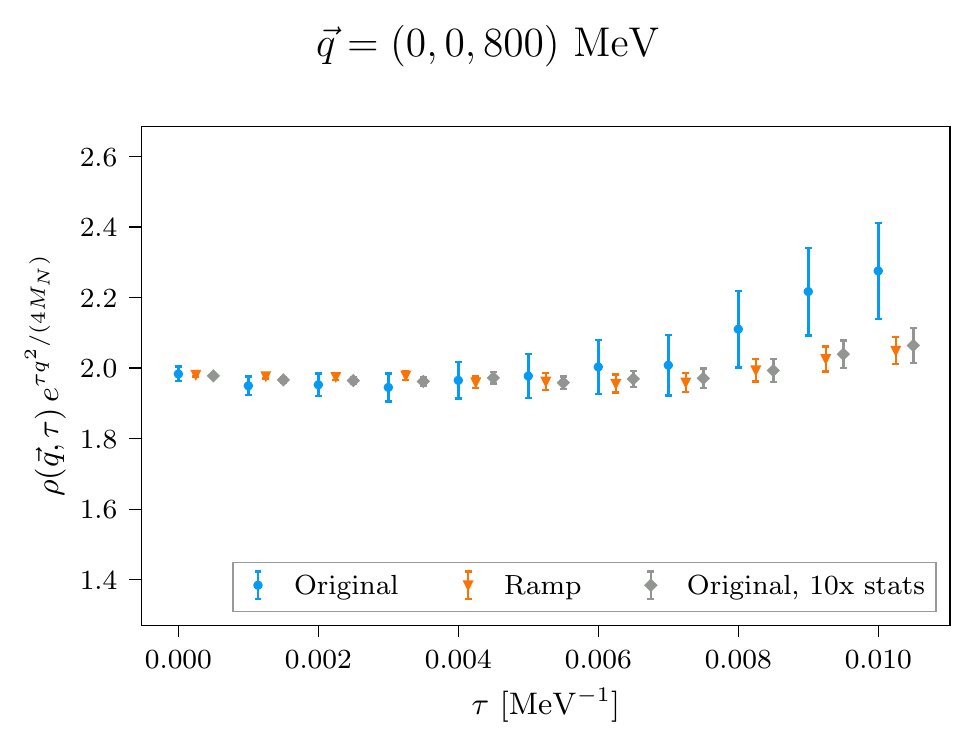}
	\includegraphics[width=0.47\textwidth]{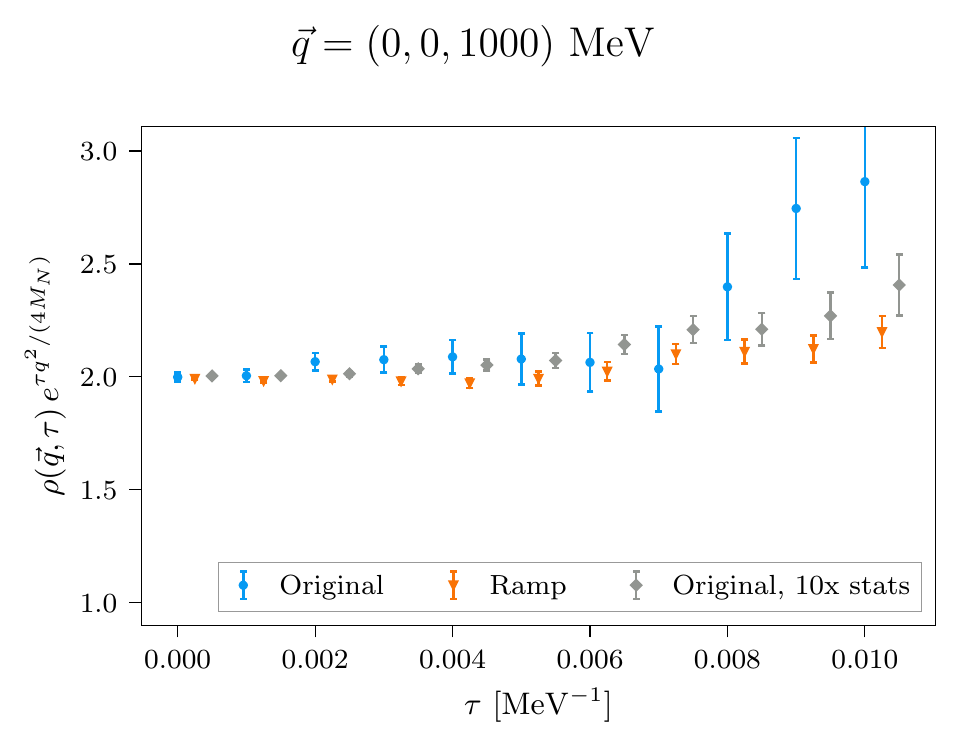}
	\includegraphics[width=0.47\textwidth]{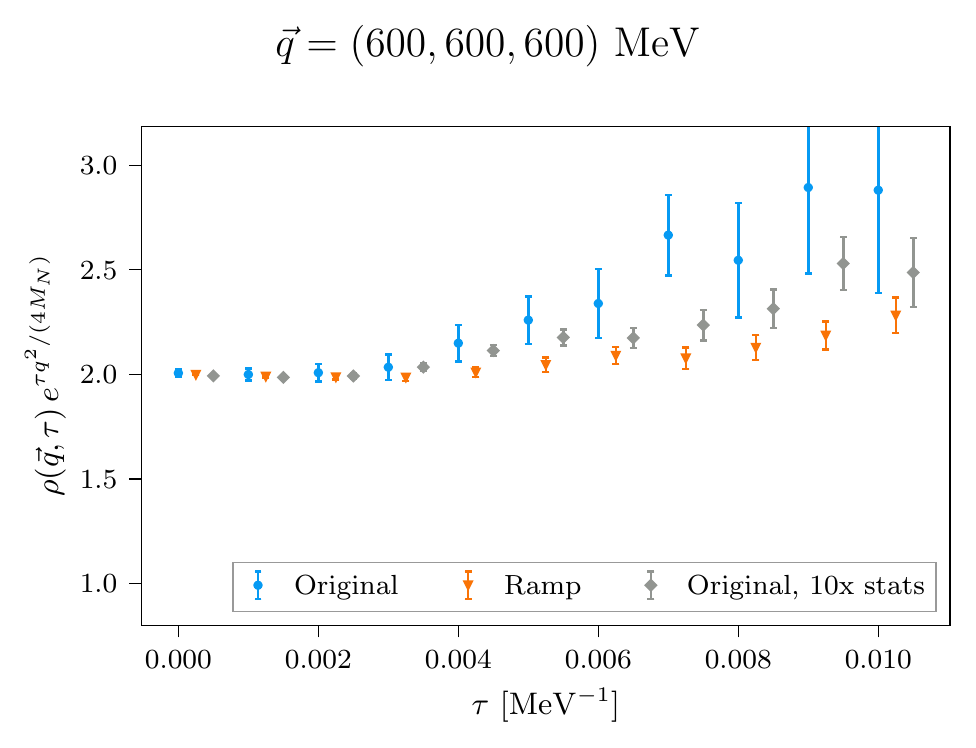}
   \caption{Comparison of GFMC results for Euclidean response functions without contour deformation (blue) and with a one-parameter ramp contour in which the coordinate integration contours are shifted in the imaginary direction by Eq.~\eqref{eq:ramp} (red) using $N=5,000$ walkers. Shift magnitudes of $\lambda =  \pm 0.64\text{ MeV}^{-1}$ are used for all $\vec{q}$ except for $\vec{q} = (600,600,600)$ MeV where $\lambda =  \pm 0.56\text{ MeV}^{-1}$ was found to give more optimal variance reduction. Higher-statistics results using $N=50,000$ walkers without contour deformation are also shown for reference (gray). \label{fig:response_comparison_small} }
\end{figure*}
\begin{figure*}[p]
	\includegraphics[width=0.47\textwidth]{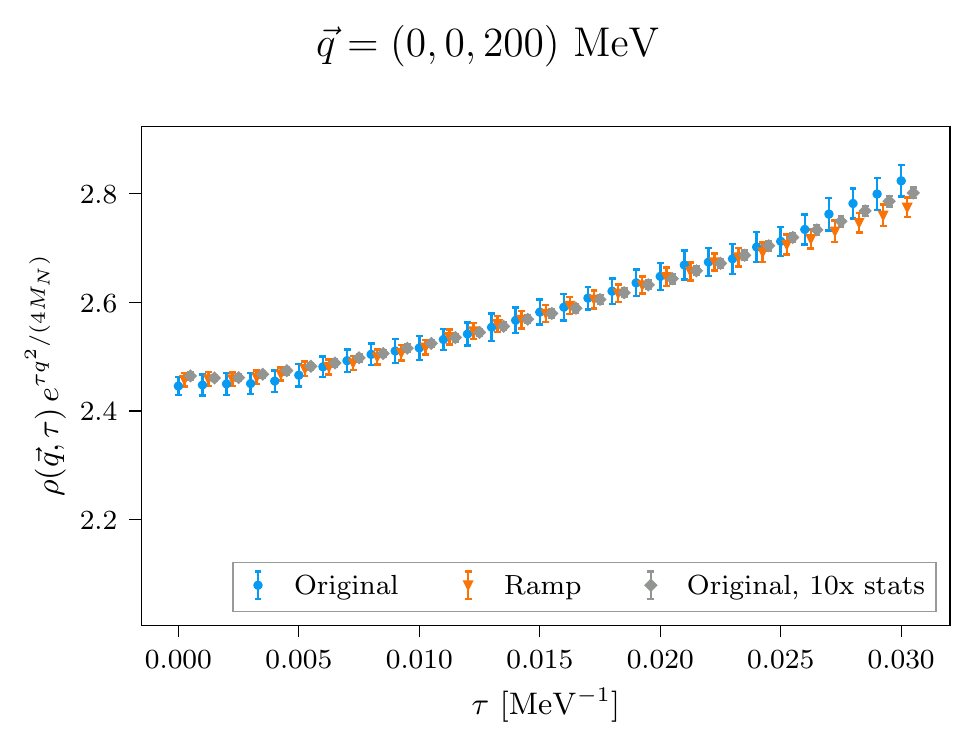}
	\includegraphics[width=0.47\textwidth]{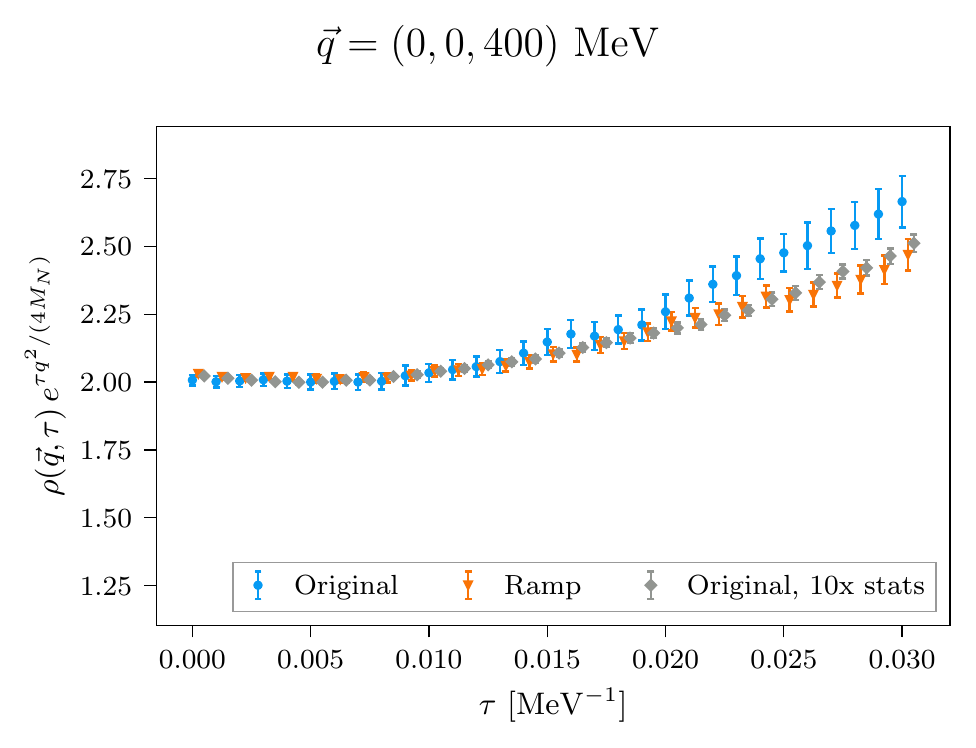}
	\includegraphics[width=0.47\textwidth]{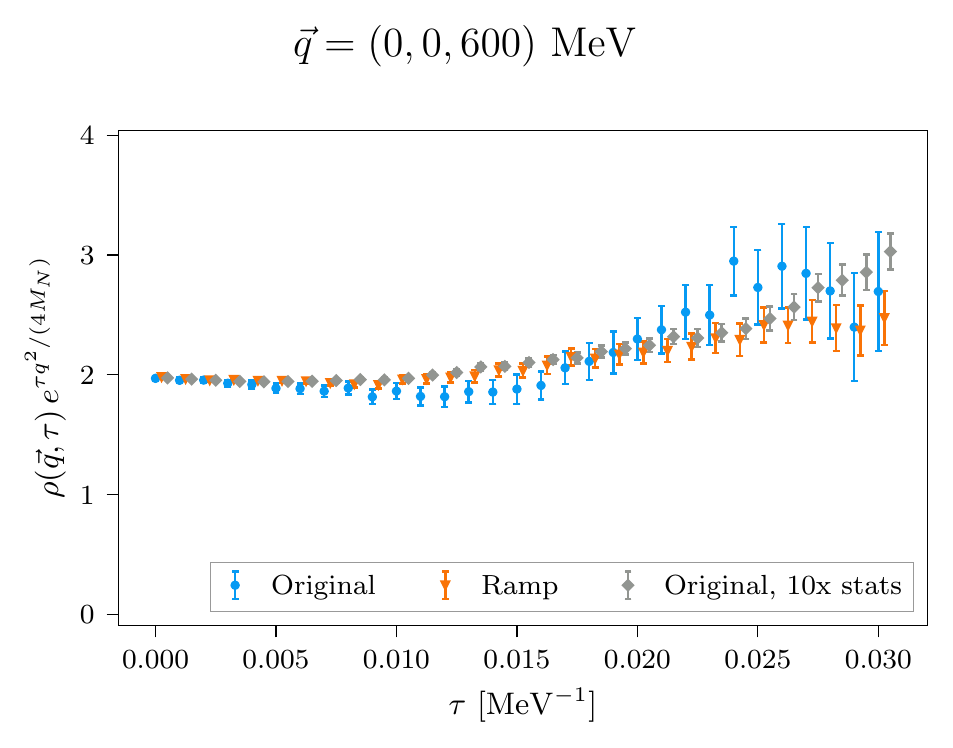}
	\includegraphics[width=0.47\textwidth]{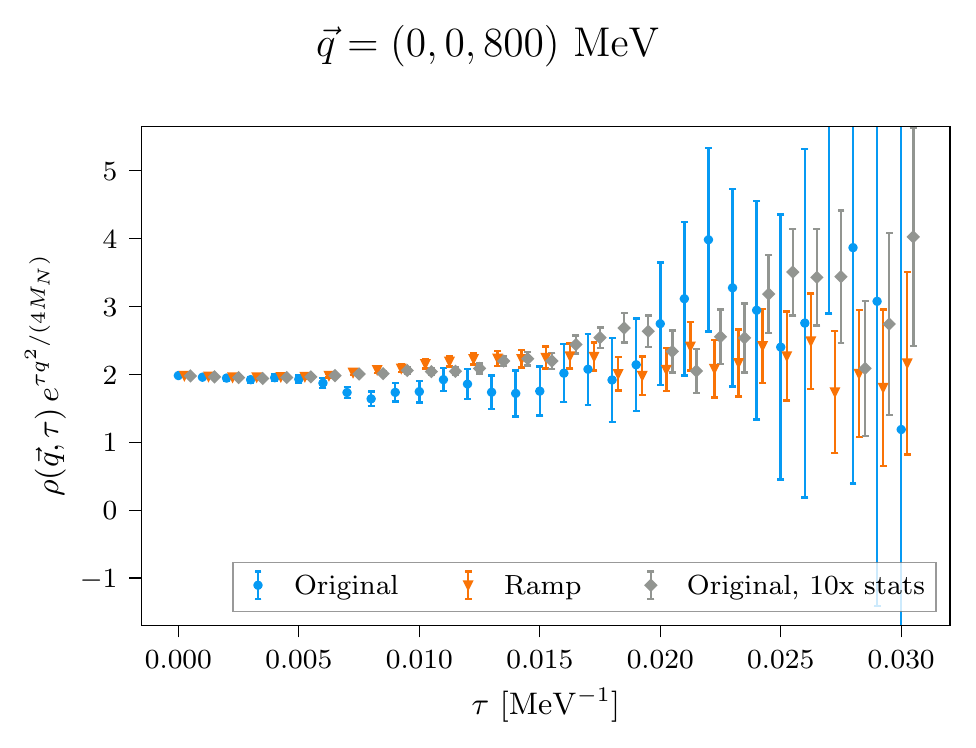}
	\includegraphics[width=0.47\textwidth]{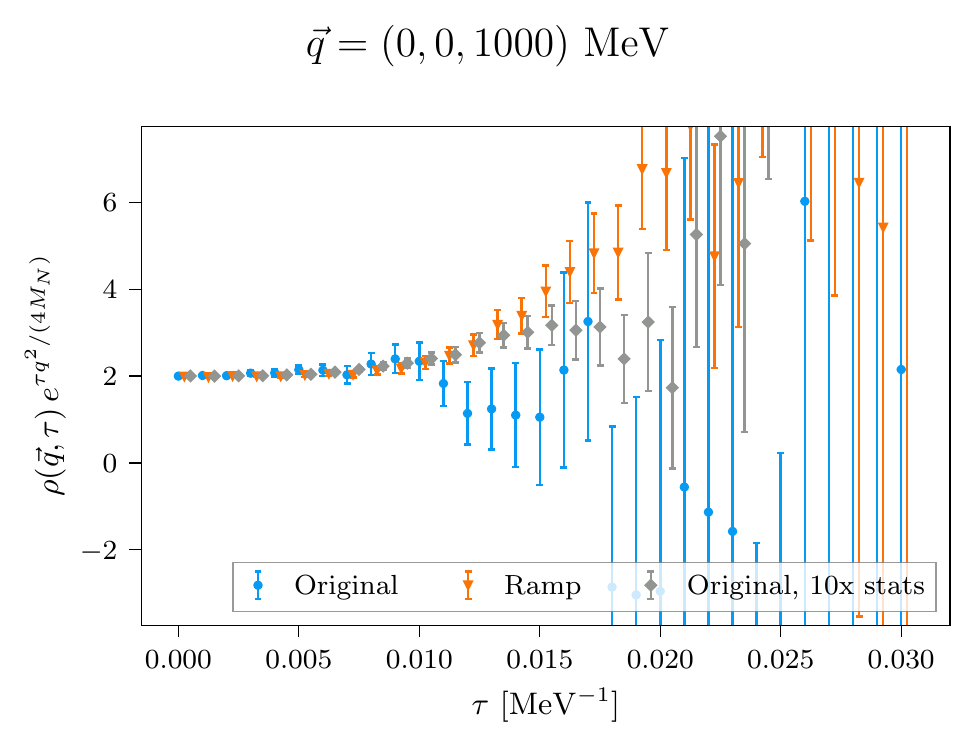}
	\includegraphics[width=0.47\textwidth]{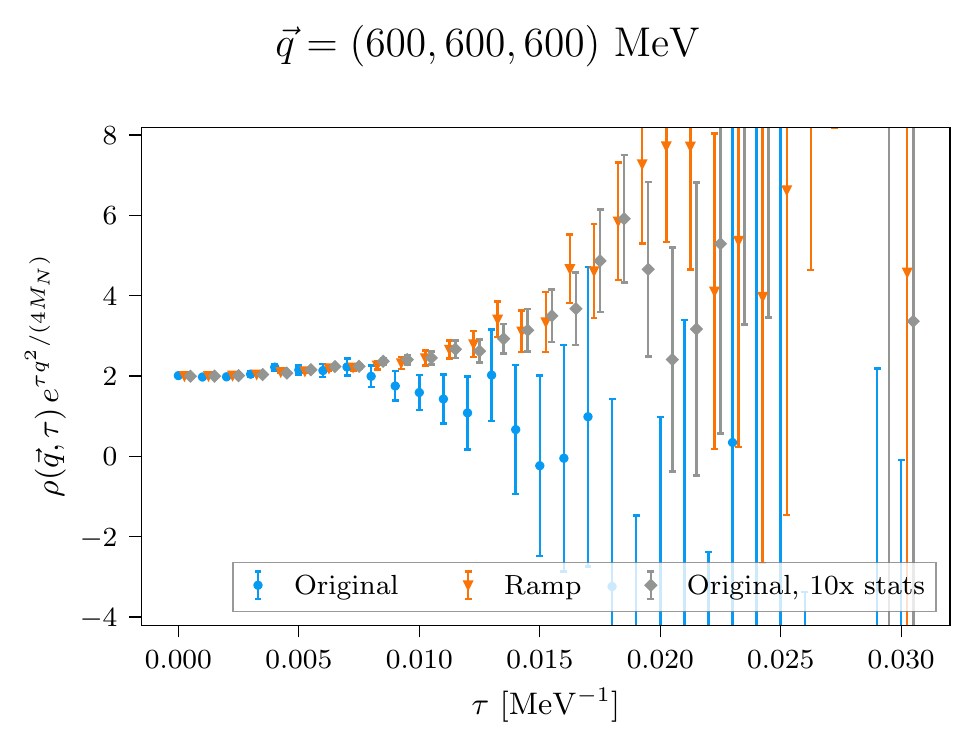}
   \caption{Comparison of GFMC results for Euclidean response functions  with and without contour deformation analogous to Fig.~\ref{fig:response_comparison_small} but with $\tau \leq 0.03 \text{ MeV}^{-1}$.  \label{fig:response_comparison_medium}  }
\end{figure*}
\begin{figure*}[!ht]
	\includegraphics[width=0.47\textwidth]{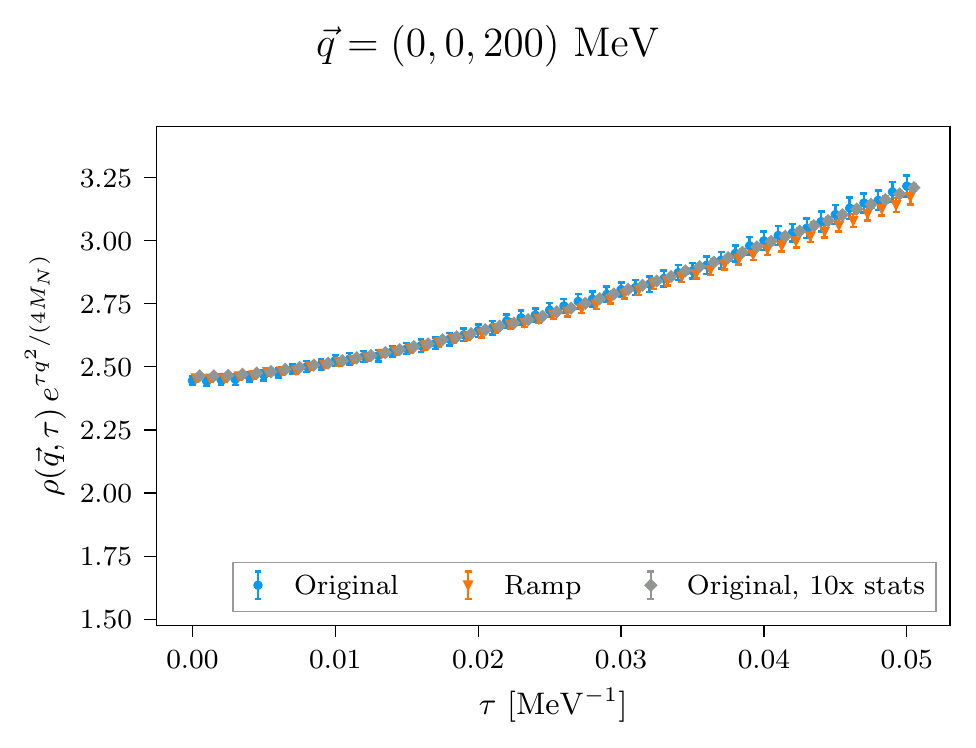}
	\includegraphics[width=0.47\textwidth]{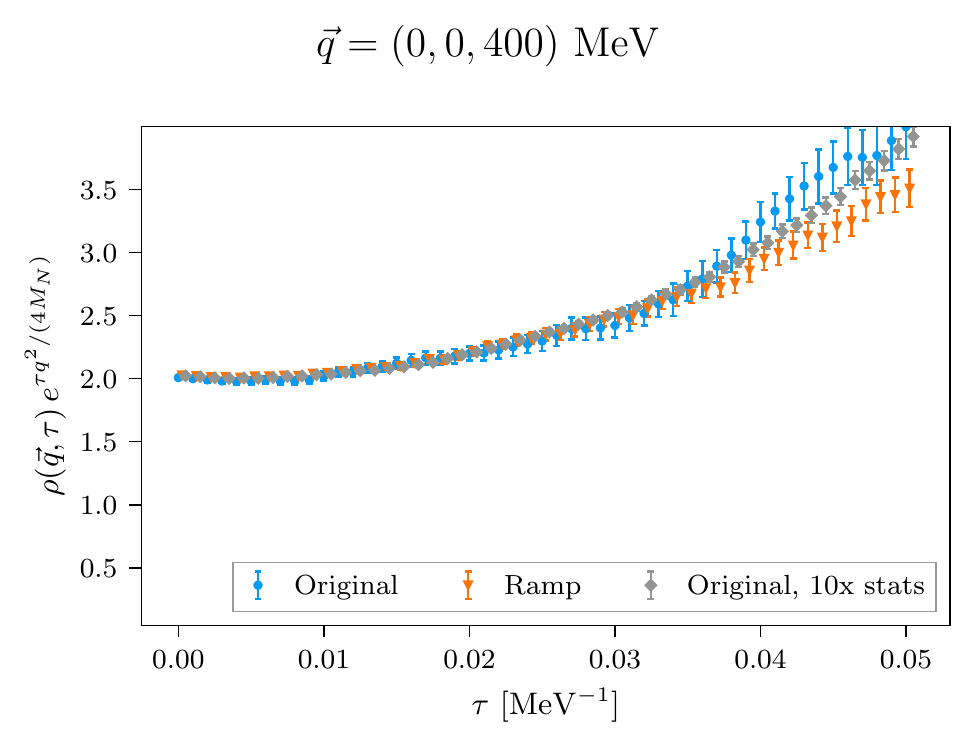}
	\includegraphics[width=0.47\textwidth]{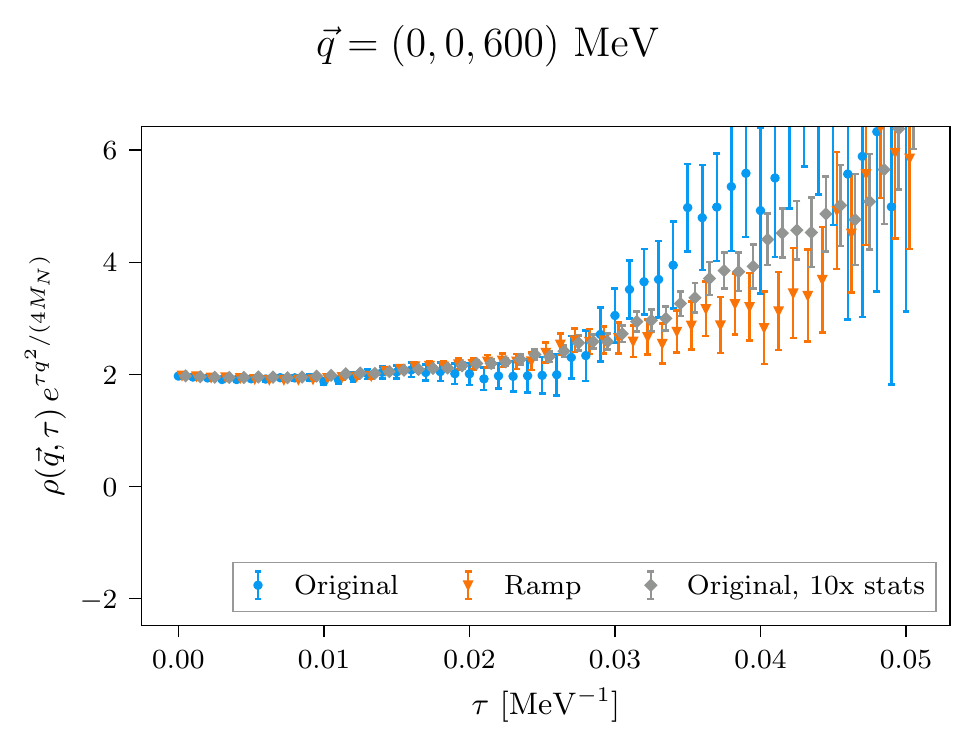}
	\includegraphics[width=0.47\textwidth]{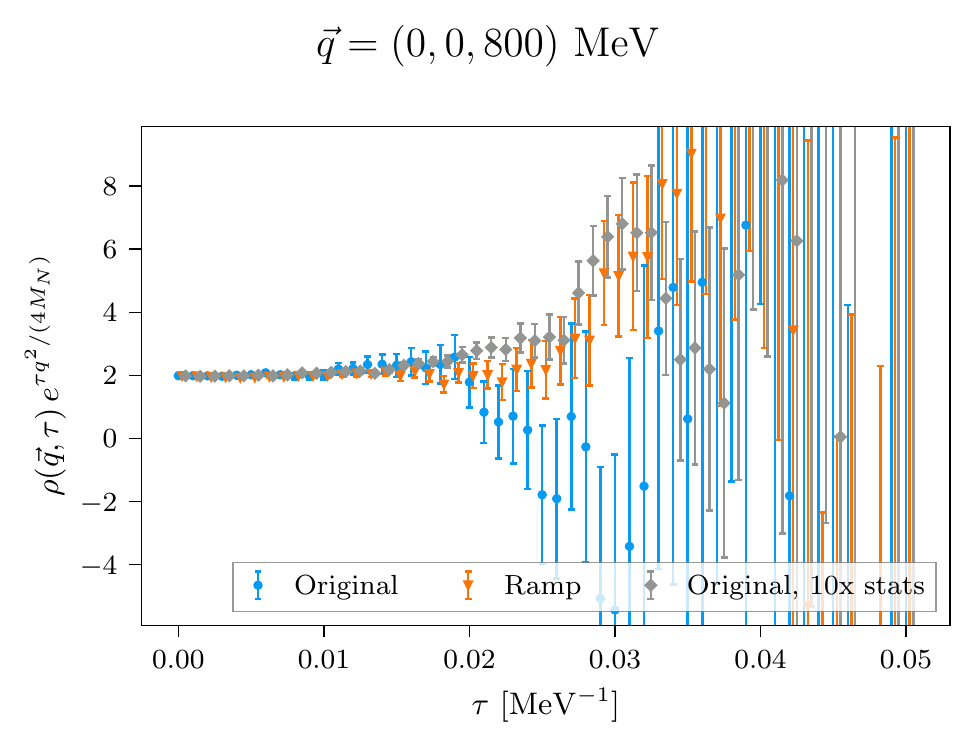}
	\includegraphics[width=0.47\textwidth]{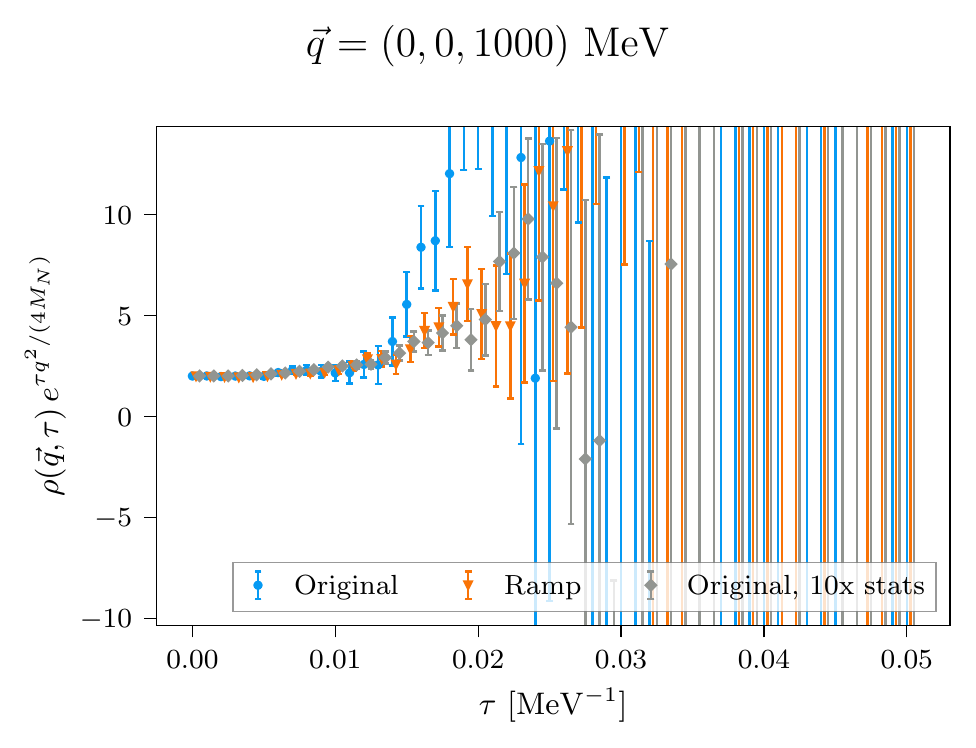}
	\includegraphics[width=0.47\textwidth]{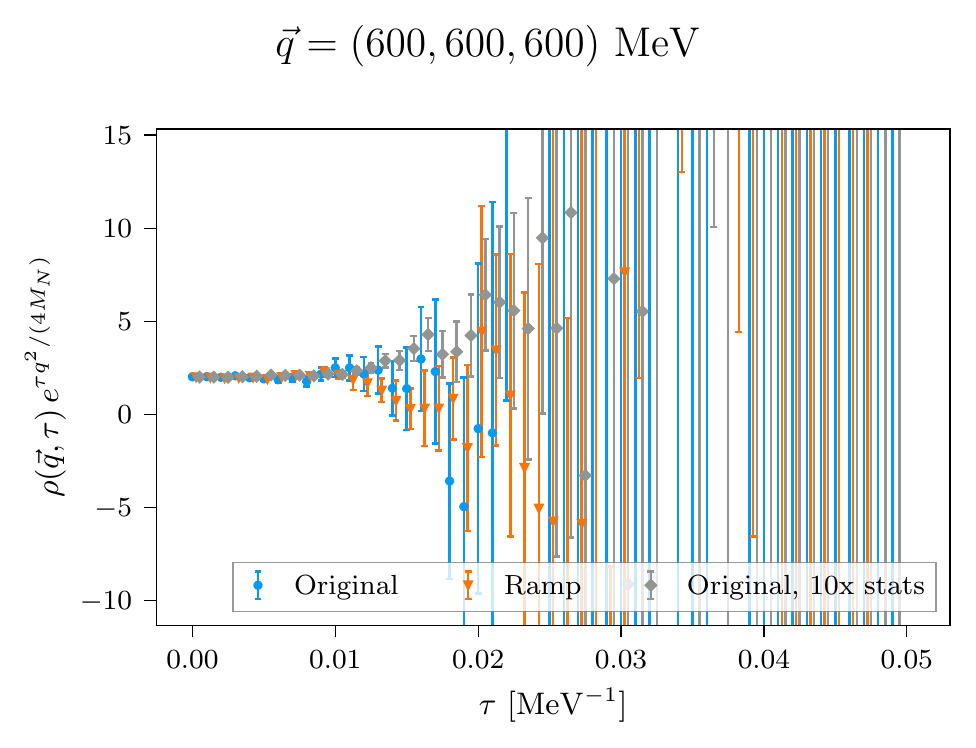}
   \caption{Comparison of GFMC results for Euclidean response functions  with and without contour deformation analogous to Fig.~\ref{fig:response_comparison_small} but with $\tau \leq 0.05 \text{ MeV}^{-1}$.  \label{fig:response_comparison_large}  }
\end{figure*}

The dependence of contour deformation results on $\vec{q}$ and the total length of the GFMC evolution $N \delta \tau$ is shown in Figs.~\ref{fig:response_comparison_small}--\ref{fig:response_comparison_large}.
There is a clear increase in the level of variance reduction achieved with increasing $|\vec{q}|$, for example ranging from approximately $2\times$ variance reduction for $|\vec{q}| = 200$ MeV to approximately $10\times$ variance reduction for $|\vec{q}| = 1000$ MeV for the intermediate choice of $N \delta \tau  = 0.03$ MeV$^{-1}$.
Very similar results are achieved for $\vec{q} = (600,600,600)$ MeV and $\vec{q} = (0,0,1000)$ MeV, which differ in $|\vec{q}|$ by only a few percent, confirming the expected approximate scaling of variance reduction with $|\vec{q}|$ rather than a different function of its components.
On the other hand, for fixed $\vec{q}$ and $\tau$, the variance reduction is most significant for the smallest choice of $N \dt = 0.01 \, \mathrm{MeV}^{-1}$, while smaller effects are seen for the larger choices of the total GFMC time $N \dt = 0.03$--$0.05 \, \mathrm{MeV}^{-1}$.
This feature arises in the optimal one-parameter ramp contours as well as the optimal generic Cartesian constant shifts $\vec{\lambda}_{ij}(\vec{q},\tau)$ obtained with the training procedures described above.
Although the need to choose $N \delta \tau$ in order to maximize the signal-to-noise of smaller $\tau \leq N \delta \tau$ is somewhat undesirable,
significant improvements are gained for the range of $N\dt \sim 0.01 - 0.05$ MeV$^{-1} \sim 2-10$ fm studied here, which are large enough to be relevant for the extraction for response functions used in scattering cross-section calculations~\cite{Carlson:2001mp,Lovato:2016gkq,Lovato:2017cux,Lovato:2020kba}.

The optimal $\lambda$ minimizing the variance of the ramp contour for particular $\vec{q}$ and $N \dt$ is found to be approximately $\lambda \approx 0.64 \, \mathrm{MeV}^{-1}$ over a broad range of $|\vec{q}| \lesssim 800 \, \mathrm{MeV}^{-1}$ and for all choices of $N \dt$ studied here.
Deformed contours using smaller values of $0 < \lambda < 0.64\,\mathrm{MeV}^{-1}$ lead to mildly larger variance than the optimal contour but in all cases have somewhat smaller variance than the original contour, which corresponds to $\lambda = 0$.
Conversely, the variance begins increasing rapidly for $\lambda > 0.64\,\mathrm{MeV}^{-1}$ and within roughly $0.05\,\mathrm{MeV}^{-1}$ becomes larger than the variance of the original contour.
The optimal value of $\lambda$ decreases mildly for the largest choices of $|\vec{q}|$, and in particular $\vec{q} = (600,600,600)$ MeV achieves optimal variance reduction with $\lambda = 0.56\,\mathrm{MeV}^{-1}$ and has larger variance than the original contour for $\lambda = 0.64\,\mathrm{MeV}^{-1}$.
The approximately constant scaling of the optimal $\lambda$ with $N\delta \tau$ means that the slope of the ``ramp'' is shallower for larger $N \delta \tau$, explaining the feature that the level of variance reduction achieved by the optimal contour deformations studied here are found to be proportional to $1/(N\delta \tau)$ for fixed $\vec{q}$. This result can be contrasted against the expectation from the signal-to-noise analysis that $\lambda \sim N\delta \tau$ with a reduction of variance independent of $N\dt$.

\begin{figure*}[]
\includegraphics[width=0.47\textwidth]{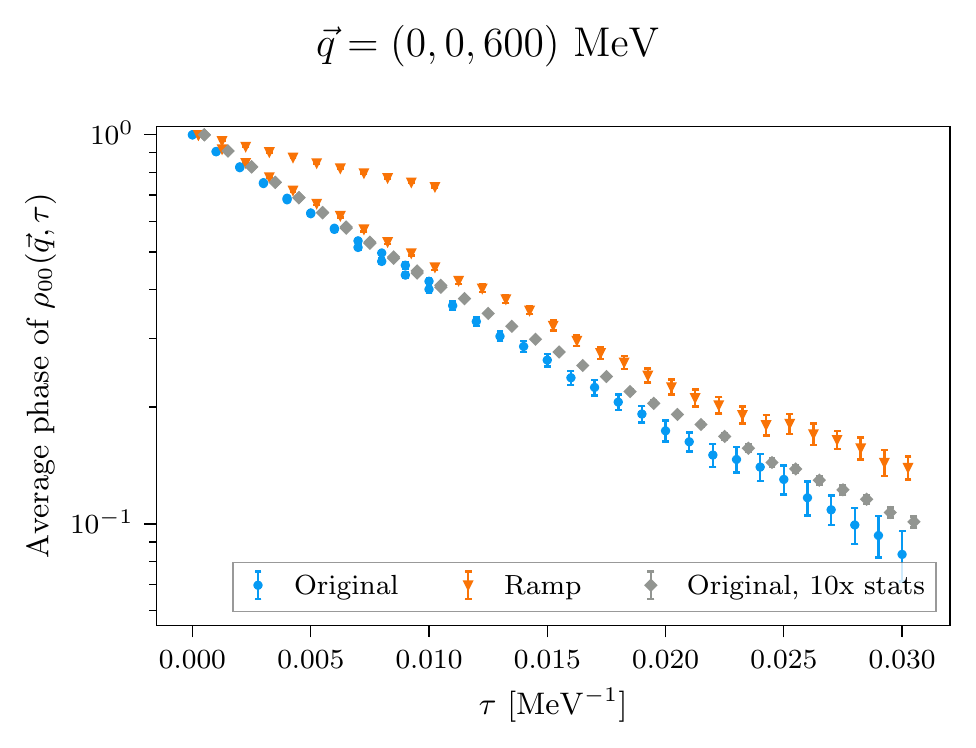}
\includegraphics[width=0.47\textwidth]{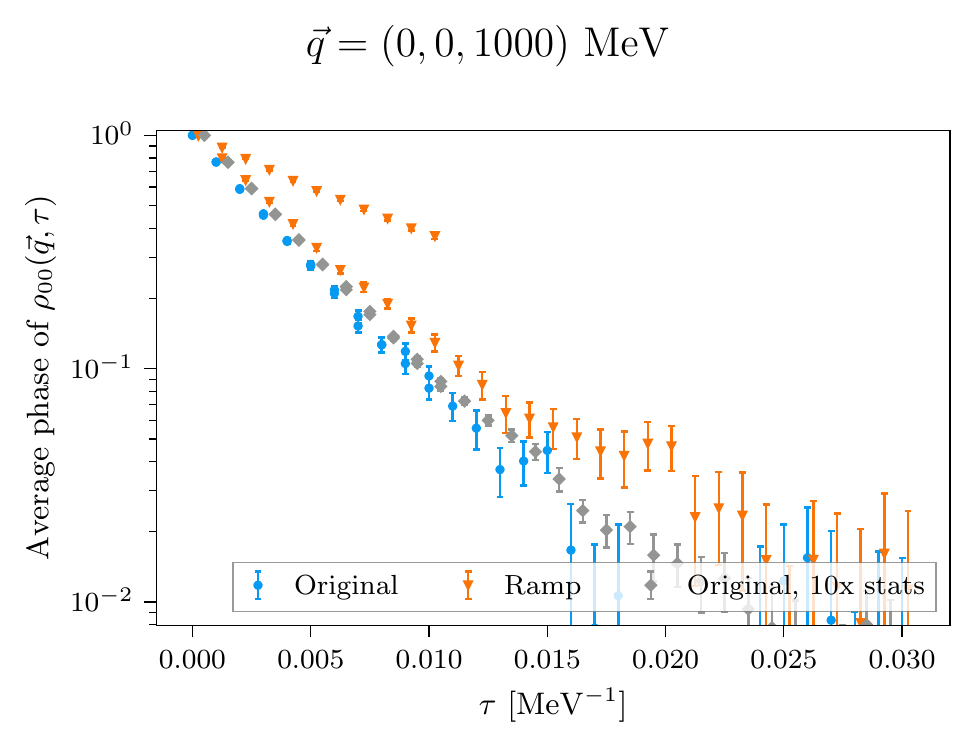}
   \caption{Comparison of the average phase factors of Euclidean response function calculations without contour deformation using $N=5,000$ walkers (blue) and $N=50,000$ walkers (gray) with the average phase factors for contour-deformed calculations using the same one-parameter ramp contours used in Figs.~\ref{fig:response_comparison_small}--\ref{fig:response_comparison_large}  using $N=5,000$ walkers  (red). Small average phase factors indicate the presence of a severe sign problem leading to a signal-to-noise problem. Results using $\tau \leq 0.01 \text{ MeV}^{-1}$ and $\tau \leq 0.03 \text{ MeV}^{-1}$ are overlayed to contrast the behavior of deformed results using contours optimized for different $\tau$. \label{fig:response_comparison_phase} }
\end{figure*}

To verify that the observed variance reduction actually arises from decreases in phase fluctuations of the Fourier transform factors $e^{i\vec{q}\cdot (\vec{r}^N_i - \vec{r}^0_j)}$, the average phase $\left< \rho_{ij}(\vec{q},\tau)/|\rho_{ij}(\vec{q},\tau)| \right>$ is shown for several values of $\vec{q}$ and $N\delta \tau$ in Fig~\ref{fig:response_comparison_phase}. Larger values of this quantity indicate reduced phase fluctuations.
As expected from the arguments motivating the constant Cartesian shift above, the larger $\vec{\lambda}$ corresponding to smaller $N\delta \tau$ also have exponentially larger average phases values and exponentially slower signal-to-noise degradation with $\tau$.
Since these arguments are applicable to generic response functions in larger nuclei, path integral contour deformations might be able to achieve significant improvements with respect to sign and signal-to-noise problems for electroweak and other response functions using relatively simple vertical Cartesian deformations.

\section{Outlook}
In this work, we have developed contour-deformation techniques suitable to mitigate the fermion sign problem in GFMC calculations of light nuclei. We have limited this initial analysis to the deuteron, considering both static and dynamic observables.

First, we considered the expectation value of a Hamiltonian that includes a nucleon-nucleon potential with a substantial tensor component, which generates a complicated phase structure in the ground-state wave function. Consequently, sophisticated importance-sampling wave functions are required to control the GFMC sign problem. As an alternative strategy, we utilized simplified importance-sampling wave functions and applied a spherical contour deformation to reduce the exponentially-growing variance of the energy expectation value. For large values of imaginary times, critical to determining the converged value of the ground-state energy of the system, we observe a reduction of about $20$--$30\%$ of the original statistical error. While the gain appears to be modest, the statistical noise was already small and not strongly affected by a sign problem in this case. 

As a follow-up of this work, we plan to carry out GFMC calculations of nuclei up to $^{12}$C, applying more sophisticated contour transformations beyond the spherical ones. Concurrently, we will extend the application of the methodology developed in this work to improve the accuracy of AFDMC calculations of nuclei with up to $A\sim 20$ nucleons and infinite neutron matter. To maintain a polynomial scaling with $A$, the latter are based on less accurate importance-sampling wave functions than the GFMC ones and are affected by a sizable fermion sign problem, suggesting that contour deformations may have a larger impact for AFDMC.

We also studied the utility of contour deformations for calculations of the density response of the deuteron for different momentum transfer values. This quantity is sensitive to the real-time dynamics of the system and displays a more severe signal-to-noise problem than ground-state observables. Since the imaginary-time evolution operator appears between the current insertions, the Euclidean density response is sensitive to the phase structure of the excited states of the nucleus, which is even more complicated than the ground-state one. For this observable, we have shown that a constant deformation in the Cartesian coordinates of the nucleons brings about a significant reduction of the statistical noise. This reduction becomes more critical for larger imaginary times and momentum transfer values. 

Response function results using contour deformations are consistent with undeformed results obtained using ten times more statistics, validating their unbiasedness, and in some cases are as or more precise.
Although variance reduction decreases with the extent of imaginary-time evolution, significant signal-to-noise improvements are found at phemonenologically relevant imaginary times in this proof-of-principle study of the deuteron.
The phase fluctuation arguments motivating the form of this deformation extend straightforwardly to larger nuclei, though the level of variance reduction that can be achieved in larger systems remains to be studied in future work.
The contour-deformation techniques developed in this work may therefore be crucial to reducing the computational cost of response function calculations. As such, they may enable extensive studies of the responses of nuclei such as $^{12}$C using different nuclear Hamiltonians and consistent electroweak currents. 
These studies are needed to assess the theoretical error of calculating lepton-nucleus inclusive cross sections, a fundamental ingredient for the analysis of neutrino oscillation experiments~\cite{Alvarez-Ruso:2014bla,DUNE:2015lol,NuSTEC:2017hzk,Meyer:2022mix,Ruso:2022qes,Simons:2022ltq}.
Furthermore, implementing contour-deformation techniques for the AFDMC method will allow extending these studies to larger nuclei such as $^{16}$O and $^{40}$Ca. Since the AFDMC samples both the spatial coordinates and the spin-isospin degrees of freedom of the nucleons, a generalization of the contour deformation will be required; analogous methods applied to auxiliary field Monte Carlo for electronic structure~\cite{Rom:1997,Rom:1998,Baer:1998,Baer:2000,Baer:2000b} may provide a useful starting point.
In addition to lepton-nucleus scattering, accurate calculations of imaginary-time propagators are relevant to estimate radiative corrections in superallowed $\beta$-decay rates, which are utilized to determine with high-precision the $V_{ud}$ element of the Cabibbo-Kobayashi-Maskawa mixing matrix~\cite{Seng:2018qru,Hardy:2020qwl}. Hence, contour deformations are expected to play a critical role to carry out such high-precision studies using the GFMC method.

\begin{acknowledgments}
We thank Scott Lawrence, Yin Lin, and Neill Warrington for helpful comments. This manuscript has been authored by Fermi Research Alliance, LLC under Contract No.\ DE-AC02-07CH11359 with the U.S.\ Department of Energy, Office of Science, Office of High Energy Physics. GK is supported by funding from the Schweizerischer Nationalfonds, under grant agreement number 200020\_200424. AL is supported by the U.S. Department of Energy, Office of Science, Office of Nuclear Physics, under contracts DE-AC02-06CH11357, by the 2020 DOE Early Career Award, by the NUCLEI SciDAC program, and Argonne LDRD awards.
\end{acknowledgments}

\bibliography{main}

\begin{thebibliography}{86}%
\makeatletter
\providecommand \@ifxundefined [1]{%
 \@ifx{#1\undefined}
}%
\providecommand \@ifnum [1]{%
 \ifnum #1\expandafter \@firstoftwo
 \else \expandafter \@secondoftwo
 \fi
}%
\providecommand \@ifx [1]{%
 \ifx #1\expandafter \@firstoftwo
 \else \expandafter \@secondoftwo
 \fi
}%
\providecommand \natexlab [1]{#1}%
\providecommand \enquote  [1]{``#1''}%
\providecommand \bibnamefont  [1]{#1}%
\providecommand \bibfnamefont [1]{#1}%
\providecommand \citenamefont [1]{#1}%
\providecommand \href@noop [0]{\@secondoftwo}%
\providecommand \href [0]{\begingroup \@sanitize@url \@href}%
\providecommand \@href[1]{\@@startlink{#1}\@@href}%
\providecommand \@@href[1]{\endgroup#1\@@endlink}%
\providecommand \@sanitize@url [0]{\catcode `\\12\catcode `\$12\catcode
  `\&12\catcode `\#12\catcode `\^12\catcode `\_12\catcode `\%12\relax}%
\providecommand \@@startlink[1]{}%
\providecommand \@@endlink[0]{}%
\providecommand \url  [0]{\begingroup\@sanitize@url \@url }%
\providecommand \@url [1]{\endgroup\@href {#1}{\urlprefix }}%
\providecommand \urlprefix  [0]{URL }%
\providecommand \Eprint [0]{\href }%
\providecommand \doibase [0]{http://dx.doi.org/}%
\providecommand \selectlanguage [0]{\@gobble}%
\providecommand \bibinfo  [0]{\@secondoftwo}%
\providecommand \bibfield  [0]{\@secondoftwo}%
\providecommand \translation [1]{[#1]}%
\providecommand \BibitemOpen [0]{}%
\providecommand \bibitemStop [0]{}%
\providecommand \bibitemNoStop [0]{.\EOS\space}%
\providecommand \EOS [0]{\spacefactor3000\relax}%
\providecommand \BibitemShut  [1]{\csname bibitem#1\endcsname}%
\let\auto@bib@innerbib\@empty
%</preamble>
\bibitem [{\citenamefont {Hergert}(2020)}]{Hergert:2020bxy}%
  \BibitemOpen
  \bibfield  {author} {\bibinfo {author} {\bibfnamefont {H.}~\bibnamefont
  {Hergert}},\ }\href {\doibase 10.3389/fphy.2020.00379} {\bibfield  {journal}
  {\bibinfo  {journal} {Front. in Phys.}\ }\textbf {\bibinfo {volume} {8}},\
  \bibinfo {pages} {379} (\bibinfo {year} {2020})},\ \Eprint
  {http://arxiv.org/abs/2008.05061} {arXiv:2008.05061 [nucl-th]} \BibitemShut
  {NoStop}%
\bibitem [{\citenamefont {Carlson}\ \emph {et~al.}(2015)\citenamefont
  {Carlson}, \citenamefont {Gandolfi}, \citenamefont {Pederiva}, \citenamefont
  {Pieper}, \citenamefont {Schiavilla}, \citenamefont {Schmidt},\ and\
  \citenamefont {Wiringa}}]{Carlson:2014vla}%
  \BibitemOpen
  \bibfield  {author} {\bibinfo {author} {\bibfnamefont {J.}~\bibnamefont
  {Carlson}}, \bibinfo {author} {\bibfnamefont {S.}~\bibnamefont {Gandolfi}},
  \bibinfo {author} {\bibfnamefont {F.}~\bibnamefont {Pederiva}}, \bibinfo
  {author} {\bibfnamefont {S.~C.}\ \bibnamefont {Pieper}}, \bibinfo {author}
  {\bibfnamefont {R.}~\bibnamefont {Schiavilla}}, \bibinfo {author}
  {\bibfnamefont {K.~E.}\ \bibnamefont {Schmidt}}, \ and\ \bibinfo {author}
  {\bibfnamefont {R.~B.}\ \bibnamefont {Wiringa}},\ }\href {\doibase
  10.1103/RevModPhys.87.1067} {\bibfield  {journal} {\bibinfo  {journal} {Rev.
  Mod. Phys.}\ }\textbf {\bibinfo {volume} {87}},\ \bibinfo {pages} {1067}
  (\bibinfo {year} {2015})},\ \Eprint {http://arxiv.org/abs/1412.3081}
  {arXiv:1412.3081 [nucl-th]} \BibitemShut {NoStop}%
\bibitem [{\citenamefont {Gandolfi}\ \emph {et~al.}(2020)\citenamefont
  {Gandolfi}, \citenamefont {Lonardoni}, \citenamefont {Lovato},\ and\
  \citenamefont {Piarulli}}]{Gandolfi:2020pbj}%
  \BibitemOpen
  \bibfield  {author} {\bibinfo {author} {\bibfnamefont {S.}~\bibnamefont
  {Gandolfi}}, \bibinfo {author} {\bibfnamefont {D.}~\bibnamefont {Lonardoni}},
  \bibinfo {author} {\bibfnamefont {A.}~\bibnamefont {Lovato}}, \ and\ \bibinfo
  {author} {\bibfnamefont {M.}~\bibnamefont {Piarulli}},\ }\href {\doibase
  10.3389/fphy.2020.00117} {\bibfield  {journal} {\bibinfo  {journal} {Front.
  in Phys.}\ }\textbf {\bibinfo {volume} {8}},\ \bibinfo {pages} {117}
  (\bibinfo {year} {2020})},\ \Eprint {http://arxiv.org/abs/2001.01374}
  {arXiv:2001.01374 [nucl-th]} \BibitemShut {NoStop}%
\bibitem [{\citenamefont {Schmidt}\ and\ \citenamefont
  {Fantoni}(1999)}]{Schmidt:1999lik}%
  \BibitemOpen
  \bibfield  {author} {\bibinfo {author} {\bibfnamefont {K.~E.}\ \bibnamefont
  {Schmidt}}\ and\ \bibinfo {author} {\bibfnamefont {S.}~\bibnamefont
  {Fantoni}},\ }\href {\doibase 10.1016/S0370-2693(98)01522-6} {\bibfield
  {journal} {\bibinfo  {journal} {Phys. Lett. B}\ }\textbf {\bibinfo {volume}
  {446}},\ \bibinfo {pages} {99} (\bibinfo {year} {1999})}\BibitemShut
  {NoStop}%
\bibitem [{\citenamefont {{Anderson}}(1976)}]{Anderson:1976JChPh}%
  \BibitemOpen
  \bibfield  {author} {\bibinfo {author} {\bibfnamefont {J.~B.}\ \bibnamefont
  {{Anderson}}},\ }\href {\doibase 10.1063/1.432868} {\bibfield  {journal}
  {\bibinfo  {journal} {\jcp}\ }\textbf {\bibinfo {volume} {65}},\ \bibinfo
  {pages} {4121} (\bibinfo {year} {1976})}\BibitemShut {NoStop}%
\bibitem [{\citenamefont {{Ceperley}}\ and\ \citenamefont
  {{Alder}}(1984)}]{Ceperley:1984JChPh}%
  \BibitemOpen
  \bibfield  {author} {\bibinfo {author} {\bibfnamefont {D.~M.}\ \bibnamefont
  {{Ceperley}}}\ and\ \bibinfo {author} {\bibfnamefont {B.~J.}\ \bibnamefont
  {{Alder}}},\ }\href {\doibase 10.1063/1.447637} {\bibfield  {journal}
  {\bibinfo  {journal} {\jcp}\ }\textbf {\bibinfo {volume} {81}},\ \bibinfo
  {pages} {5833} (\bibinfo {year} {1984})}\BibitemShut {NoStop}%
\bibitem [{\citenamefont {Pudliner}\ \emph {et~al.}(1997)\citenamefont
  {Pudliner}, \citenamefont {Pandharipande}, \citenamefont {Carlson},
  \citenamefont {Pieper},\ and\ \citenamefont {Wiringa}}]{Pudliner:1997ck}%
  \BibitemOpen
  \bibfield  {author} {\bibinfo {author} {\bibfnamefont {B.~S.}\ \bibnamefont
  {Pudliner}}, \bibinfo {author} {\bibfnamefont {V.~R.}\ \bibnamefont
  {Pandharipande}}, \bibinfo {author} {\bibfnamefont {J.}~\bibnamefont
  {Carlson}}, \bibinfo {author} {\bibfnamefont {S.~C.}\ \bibnamefont {Pieper}},
  \ and\ \bibinfo {author} {\bibfnamefont {R.~B.}\ \bibnamefont {Wiringa}},\
  }\href {\doibase 10.1103/PhysRevC.56.1720} {\bibfield  {journal} {\bibinfo
  {journal} {Phys. Rev. C}\ }\textbf {\bibinfo {volume} {56}},\ \bibinfo
  {pages} {1720} (\bibinfo {year} {1997})},\ \Eprint
  {http://arxiv.org/abs/nucl-th/9705009} {arXiv:nucl-th/9705009} \BibitemShut
  {NoStop}%
\bibitem [{\citenamefont {Piarulli}\ \emph {et~al.}(2020)\citenamefont
  {Piarulli}, \citenamefont {Bombaci}, \citenamefont {Logoteta}, \citenamefont
  {Lovato},\ and\ \citenamefont {Wiringa}}]{Piarulli:2019pfq}%
  \BibitemOpen
  \bibfield  {author} {\bibinfo {author} {\bibfnamefont {M.}~\bibnamefont
  {Piarulli}}, \bibinfo {author} {\bibfnamefont {I.}~\bibnamefont {Bombaci}},
  \bibinfo {author} {\bibfnamefont {D.}~\bibnamefont {Logoteta}}, \bibinfo
  {author} {\bibfnamefont {A.}~\bibnamefont {Lovato}}, \ and\ \bibinfo {author}
  {\bibfnamefont {R.~B.}\ \bibnamefont {Wiringa}},\ }\href {\doibase
  10.1103/PhysRevC.101.045801} {\bibfield  {journal} {\bibinfo  {journal}
  {Phys. Rev. C}\ }\textbf {\bibinfo {volume} {101}},\ \bibinfo {pages}
  {045801} (\bibinfo {year} {2020})},\ \Eprint
  {http://arxiv.org/abs/1908.04426} {arXiv:1908.04426 [nucl-th]} \BibitemShut
  {NoStop}%
\bibitem [{\citenamefont {Carlson}\ \emph {et~al.}(2002)\citenamefont
  {Carlson}, \citenamefont {Jourdan}, \citenamefont {Schiavilla},\ and\
  \citenamefont {Sick}}]{Carlson:2001mp}%
  \BibitemOpen
  \bibfield  {author} {\bibinfo {author} {\bibfnamefont {J.}~\bibnamefont
  {Carlson}}, \bibinfo {author} {\bibfnamefont {J.}~\bibnamefont {Jourdan}},
  \bibinfo {author} {\bibfnamefont {R.}~\bibnamefont {Schiavilla}}, \ and\
  \bibinfo {author} {\bibfnamefont {I.}~\bibnamefont {Sick}},\ }\href {\doibase
  10.1103/PhysRevC.65.024002} {\bibfield  {journal} {\bibinfo  {journal} {Phys.
  Rev. C}\ }\textbf {\bibinfo {volume} {65}},\ \bibinfo {pages} {024002}
  (\bibinfo {year} {2002})},\ \Eprint {http://arxiv.org/abs/nucl-th/0106047}
  {arXiv:nucl-th/0106047} \BibitemShut {NoStop}%
\bibitem [{\citenamefont {Lovato}\ \emph {et~al.}(2016)\citenamefont {Lovato},
  \citenamefont {Gandolfi}, \citenamefont {Carlson}, \citenamefont {Pieper},\
  and\ \citenamefont {Schiavilla}}]{Lovato:2016gkq}%
  \BibitemOpen
  \bibfield  {author} {\bibinfo {author} {\bibfnamefont {A.}~\bibnamefont
  {Lovato}}, \bibinfo {author} {\bibfnamefont {S.}~\bibnamefont {Gandolfi}},
  \bibinfo {author} {\bibfnamefont {J.}~\bibnamefont {Carlson}}, \bibinfo
  {author} {\bibfnamefont {S.~C.}\ \bibnamefont {Pieper}}, \ and\ \bibinfo
  {author} {\bibfnamefont {R.}~\bibnamefont {Schiavilla}},\ }\href {\doibase
  10.1103/PhysRevLett.117.082501} {\bibfield  {journal} {\bibinfo  {journal}
  {Phys. Rev. Lett.}\ }\textbf {\bibinfo {volume} {117}},\ \bibinfo {pages}
  {082501} (\bibinfo {year} {2016})},\ \Eprint
  {http://arxiv.org/abs/1605.00248} {arXiv:1605.00248 [nucl-th]} \BibitemShut
  {NoStop}%
\bibitem [{\citenamefont {Lovato}\ \emph {et~al.}(2018)\citenamefont {Lovato},
  \citenamefont {Gandolfi}, \citenamefont {Carlson}, \citenamefont {Lusk},
  \citenamefont {Pieper},\ and\ \citenamefont {Schiavilla}}]{Lovato:2017cux}%
  \BibitemOpen
  \bibfield  {author} {\bibinfo {author} {\bibfnamefont {A.}~\bibnamefont
  {Lovato}}, \bibinfo {author} {\bibfnamefont {S.}~\bibnamefont {Gandolfi}},
  \bibinfo {author} {\bibfnamefont {J.}~\bibnamefont {Carlson}}, \bibinfo
  {author} {\bibfnamefont {E.}~\bibnamefont {Lusk}}, \bibinfo {author}
  {\bibfnamefont {S.~C.}\ \bibnamefont {Pieper}}, \ and\ \bibinfo {author}
  {\bibfnamefont {R.}~\bibnamefont {Schiavilla}},\ }\href {\doibase
  10.1103/PhysRevC.97.022502} {\bibfield  {journal} {\bibinfo  {journal} {Phys.
  Rev. C}\ }\textbf {\bibinfo {volume} {97}},\ \bibinfo {pages} {022502}
  (\bibinfo {year} {2018})},\ \Eprint {http://arxiv.org/abs/1711.02047}
  {arXiv:1711.02047 [nucl-th]} \BibitemShut {NoStop}%
\bibitem [{\citenamefont {Lovato}\ \emph {et~al.}(2020)\citenamefont {Lovato},
  \citenamefont {Carlson}, \citenamefont {Gandolfi}, \citenamefont {Rocco},\
  and\ \citenamefont {Schiavilla}}]{Lovato:2020kba}%
  \BibitemOpen
  \bibfield  {author} {\bibinfo {author} {\bibfnamefont {A.}~\bibnamefont
  {Lovato}}, \bibinfo {author} {\bibfnamefont {J.}~\bibnamefont {Carlson}},
  \bibinfo {author} {\bibfnamefont {S.}~\bibnamefont {Gandolfi}}, \bibinfo
  {author} {\bibfnamefont {N.}~\bibnamefont {Rocco}}, \ and\ \bibinfo {author}
  {\bibfnamefont {R.}~\bibnamefont {Schiavilla}},\ }\href {\doibase
  10.1103/PhysRevX.10.031068} {\bibfield  {journal} {\bibinfo  {journal} {Phys.
  Rev. X}\ }\textbf {\bibinfo {volume} {10}},\ \bibinfo {pages} {031068}
  (\bibinfo {year} {2020})},\ \Eprint {http://arxiv.org/abs/2003.07710}
  {arXiv:2003.07710 [nucl-th]} \BibitemShut {NoStop}%
\bibitem [{\citenamefont {Lovato}\ \emph {et~al.}(2019)\citenamefont {Lovato},
  \citenamefont {Rocco},\ and\ \citenamefont {Schiavilla}}]{Lovato:2019fiw}%
  \BibitemOpen
  \bibfield  {author} {\bibinfo {author} {\bibfnamefont {A.}~\bibnamefont
  {Lovato}}, \bibinfo {author} {\bibfnamefont {N.}~\bibnamefont {Rocco}}, \
  and\ \bibinfo {author} {\bibfnamefont {R.}~\bibnamefont {Schiavilla}},\
  }\href {\doibase 10.1103/PhysRevC.100.035502} {\bibfield  {journal} {\bibinfo
   {journal} {Phys. Rev. C}\ }\textbf {\bibinfo {volume} {100}},\ \bibinfo
  {pages} {035502} (\bibinfo {year} {2019})},\ \Eprint
  {http://arxiv.org/abs/1903.08078} {arXiv:1903.08078 [nucl-th]} \BibitemShut
  {NoStop}%
\bibitem [{\citenamefont {Raghavan}\ \emph {et~al.}(2021)\citenamefont
  {Raghavan}, \citenamefont {Balaprakash}, \citenamefont {Lovato},
  \citenamefont {Rocco},\ and\ \citenamefont {Wild}}]{Raghavan:2020bze}%
  \BibitemOpen
  \bibfield  {author} {\bibinfo {author} {\bibfnamefont {K.}~\bibnamefont
  {Raghavan}}, \bibinfo {author} {\bibfnamefont {P.}~\bibnamefont
  {Balaprakash}}, \bibinfo {author} {\bibfnamefont {A.}~\bibnamefont {Lovato}},
  \bibinfo {author} {\bibfnamefont {N.}~\bibnamefont {Rocco}}, \ and\ \bibinfo
  {author} {\bibfnamefont {S.~M.}\ \bibnamefont {Wild}},\ }\href {\doibase
  10.1103/PhysRevC.103.035502} {\bibfield  {journal} {\bibinfo  {journal}
  {Phys. Rev. C}\ }\textbf {\bibinfo {volume} {103}},\ \bibinfo {pages}
  {035502} (\bibinfo {year} {2021})},\ \Eprint
  {http://arxiv.org/abs/2010.12703} {arXiv:2010.12703 [nucl-th]} \BibitemShut
  {NoStop}%
\bibitem [{\citenamefont {Alvarez-Ruso}\ \emph {et~al.}(2014)\citenamefont
  {Alvarez-Ruso}, \citenamefont {Hayato},\ and\ \citenamefont
  {Nieves}}]{Alvarez-Ruso:2014bla}%
  \BibitemOpen
  \bibfield  {author} {\bibinfo {author} {\bibfnamefont {L.}~\bibnamefont
  {Alvarez-Ruso}}, \bibinfo {author} {\bibfnamefont {Y.}~\bibnamefont
  {Hayato}}, \ and\ \bibinfo {author} {\bibfnamefont {J.}~\bibnamefont
  {Nieves}},\ }\href {\doibase 10.1088/1367-2630/16/7/075015} {\bibfield
  {journal} {\bibinfo  {journal} {New J. Phys.}\ }\textbf {\bibinfo {volume}
  {16}},\ \bibinfo {pages} {075015} (\bibinfo {year} {2014})},\ \Eprint
  {http://arxiv.org/abs/1403.2673} {arXiv:1403.2673 [hep-ph]} \BibitemShut
  {NoStop}%
\bibitem [{\citenamefont {Acciarri}\ \emph {et~al.}(2015)\citenamefont
  {Acciarri} \emph {et~al.}}]{DUNE:2015lol}%
  \BibitemOpen
  \bibfield  {author} {\bibinfo {author} {\bibfnamefont {R.}~\bibnamefont
  {Acciarri}} \emph {et~al.} (\bibinfo {collaboration} {DUNE}),\ }\href@noop {}
  {\  (\bibinfo {year} {2015})},\ \Eprint {http://arxiv.org/abs/1512.06148}
  {arXiv:1512.06148 [physics.ins-det]} \BibitemShut {NoStop}%
\bibitem [{\citenamefont {Alvarez-Ruso}\ \emph {et~al.}(2018)\citenamefont
  {Alvarez-Ruso} \emph {et~al.}}]{NuSTEC:2017hzk}%
  \BibitemOpen
  \bibfield  {author} {\bibinfo {author} {\bibfnamefont {L.}~\bibnamefont
  {Alvarez-Ruso}} \emph {et~al.} (\bibinfo {collaboration} {NuSTEC}),\ }\href
  {\doibase 10.1016/j.ppnp.2018.01.006} {\bibfield  {journal} {\bibinfo
  {journal} {Prog. Part. Nucl. Phys.}\ }\textbf {\bibinfo {volume} {100}},\
  \bibinfo {pages} {1} (\bibinfo {year} {2018})},\ \Eprint
  {http://arxiv.org/abs/1706.03621} {arXiv:1706.03621 [hep-ph]} \BibitemShut
  {NoStop}%
\bibitem [{\citenamefont {Meyer}\ \emph {et~al.}(2022)\citenamefont {Meyer},
  \citenamefont {Walker-Loud},\ and\ \citenamefont
  {Wilkinson}}]{Meyer:2022mix}%
  \BibitemOpen
  \bibfield  {author} {\bibinfo {author} {\bibfnamefont {A.~S.}\ \bibnamefont
  {Meyer}}, \bibinfo {author} {\bibfnamefont {A.}~\bibnamefont {Walker-Loud}},
  \ and\ \bibinfo {author} {\bibfnamefont {C.}~\bibnamefont {Wilkinson}},\
  }\href {\doibase 10.1146/annurev-nucl-010622-120608} {\  (\bibinfo {year}
  {2022}),\ 10.1146/annurev-nucl-010622-120608},\ \Eprint
  {http://arxiv.org/abs/2201.01839} {arXiv:2201.01839 [hep-lat]} \BibitemShut
  {NoStop}%
\bibitem [{\citenamefont {Ruso}\ \emph {et~al.}(2022)\citenamefont {Ruso} \emph
  {et~al.}}]{Ruso:2022qes}%
  \BibitemOpen
  \bibfield  {author} {\bibinfo {author} {\bibfnamefont {L.~A.}\ \bibnamefont
  {Ruso}} \emph {et~al.},\ }\href@noop {} {\  (\bibinfo {year} {2022})},\
  \Eprint {http://arxiv.org/abs/2203.09030} {arXiv:2203.09030 [hep-ph]}
  \BibitemShut {NoStop}%
\bibitem [{\citenamefont {Simons}\ \emph {et~al.}(2022)\citenamefont {Simons},
  \citenamefont {Steinberg}, \citenamefont {Lovato}, \citenamefont {Meurice},
  \citenamefont {Rocco},\ and\ \citenamefont {Wagman}}]{Simons:2022ltq}%
  \BibitemOpen
  \bibfield  {author} {\bibinfo {author} {\bibfnamefont {D.}~\bibnamefont
  {Simons}}, \bibinfo {author} {\bibfnamefont {N.}~\bibnamefont {Steinberg}},
  \bibinfo {author} {\bibfnamefont {A.}~\bibnamefont {Lovato}}, \bibinfo
  {author} {\bibfnamefont {Y.}~\bibnamefont {Meurice}}, \bibinfo {author}
  {\bibfnamefont {N.}~\bibnamefont {Rocco}}, \ and\ \bibinfo {author}
  {\bibfnamefont {M.}~\bibnamefont {Wagman}},\ }\href@noop {} {\  (\bibinfo
  {year} {2022})},\ \Eprint {http://arxiv.org/abs/2210.02455} {arXiv:2210.02455
  [hep-ph]} \BibitemShut {NoStop}%
\bibitem [{\citenamefont {Cristoforetti}\ \emph {et~al.}(2012)\citenamefont
  {Cristoforetti}, \citenamefont {Di~Renzo},\ and\ \citenamefont
  {Scorzato}}]{Cristoforetti:2012su}%
  \BibitemOpen
  \bibfield  {author} {\bibinfo {author} {\bibfnamefont {M.}~\bibnamefont
  {Cristoforetti}}, \bibinfo {author} {\bibfnamefont {F.}~\bibnamefont
  {Di~Renzo}}, \ and\ \bibinfo {author} {\bibfnamefont {L.}~\bibnamefont
  {Scorzato}} (\bibinfo {collaboration} {AuroraScience}),\ }\href {\doibase
  10.1103/PhysRevD.86.074506} {\bibfield  {journal} {\bibinfo  {journal} {Phys.
  Rev. D}\ }\textbf {\bibinfo {volume} {86}},\ \bibinfo {pages} {074506}
  (\bibinfo {year} {2012})},\ \Eprint {http://arxiv.org/abs/1205.3996}
  {arXiv:1205.3996 [hep-lat]} \BibitemShut {NoStop}%
\bibitem [{\citenamefont {Aarts}(2013)}]{Aarts:2013fpa}%
  \BibitemOpen
  \bibfield  {author} {\bibinfo {author} {\bibfnamefont {G.}~\bibnamefont
  {Aarts}},\ }\href {\doibase 10.1103/PhysRevD.88.094501} {\bibfield  {journal}
  {\bibinfo  {journal} {Phys. Rev. D}\ }\textbf {\bibinfo {volume} {88}},\
  \bibinfo {pages} {094501} (\bibinfo {year} {2013})},\ \Eprint
  {http://arxiv.org/abs/1308.4811} {arXiv:1308.4811 [hep-lat]} \BibitemShut
  {NoStop}%
\bibitem [{\citenamefont {Mukherjee}\ \emph {et~al.}(2013)\citenamefont
  {Mukherjee}, \citenamefont {Cristoforetti},\ and\ \citenamefont
  {Scorzato}}]{Mukherjee:2013aga}%
  \BibitemOpen
  \bibfield  {author} {\bibinfo {author} {\bibfnamefont {A.}~\bibnamefont
  {Mukherjee}}, \bibinfo {author} {\bibfnamefont {M.}~\bibnamefont
  {Cristoforetti}}, \ and\ \bibinfo {author} {\bibfnamefont {L.}~\bibnamefont
  {Scorzato}},\ }\href {\doibase 10.1103/PhysRevD.88.051502} {\bibfield
  {journal} {\bibinfo  {journal} {Phys. Rev. D}\ }\textbf {\bibinfo {volume}
  {88}},\ \bibinfo {pages} {051502} (\bibinfo {year} {2013})},\ \Eprint
  {http://arxiv.org/abs/1308.0233} {arXiv:1308.0233 [physics.comp-ph]}
  \BibitemShut {NoStop}%
\bibitem [{\citenamefont {Aarts}\ \emph {et~al.}(2014)\citenamefont {Aarts},
  \citenamefont {Bongiovanni}, \citenamefont {Seiler},\ and\ \citenamefont
  {Sexty}}]{Aarts:2014nxa}%
  \BibitemOpen
  \bibfield  {author} {\bibinfo {author} {\bibfnamefont {G.}~\bibnamefont
  {Aarts}}, \bibinfo {author} {\bibfnamefont {L.}~\bibnamefont {Bongiovanni}},
  \bibinfo {author} {\bibfnamefont {E.}~\bibnamefont {Seiler}}, \ and\ \bibinfo
  {author} {\bibfnamefont {D.}~\bibnamefont {Sexty}},\ }\href {\doibase
  10.1007/JHEP10(2014)159} {\bibfield  {journal} {\bibinfo  {journal} {JHEP}\
  }\textbf {\bibinfo {volume} {10}},\ \bibinfo {pages} {159} (\bibinfo {year}
  {2014})},\ \Eprint {http://arxiv.org/abs/1407.2090} {arXiv:1407.2090
  [hep-lat]} \BibitemShut {NoStop}%
\bibitem [{\citenamefont {Schmidt}\ and\ \citenamefont
  {Ziesch\'e}(2017)}]{Schmidt:2017gvu}%
  \BibitemOpen
  \bibfield  {author} {\bibinfo {author} {\bibfnamefont {C.}~\bibnamefont
  {Schmidt}}\ and\ \bibinfo {author} {\bibfnamefont {F.}~\bibnamefont
  {Ziesch\'e}},\ }\href {\doibase 10.22323/1.256.0076} {\bibfield  {journal}
  {\bibinfo  {journal} {PoS}\ }\textbf {\bibinfo {volume} {LATTICE2016}},\
  \bibinfo {pages} {076} (\bibinfo {year} {2017})},\ \Eprint
  {http://arxiv.org/abs/1701.08959} {arXiv:1701.08959 [hep-lat]} \BibitemShut
  {NoStop}%
\bibitem [{\citenamefont {Di~Renzo}\ and\ \citenamefont
  {Eruzzi}(2018)}]{DiRenzo:2017igr}%
  \BibitemOpen
  \bibfield  {author} {\bibinfo {author} {\bibfnamefont {F.}~\bibnamefont
  {Di~Renzo}}\ and\ \bibinfo {author} {\bibfnamefont {G.}~\bibnamefont
  {Eruzzi}},\ }\href {\doibase 10.1103/PhysRevD.97.014503} {\bibfield
  {journal} {\bibinfo  {journal} {Phys. Rev. D}\ }\textbf {\bibinfo {volume}
  {97}},\ \bibinfo {pages} {014503} (\bibinfo {year} {2018})},\ \Eprint
  {http://arxiv.org/abs/1709.10468} {arXiv:1709.10468 [hep-lat]} \BibitemShut
  {NoStop}%
\bibitem [{\citenamefont {Kashiwa}\ \emph
  {et~al.}(2019{\natexlab{a}})\citenamefont {Kashiwa}, \citenamefont {Mori},\
  and\ \citenamefont {Ohnishi}}]{Kashiwa:2018vxr}%
  \BibitemOpen
  \bibfield  {author} {\bibinfo {author} {\bibfnamefont {K.}~\bibnamefont
  {Kashiwa}}, \bibinfo {author} {\bibfnamefont {Y.}~\bibnamefont {Mori}}, \
  and\ \bibinfo {author} {\bibfnamefont {A.}~\bibnamefont {Ohnishi}},\ }\href
  {\doibase 10.1103/PhysRevD.99.014033} {\bibfield  {journal} {\bibinfo
  {journal} {Phys. Rev. D}\ }\textbf {\bibinfo {volume} {99}},\ \bibinfo
  {pages} {014033} (\bibinfo {year} {2019}{\natexlab{a}})},\ \Eprint
  {http://arxiv.org/abs/1805.08940} {arXiv:1805.08940 [hep-ph]} \BibitemShut
  {NoStop}%
\bibitem [{\citenamefont {Alexandru}\ \emph
  {et~al.}(2018{\natexlab{a}})\citenamefont {Alexandru}, \citenamefont
  {Ba\c{s}ar}, \citenamefont {Bedaque}, \citenamefont {Lamm},\ and\
  \citenamefont {Lawrence}}]{Alexandru:2018ngw}%
  \BibitemOpen
  \bibfield  {author} {\bibinfo {author} {\bibfnamefont {A.}~\bibnamefont
  {Alexandru}}, \bibinfo {author} {\bibfnamefont {G.}~\bibnamefont
  {Ba\c{s}ar}}, \bibinfo {author} {\bibfnamefont {P.~F.}\ \bibnamefont
  {Bedaque}}, \bibinfo {author} {\bibfnamefont {H.}~\bibnamefont {Lamm}}, \
  and\ \bibinfo {author} {\bibfnamefont {S.}~\bibnamefont {Lawrence}},\ }\href
  {\doibase 10.1103/PhysRevD.98.034506} {\bibfield  {journal} {\bibinfo
  {journal} {Phys. Rev. D}\ }\textbf {\bibinfo {volume} {98}},\ \bibinfo
  {pages} {034506} (\bibinfo {year} {2018}{\natexlab{a}})},\ \Eprint
  {http://arxiv.org/abs/1807.02027} {arXiv:1807.02027 [hep-lat]} \BibitemShut
  {NoStop}%
\bibitem [{\citenamefont {Kashiwa}\ \emph
  {et~al.}(2019{\natexlab{b}})\citenamefont {Kashiwa}, \citenamefont {Mori},\
  and\ \citenamefont {Ohnishi}}]{Kashiwa:2019lkv}%
  \BibitemOpen
  \bibfield  {author} {\bibinfo {author} {\bibfnamefont {K.}~\bibnamefont
  {Kashiwa}}, \bibinfo {author} {\bibfnamefont {Y.}~\bibnamefont {Mori}}, \
  and\ \bibinfo {author} {\bibfnamefont {A.}~\bibnamefont {Ohnishi}},\ }\href
  {\doibase 10.1103/PhysRevD.99.114005} {\bibfield  {journal} {\bibinfo
  {journal} {Phys. Rev. D}\ }\textbf {\bibinfo {volume} {99}},\ \bibinfo
  {pages} {114005} (\bibinfo {year} {2019}{\natexlab{b}})},\ \Eprint
  {http://arxiv.org/abs/1903.03679} {arXiv:1903.03679 [hep-lat]} \BibitemShut
  {NoStop}%
\bibitem [{\citenamefont {Detmold}\ \emph {et~al.}(2020)\citenamefont
  {Detmold}, \citenamefont {Kanwar}, \citenamefont {Wagman},\ and\
  \citenamefont {Warrington}}]{Detmold:2020ncp}%
  \BibitemOpen
  \bibfield  {author} {\bibinfo {author} {\bibfnamefont {W.}~\bibnamefont
  {Detmold}}, \bibinfo {author} {\bibfnamefont {G.}~\bibnamefont {Kanwar}},
  \bibinfo {author} {\bibfnamefont {M.~L.}\ \bibnamefont {Wagman}}, \ and\
  \bibinfo {author} {\bibfnamefont {N.~C.}\ \bibnamefont {Warrington}},\ }\href
  {\doibase 10.1103/PhysRevD.102.014514} {\bibfield  {journal} {\bibinfo
  {journal} {Phys. Rev. D}\ }\textbf {\bibinfo {volume} {102}},\ \bibinfo
  {pages} {014514} (\bibinfo {year} {2020})},\ \Eprint
  {http://arxiv.org/abs/2003.05914} {arXiv:2003.05914 [hep-lat]} \BibitemShut
  {NoStop}%
\bibitem [{\citenamefont {Pawlowski}\ \emph {et~al.}(2021)\citenamefont
  {Pawlowski}, \citenamefont {Scherzer}, \citenamefont {Schmidt}, \citenamefont
  {Ziegler},\ and\ \citenamefont {Ziesch\'e}}]{Pawlowski:2021bbu}%
  \BibitemOpen
  \bibfield  {author} {\bibinfo {author} {\bibfnamefont {J.~M.}\ \bibnamefont
  {Pawlowski}}, \bibinfo {author} {\bibfnamefont {M.}~\bibnamefont {Scherzer}},
  \bibinfo {author} {\bibfnamefont {C.}~\bibnamefont {Schmidt}}, \bibinfo
  {author} {\bibfnamefont {F.~P.~G.}\ \bibnamefont {Ziegler}}, \ and\ \bibinfo
  {author} {\bibfnamefont {F.}~\bibnamefont {Ziesch\'e}},\ }\href@noop {} {\
  (\bibinfo {year} {2021})},\ \Eprint {http://arxiv.org/abs/2101.03938}
  {arXiv:2101.03938 [hep-lat]} \BibitemShut {NoStop}%
\bibitem [{\citenamefont {Detmold}\ \emph {et~al.}(2021)\citenamefont
  {Detmold}, \citenamefont {Kanwar}, \citenamefont {Lamm}, \citenamefont
  {Wagman},\ and\ \citenamefont {Warrington}}]{Detmold:2021ulb}%
  \BibitemOpen
  \bibfield  {author} {\bibinfo {author} {\bibfnamefont {W.}~\bibnamefont
  {Detmold}}, \bibinfo {author} {\bibfnamefont {G.}~\bibnamefont {Kanwar}},
  \bibinfo {author} {\bibfnamefont {H.}~\bibnamefont {Lamm}}, \bibinfo {author}
  {\bibfnamefont {M.~L.}\ \bibnamefont {Wagman}}, \ and\ \bibinfo {author}
  {\bibfnamefont {N.~C.}\ \bibnamefont {Warrington}},\ }\href {\doibase
  10.1103/PhysRevD.103.094517} {\bibfield  {journal} {\bibinfo  {journal}
  {Phys. Rev. D}\ }\textbf {\bibinfo {volume} {103}},\ \bibinfo {pages}
  {094517} (\bibinfo {year} {2021})},\ \Eprint
  {http://arxiv.org/abs/2101.12668} {arXiv:2101.12668 [hep-lat]} \BibitemShut
  {NoStop}%
\bibitem [{\citenamefont {Kanwar}\ and\ \citenamefont
  {Wagman}(2021)}]{Kanwar:2021tkd}%
  \BibitemOpen
  \bibfield  {author} {\bibinfo {author} {\bibfnamefont {G.}~\bibnamefont
  {Kanwar}}\ and\ \bibinfo {author} {\bibfnamefont {M.~L.}\ \bibnamefont
  {Wagman}},\ }\href {\doibase 10.1103/PhysRevD.104.014513} {\bibfield
  {journal} {\bibinfo  {journal} {Phys. Rev. D}\ }\textbf {\bibinfo {volume}
  {104}},\ \bibinfo {pages} {014513} (\bibinfo {year} {2021})},\ \Eprint
  {http://arxiv.org/abs/2103.02602} {arXiv:2103.02602 [hep-lat]} \BibitemShut
  {NoStop}%
\bibitem [{\citenamefont {Cristoforetti}\ \emph {et~al.}(2013)\citenamefont
  {Cristoforetti}, \citenamefont {Di~Renzo}, \citenamefont {Mukherjee},\ and\
  \citenamefont {Scorzato}}]{Cristoforetti:2013wha}%
  \BibitemOpen
  \bibfield  {author} {\bibinfo {author} {\bibfnamefont {M.}~\bibnamefont
  {Cristoforetti}}, \bibinfo {author} {\bibfnamefont {F.}~\bibnamefont
  {Di~Renzo}}, \bibinfo {author} {\bibfnamefont {A.}~\bibnamefont {Mukherjee}},
  \ and\ \bibinfo {author} {\bibfnamefont {L.}~\bibnamefont {Scorzato}},\
  }\href {\doibase 10.1103/PhysRevD.88.051501} {\bibfield  {journal} {\bibinfo
  {journal} {Phys. Rev. D}\ }\textbf {\bibinfo {volume} {88}},\ \bibinfo
  {pages} {051501} (\bibinfo {year} {2013})},\ \Eprint
  {http://arxiv.org/abs/1303.7204} {arXiv:1303.7204 [hep-lat]} \BibitemShut
  {NoStop}%
\bibitem [{\citenamefont {Fujii}\ \emph {et~al.}(2013)\citenamefont {Fujii},
  \citenamefont {Honda}, \citenamefont {Kato}, \citenamefont {Kikukawa},
  \citenamefont {Komatsu},\ and\ \citenamefont {Sano}}]{Fujii:2013sra}%
  \BibitemOpen
  \bibfield  {author} {\bibinfo {author} {\bibfnamefont {H.}~\bibnamefont
  {Fujii}}, \bibinfo {author} {\bibfnamefont {D.}~\bibnamefont {Honda}},
  \bibinfo {author} {\bibfnamefont {M.}~\bibnamefont {Kato}}, \bibinfo {author}
  {\bibfnamefont {Y.}~\bibnamefont {Kikukawa}}, \bibinfo {author}
  {\bibfnamefont {S.}~\bibnamefont {Komatsu}}, \ and\ \bibinfo {author}
  {\bibfnamefont {T.}~\bibnamefont {Sano}},\ }\href {\doibase
  10.1007/JHEP10(2013)147} {\bibfield  {journal} {\bibinfo  {journal} {JHEP}\
  }\textbf {\bibinfo {volume} {10}},\ \bibinfo {pages} {147} (\bibinfo {year}
  {2013})},\ \Eprint {http://arxiv.org/abs/1309.4371} {arXiv:1309.4371
  [hep-lat]} \BibitemShut {NoStop}%
\bibitem [{\citenamefont {Cristoforetti}\ \emph {et~al.}(2014)\citenamefont
  {Cristoforetti}, \citenamefont {Di~Renzo}, \citenamefont {Eruzzi},
  \citenamefont {Mukherjee}, \citenamefont {Schmidt}, \citenamefont
  {Scorzato},\ and\ \citenamefont {Torrero}}]{Cristoforetti:2014gsa}%
  \BibitemOpen
  \bibfield  {author} {\bibinfo {author} {\bibfnamefont {M.}~\bibnamefont
  {Cristoforetti}}, \bibinfo {author} {\bibfnamefont {F.}~\bibnamefont
  {Di~Renzo}}, \bibinfo {author} {\bibfnamefont {G.}~\bibnamefont {Eruzzi}},
  \bibinfo {author} {\bibfnamefont {A.}~\bibnamefont {Mukherjee}}, \bibinfo
  {author} {\bibfnamefont {C.}~\bibnamefont {Schmidt}}, \bibinfo {author}
  {\bibfnamefont {L.}~\bibnamefont {Scorzato}}, \ and\ \bibinfo {author}
  {\bibfnamefont {C.}~\bibnamefont {Torrero}},\ }\href {\doibase
  10.1103/PhysRevD.89.114505} {\bibfield  {journal} {\bibinfo  {journal} {Phys.
  Rev. D}\ }\textbf {\bibinfo {volume} {89}},\ \bibinfo {pages} {114505}
  (\bibinfo {year} {2014})},\ \Eprint {http://arxiv.org/abs/1403.5637}
  {arXiv:1403.5637 [hep-lat]} \BibitemShut {NoStop}%
\bibitem [{\citenamefont {Alexandru}\ \emph
  {et~al.}(2016{\natexlab{a}})\citenamefont {Alexandru}, \citenamefont
  {Basar},\ and\ \citenamefont {Bedaque}}]{Alexandru:2015xva}%
  \BibitemOpen
  \bibfield  {author} {\bibinfo {author} {\bibfnamefont {A.}~\bibnamefont
  {Alexandru}}, \bibinfo {author} {\bibfnamefont {G.}~\bibnamefont {Basar}}, \
  and\ \bibinfo {author} {\bibfnamefont {P.}~\bibnamefont {Bedaque}},\ }\href
  {\doibase 10.1103/PhysRevD.93.014504} {\bibfield  {journal} {\bibinfo
  {journal} {Phys. Rev. D}\ }\textbf {\bibinfo {volume} {93}},\ \bibinfo
  {pages} {014504} (\bibinfo {year} {2016}{\natexlab{a}})},\ \Eprint
  {http://arxiv.org/abs/1510.03258} {arXiv:1510.03258 [hep-lat]} \BibitemShut
  {NoStop}%
\bibitem [{\citenamefont {Alexandru}\ \emph
  {et~al.}(2016{\natexlab{b}})\citenamefont {Alexandru}, \citenamefont {Basar},
  \citenamefont {Bedaque}, \citenamefont {Ridgway},\ and\ \citenamefont
  {Warrington}}]{Alexandru:2015sua}%
  \BibitemOpen
  \bibfield  {author} {\bibinfo {author} {\bibfnamefont {A.}~\bibnamefont
  {Alexandru}}, \bibinfo {author} {\bibfnamefont {G.}~\bibnamefont {Basar}},
  \bibinfo {author} {\bibfnamefont {P.~F.}\ \bibnamefont {Bedaque}}, \bibinfo
  {author} {\bibfnamefont {G.~W.}\ \bibnamefont {Ridgway}}, \ and\ \bibinfo
  {author} {\bibfnamefont {N.~C.}\ \bibnamefont {Warrington}},\ }\href
  {\doibase 10.1007/JHEP05(2016)053} {\bibfield  {journal} {\bibinfo  {journal}
  {JHEP}\ }\textbf {\bibinfo {volume} {05}},\ \bibinfo {pages} {053} (\bibinfo
  {year} {2016}{\natexlab{b}})},\ \Eprint {http://arxiv.org/abs/1512.08764}
  {arXiv:1512.08764 [hep-lat]} \BibitemShut {NoStop}%
\bibitem [{\citenamefont {Fujii}\ \emph {et~al.}(2015)\citenamefont {Fujii},
  \citenamefont {Kamata},\ and\ \citenamefont {Kikukawa}}]{Fujii:2015vha}%
  \BibitemOpen
  \bibfield  {author} {\bibinfo {author} {\bibfnamefont {H.}~\bibnamefont
  {Fujii}}, \bibinfo {author} {\bibfnamefont {S.}~\bibnamefont {Kamata}}, \
  and\ \bibinfo {author} {\bibfnamefont {Y.}~\bibnamefont {Kikukawa}},\ }\href
  {\doibase 10.1007/JHEP12(2015)125} {\bibfield  {journal} {\bibinfo  {journal}
  {JHEP}\ }\textbf {\bibinfo {volume} {12}},\ \bibinfo {pages} {125} (\bibinfo
  {year} {2015})},\ \bibinfo {note} {[Erratum: JHEP 09, 172 (2016)]},\ \Eprint
  {http://arxiv.org/abs/1509.09141} {arXiv:1509.09141 [hep-lat]} \BibitemShut
  {NoStop}%
\bibitem [{\citenamefont {Alexandru}\ \emph
  {et~al.}(2016{\natexlab{c}})\citenamefont {Alexandru}, \citenamefont {Basar},
  \citenamefont {Bedaque}, \citenamefont {Vartak},\ and\ \citenamefont
  {Warrington}}]{Alexandru:2016gsd}%
  \BibitemOpen
  \bibfield  {author} {\bibinfo {author} {\bibfnamefont {A.}~\bibnamefont
  {Alexandru}}, \bibinfo {author} {\bibfnamefont {G.}~\bibnamefont {Basar}},
  \bibinfo {author} {\bibfnamefont {P.~F.}\ \bibnamefont {Bedaque}}, \bibinfo
  {author} {\bibfnamefont {S.}~\bibnamefont {Vartak}}, \ and\ \bibinfo {author}
  {\bibfnamefont {N.~C.}\ \bibnamefont {Warrington}},\ }\href {\doibase
  10.1103/PhysRevLett.117.081602} {\bibfield  {journal} {\bibinfo  {journal}
  {Phys. Rev. Lett.}\ }\textbf {\bibinfo {volume} {117}},\ \bibinfo {pages}
  {081602} (\bibinfo {year} {2016}{\natexlab{c}})},\ \Eprint
  {http://arxiv.org/abs/1605.08040} {arXiv:1605.08040 [hep-lat]} \BibitemShut
  {NoStop}%
\bibitem [{\citenamefont {Alexandru}\ \emph
  {et~al.}(2017{\natexlab{a}})\citenamefont {Alexandru}, \citenamefont {Basar},
  \citenamefont {Bedaque}, \citenamefont {Ridgway},\ and\ \citenamefont
  {Warrington}}]{Alexandru:2016ejd}%
  \BibitemOpen
  \bibfield  {author} {\bibinfo {author} {\bibfnamefont {A.}~\bibnamefont
  {Alexandru}}, \bibinfo {author} {\bibfnamefont {G.}~\bibnamefont {Basar}},
  \bibinfo {author} {\bibfnamefont {P.~F.}\ \bibnamefont {Bedaque}}, \bibinfo
  {author} {\bibfnamefont {G.~W.}\ \bibnamefont {Ridgway}}, \ and\ \bibinfo
  {author} {\bibfnamefont {N.~C.}\ \bibnamefont {Warrington}},\ }\href
  {\doibase 10.1103/PhysRevD.95.014502} {\bibfield  {journal} {\bibinfo
  {journal} {Phys. Rev. D}\ }\textbf {\bibinfo {volume} {95}},\ \bibinfo
  {pages} {014502} (\bibinfo {year} {2017}{\natexlab{a}})},\ \Eprint
  {http://arxiv.org/abs/1609.01730} {arXiv:1609.01730 [hep-lat]} \BibitemShut
  {NoStop}%
\bibitem [{\citenamefont {Alexandru}\ \emph
  {et~al.}(2017{\natexlab{b}})\citenamefont {Alexandru}, \citenamefont
  {Bedaque}, \citenamefont {Lamm},\ and\ \citenamefont
  {Lawrence}}]{Alexandru:2017czx}%
  \BibitemOpen
  \bibfield  {author} {\bibinfo {author} {\bibfnamefont {A.}~\bibnamefont
  {Alexandru}}, \bibinfo {author} {\bibfnamefont {P.~F.}\ \bibnamefont
  {Bedaque}}, \bibinfo {author} {\bibfnamefont {H.}~\bibnamefont {Lamm}}, \
  and\ \bibinfo {author} {\bibfnamefont {S.}~\bibnamefont {Lawrence}},\ }\href
  {\doibase 10.1103/PhysRevD.96.094505} {\bibfield  {journal} {\bibinfo
  {journal} {Phys. Rev. D}\ }\textbf {\bibinfo {volume} {96}},\ \bibinfo
  {pages} {094505} (\bibinfo {year} {2017}{\natexlab{b}})},\ \Eprint
  {http://arxiv.org/abs/1709.01971} {arXiv:1709.01971 [hep-lat]} \BibitemShut
  {NoStop}%
\bibitem [{\citenamefont {Alexandru}\ \emph
  {et~al.}(2017{\natexlab{c}})\citenamefont {Alexandru}, \citenamefont {Basar},
  \citenamefont {Bedaque},\ and\ \citenamefont {Ridgway}}]{Alexandru:2017lqr}%
  \BibitemOpen
  \bibfield  {author} {\bibinfo {author} {\bibfnamefont {A.}~\bibnamefont
  {Alexandru}}, \bibinfo {author} {\bibfnamefont {G.}~\bibnamefont {Basar}},
  \bibinfo {author} {\bibfnamefont {P.~F.}\ \bibnamefont {Bedaque}}, \ and\
  \bibinfo {author} {\bibfnamefont {G.~W.}\ \bibnamefont {Ridgway}},\ }\href
  {\doibase 10.1103/PhysRevD.95.114501} {\bibfield  {journal} {\bibinfo
  {journal} {Phys. Rev. D}\ }\textbf {\bibinfo {volume} {95}},\ \bibinfo
  {pages} {114501} (\bibinfo {year} {2017}{\natexlab{c}})},\ \Eprint
  {http://arxiv.org/abs/1704.06404} {arXiv:1704.06404 [hep-lat]} \BibitemShut
  {NoStop}%
\bibitem [{\citenamefont {Mori}\ \emph {et~al.}(2018)\citenamefont {Mori},
  \citenamefont {Kashiwa},\ and\ \citenamefont {Ohnishi}}]{Mori:2017nwj}%
  \BibitemOpen
  \bibfield  {author} {\bibinfo {author} {\bibfnamefont {Y.}~\bibnamefont
  {Mori}}, \bibinfo {author} {\bibfnamefont {K.}~\bibnamefont {Kashiwa}}, \
  and\ \bibinfo {author} {\bibfnamefont {A.}~\bibnamefont {Ohnishi}},\ }\href
  {\doibase 10.1093/ptep/ptx191} {\bibfield  {journal} {\bibinfo  {journal}
  {PTEP}\ }\textbf {\bibinfo {volume} {2018}},\ \bibinfo {pages} {023B04}
  (\bibinfo {year} {2018})},\ \Eprint {http://arxiv.org/abs/1709.03208}
  {arXiv:1709.03208 [hep-lat]} \BibitemShut {NoStop}%
\bibitem [{\citenamefont {Tanizaki}\ \emph {et~al.}(2017)\citenamefont
  {Tanizaki}, \citenamefont {Nishimura},\ and\ \citenamefont
  {Verbaarschot}}]{Tanizaki:2017yow}%
  \BibitemOpen
  \bibfield  {author} {\bibinfo {author} {\bibfnamefont {Y.}~\bibnamefont
  {Tanizaki}}, \bibinfo {author} {\bibfnamefont {H.}~\bibnamefont {Nishimura}},
  \ and\ \bibinfo {author} {\bibfnamefont {J.~J.~M.}\ \bibnamefont
  {Verbaarschot}},\ }\href {\doibase 10.1007/JHEP10(2017)100} {\bibfield
  {journal} {\bibinfo  {journal} {JHEP}\ }\textbf {\bibinfo {volume} {10}},\
  \bibinfo {pages} {100} (\bibinfo {year} {2017})},\ \Eprint
  {http://arxiv.org/abs/1706.03822} {arXiv:1706.03822 [hep-lat]} \BibitemShut
  {NoStop}%
\bibitem [{\citenamefont {Alexandru}\ \emph
  {et~al.}(2018{\natexlab{b}})\citenamefont {Alexandru}, \citenamefont
  {Bedaque},\ and\ \citenamefont {Warrington}}]{Alexandru:2018brw}%
  \BibitemOpen
  \bibfield  {author} {\bibinfo {author} {\bibfnamefont {A.}~\bibnamefont
  {Alexandru}}, \bibinfo {author} {\bibfnamefont {P.~F.}\ \bibnamefont
  {Bedaque}}, \ and\ \bibinfo {author} {\bibfnamefont {N.~C.}\ \bibnamefont
  {Warrington}},\ }\href {\doibase 10.1103/PhysRevD.98.054514} {\bibfield
  {journal} {\bibinfo  {journal} {Phys. Rev. D}\ }\textbf {\bibinfo {volume}
  {98}},\ \bibinfo {pages} {054514} (\bibinfo {year} {2018}{\natexlab{b}})},\
  \Eprint {http://arxiv.org/abs/1805.00125} {arXiv:1805.00125 [hep-lat]}
  \BibitemShut {NoStop}%
\bibitem [{\citenamefont {Alexandru}\ \emph
  {et~al.}(2018{\natexlab{c}})\citenamefont {Alexandru}, \citenamefont
  {Bedaque}, \citenamefont {Lamm},\ and\ \citenamefont
  {Lawrence}}]{Alexandru:2018fqp}%
  \BibitemOpen
  \bibfield  {author} {\bibinfo {author} {\bibfnamefont {A.}~\bibnamefont
  {Alexandru}}, \bibinfo {author} {\bibfnamefont {P.~F.}\ \bibnamefont
  {Bedaque}}, \bibinfo {author} {\bibfnamefont {H.}~\bibnamefont {Lamm}}, \
  and\ \bibinfo {author} {\bibfnamefont {S.}~\bibnamefont {Lawrence}},\ }\href
  {\doibase 10.1103/PhysRevD.97.094510} {\bibfield  {journal} {\bibinfo
  {journal} {Phys. Rev. D}\ }\textbf {\bibinfo {volume} {97}},\ \bibinfo
  {pages} {094510} (\bibinfo {year} {2018}{\natexlab{c}})},\ \Eprint
  {http://arxiv.org/abs/1804.00697} {arXiv:1804.00697 [hep-lat]} \BibitemShut
  {NoStop}%
\bibitem [{\citenamefont {Alexandru}\ \emph
  {et~al.}(2018{\natexlab{d}})\citenamefont {Alexandru}, \citenamefont
  {Bedaque}, \citenamefont {Lamm}, \citenamefont {Lawrence},\ and\
  \citenamefont {Warrington}}]{Alexandru:2018ddf}%
  \BibitemOpen
  \bibfield  {author} {\bibinfo {author} {\bibfnamefont {A.}~\bibnamefont
  {Alexandru}}, \bibinfo {author} {\bibfnamefont {P.~F.}\ \bibnamefont
  {Bedaque}}, \bibinfo {author} {\bibfnamefont {H.}~\bibnamefont {Lamm}},
  \bibinfo {author} {\bibfnamefont {S.}~\bibnamefont {Lawrence}}, \ and\
  \bibinfo {author} {\bibfnamefont {N.~C.}\ \bibnamefont {Warrington}},\ }\href
  {\doibase 10.1103/PhysRevLett.121.191602} {\bibfield  {journal} {\bibinfo
  {journal} {Phys. Rev. Lett.}\ }\textbf {\bibinfo {volume} {121}},\ \bibinfo
  {pages} {191602} (\bibinfo {year} {2018}{\natexlab{d}})},\ \Eprint
  {http://arxiv.org/abs/1808.09799} {arXiv:1808.09799 [hep-lat]} \BibitemShut
  {NoStop}%
\bibitem [{\citenamefont {Mou}\ \emph {et~al.}(2019)\citenamefont {Mou},
  \citenamefont {Saffin},\ and\ \citenamefont {Tranberg}}]{Mou:2019gyl}%
  \BibitemOpen
  \bibfield  {author} {\bibinfo {author} {\bibfnamefont {Z.-G.}\ \bibnamefont
  {Mou}}, \bibinfo {author} {\bibfnamefont {P.~M.}\ \bibnamefont {Saffin}}, \
  and\ \bibinfo {author} {\bibfnamefont {A.}~\bibnamefont {Tranberg}},\ }\href
  {\doibase 10.1007/JHEP11(2019)135} {\bibfield  {journal} {\bibinfo  {journal}
  {JHEP}\ }\textbf {\bibinfo {volume} {11}},\ \bibinfo {pages} {135} (\bibinfo
  {year} {2019})},\ \Eprint {http://arxiv.org/abs/1909.02488} {arXiv:1909.02488
  [hep-th]} \BibitemShut {NoStop}%
\bibitem [{\citenamefont {Lawrence}\ and\ \citenamefont
  {Yamauchi}(2021)}]{Lawrence:2021izu}%
  \BibitemOpen
  \bibfield  {author} {\bibinfo {author} {\bibfnamefont {S.}~\bibnamefont
  {Lawrence}}\ and\ \bibinfo {author} {\bibfnamefont {Y.}~\bibnamefont
  {Yamauchi}},\ }\href {\doibase 10.1103/PhysRevD.103.114509} {\bibfield
  {journal} {\bibinfo  {journal} {Phys. Rev. D}\ }\textbf {\bibinfo {volume}
  {103}},\ \bibinfo {pages} {114509} (\bibinfo {year} {2021})},\ \Eprint
  {http://arxiv.org/abs/2101.05755} {arXiv:2101.05755 [hep-lat]} \BibitemShut
  {NoStop}%
\bibitem [{\citenamefont {Di~Renzo}\ and\ \citenamefont
  {Zambello}(2022)}]{DiRenzo:2021kcw}%
  \BibitemOpen
  \bibfield  {author} {\bibinfo {author} {\bibfnamefont {F.}~\bibnamefont
  {Di~Renzo}}\ and\ \bibinfo {author} {\bibfnamefont {K.}~\bibnamefont
  {Zambello}},\ }\href {\doibase 10.1103/PhysRevD.105.054501} {\bibfield
  {journal} {\bibinfo  {journal} {Phys. Rev. D}\ }\textbf {\bibinfo {volume}
  {105}},\ \bibinfo {pages} {054501} (\bibinfo {year} {2022})},\ \Eprint
  {http://arxiv.org/abs/2109.02511} {arXiv:2109.02511 [hep-lat]} \BibitemShut
  {NoStop}%
\bibitem [{\citenamefont {Lawrence}\ and\ \citenamefont
  {Yamauchi}(2022)}]{Lawrence:2022dba}%
  \BibitemOpen
  \bibfield  {author} {\bibinfo {author} {\bibfnamefont {S.}~\bibnamefont
  {Lawrence}}\ and\ \bibinfo {author} {\bibfnamefont {Y.}~\bibnamefont
  {Yamauchi}},\ }\href@noop {} {\  (\bibinfo {year} {2022})},\ \Eprint
  {http://arxiv.org/abs/2212.14606} {arXiv:2212.14606 [hep-lat]} \BibitemShut
  {NoStop}%
\bibitem [{\citenamefont {Mukherjee}\ and\ \citenamefont
  {Cristoforetti}(2014)}]{Mukherjee:2014hsa}%
  \BibitemOpen
  \bibfield  {author} {\bibinfo {author} {\bibfnamefont {A.}~\bibnamefont
  {Mukherjee}}\ and\ \bibinfo {author} {\bibfnamefont {M.}~\bibnamefont
  {Cristoforetti}},\ }\href {\doibase 10.1103/PhysRevB.90.035134} {\bibfield
  {journal} {\bibinfo  {journal} {Phys. Rev. B}\ }\textbf {\bibinfo {volume}
  {90}},\ \bibinfo {pages} {035134} (\bibinfo {year} {2014})},\ \Eprint
  {http://arxiv.org/abs/1403.5680} {arXiv:1403.5680 [cond-mat.str-el]}
  \BibitemShut {NoStop}%
\bibitem [{\citenamefont {Tanizaki}\ \emph {et~al.}(2016)\citenamefont
  {Tanizaki}, \citenamefont {Hidaka},\ and\ \citenamefont
  {Hayata}}]{Tanizaki:2015rda}%
  \BibitemOpen
  \bibfield  {author} {\bibinfo {author} {\bibfnamefont {Y.}~\bibnamefont
  {Tanizaki}}, \bibinfo {author} {\bibfnamefont {Y.}~\bibnamefont {Hidaka}}, \
  and\ \bibinfo {author} {\bibfnamefont {T.}~\bibnamefont {Hayata}},\ }\href
  {\doibase 10.1088/1367-2630/18/3/033002} {\bibfield  {journal} {\bibinfo
  {journal} {New J. Phys.}\ }\textbf {\bibinfo {volume} {18}},\ \bibinfo
  {pages} {033002} (\bibinfo {year} {2016})},\ \Eprint
  {http://arxiv.org/abs/1509.07146} {arXiv:1509.07146 [hep-th]} \BibitemShut
  {NoStop}%
\bibitem [{\citenamefont {Fukuma}\ \emph
  {et~al.}(2019{\natexlab{a}})\citenamefont {Fukuma}, \citenamefont
  {Matsumoto},\ and\ \citenamefont {Umeda}}]{Fukuma:2019uot}%
  \BibitemOpen
  \bibfield  {author} {\bibinfo {author} {\bibfnamefont {M.}~\bibnamefont
  {Fukuma}}, \bibinfo {author} {\bibfnamefont {N.}~\bibnamefont {Matsumoto}}, \
  and\ \bibinfo {author} {\bibfnamefont {N.}~\bibnamefont {Umeda}},\
  }\href@noop {} {\  (\bibinfo {year} {2019}{\natexlab{a}})},\ \Eprint
  {http://arxiv.org/abs/1912.13303} {arXiv:1912.13303 [hep-lat]} \BibitemShut
  {NoStop}%
\bibitem [{\citenamefont {Fukuma}\ \emph
  {et~al.}(2019{\natexlab{b}})\citenamefont {Fukuma}, \citenamefont
  {Matsumoto},\ and\ \citenamefont {Umeda}}]{Fukuma:2019wbv}%
  \BibitemOpen
  \bibfield  {author} {\bibinfo {author} {\bibfnamefont {M.}~\bibnamefont
  {Fukuma}}, \bibinfo {author} {\bibfnamefont {N.}~\bibnamefont {Matsumoto}}, \
  and\ \bibinfo {author} {\bibfnamefont {N.}~\bibnamefont {Umeda}},\ }\href
  {\doibase 10.1103/PhysRevD.100.114510} {\bibfield  {journal} {\bibinfo
  {journal} {Phys. Rev. D}\ }\textbf {\bibinfo {volume} {100}},\ \bibinfo
  {pages} {114510} (\bibinfo {year} {2019}{\natexlab{b}})},\ \Eprint
  {http://arxiv.org/abs/1906.04243} {arXiv:1906.04243 [cond-mat.str-el]}
  \BibitemShut {NoStop}%
\bibitem [{\citenamefont {Ulybyshev}\ \emph {et~al.}(2020)\citenamefont
  {Ulybyshev}, \citenamefont {Winterowd},\ and\ \citenamefont
  {Zafeiropoulos}}]{Ulybyshev:2019fte}%
  \BibitemOpen
  \bibfield  {author} {\bibinfo {author} {\bibfnamefont {M.}~\bibnamefont
  {Ulybyshev}}, \bibinfo {author} {\bibfnamefont {C.}~\bibnamefont
  {Winterowd}}, \ and\ \bibinfo {author} {\bibfnamefont {S.}~\bibnamefont
  {Zafeiropoulos}},\ }\href {\doibase 10.1103/PhysRevD.101.014508} {\bibfield
  {journal} {\bibinfo  {journal} {Phys. Rev. D}\ }\textbf {\bibinfo {volume}
  {101}},\ \bibinfo {pages} {014508} (\bibinfo {year} {2020})},\ \Eprint
  {http://arxiv.org/abs/1906.07678} {arXiv:1906.07678 [cond-mat.str-el]}
  \BibitemShut {NoStop}%
\bibitem [{\citenamefont {Mishchenko}\ \emph {et~al.}(2021)\citenamefont
  {Mishchenko}, \citenamefont {Kato},\ and\ \citenamefont
  {Motome}}]{Mishchenko:2021vnx}%
  \BibitemOpen
  \bibfield  {author} {\bibinfo {author} {\bibfnamefont {P.~A.}\ \bibnamefont
  {Mishchenko}}, \bibinfo {author} {\bibfnamefont {Y.}~\bibnamefont {Kato}}, \
  and\ \bibinfo {author} {\bibfnamefont {Y.}~\bibnamefont {Motome}},\ }\href
  {\doibase 10.1103/PhysRevD.104.074517} {\bibfield  {journal} {\bibinfo
  {journal} {Phys. Rev. D}\ }\textbf {\bibinfo {volume} {104}},\ \bibinfo
  {pages} {074517} (\bibinfo {year} {2021})},\ \Eprint
  {http://arxiv.org/abs/2106.07937} {arXiv:2106.07937 [cond-mat.str-el]}
  \BibitemShut {NoStop}%
\bibitem [{\citenamefont {Rodekamp}\ \emph {et~al.}(2022)\citenamefont
  {Rodekamp}, \citenamefont {Berkowitz}, \citenamefont {G\"antgen},
  \citenamefont {Krieg}, \citenamefont {Luu},\ and\ \citenamefont
  {Ostmeyer}}]{Rodekamp:2022xpf}%
  \BibitemOpen
  \bibfield  {author} {\bibinfo {author} {\bibfnamefont {M.}~\bibnamefont
  {Rodekamp}}, \bibinfo {author} {\bibfnamefont {E.}~\bibnamefont {Berkowitz}},
  \bibinfo {author} {\bibfnamefont {C.}~\bibnamefont {G\"antgen}}, \bibinfo
  {author} {\bibfnamefont {S.}~\bibnamefont {Krieg}}, \bibinfo {author}
  {\bibfnamefont {T.}~\bibnamefont {Luu}}, \ and\ \bibinfo {author}
  {\bibfnamefont {J.}~\bibnamefont {Ostmeyer}},\ }\href {\doibase
  10.1103/PhysRevB.106.125139} {\bibfield  {journal} {\bibinfo  {journal}
  {Phys. Rev. B}\ }\textbf {\bibinfo {volume} {106}},\ \bibinfo {pages}
  {125139} (\bibinfo {year} {2022})},\ \Eprint
  {http://arxiv.org/abs/2203.00390} {arXiv:2203.00390 [physics.comp-ph]}
  \BibitemShut {NoStop}%
\bibitem [{\citenamefont {Rom}\ \emph {et~al.}(1997)\citenamefont {Rom},
  \citenamefont {Charutz},\ and\ \citenamefont {Neuhauser}}]{Rom:1997}%
  \BibitemOpen
  \bibfield  {author} {\bibinfo {author} {\bibfnamefont {N.}~\bibnamefont
  {Rom}}, \bibinfo {author} {\bibfnamefont {D.}~\bibnamefont {Charutz}}, \ and\
  \bibinfo {author} {\bibfnamefont {D.}~\bibnamefont {Neuhauser}},\ }\href
  {\doibase https://doi.org/10.1016/S0009-2614(97)00370-9} {\bibfield
  {journal} {\bibinfo  {journal} {Chemical Physics Letters}\ }\textbf {\bibinfo
  {volume} {270}},\ \bibinfo {pages} {382} (\bibinfo {year}
  {1997})}\BibitemShut {NoStop}%
\bibitem [{\citenamefont {Rom}\ \emph {et~al.}(1998)\citenamefont {Rom},
  \citenamefont {Fattal}, \citenamefont {Gupta}, \citenamefont {Carter},\ and\
  \citenamefont {Neuhauser}}]{Rom:1998}%
  \BibitemOpen
  \bibfield  {author} {\bibinfo {author} {\bibfnamefont {N.}~\bibnamefont
  {Rom}}, \bibinfo {author} {\bibfnamefont {E.}~\bibnamefont {Fattal}},
  \bibinfo {author} {\bibfnamefont {A.~K.}\ \bibnamefont {Gupta}}, \bibinfo
  {author} {\bibfnamefont {E.~A.}\ \bibnamefont {Carter}}, \ and\ \bibinfo
  {author} {\bibfnamefont {D.}~\bibnamefont {Neuhauser}},\ }\href@noop {}
  {\bibfield  {journal} {\bibinfo  {journal} {The Journal of chemical physics}\
  }\textbf {\bibinfo {volume} {109}},\ \bibinfo {pages} {8241} (\bibinfo {year}
  {1998})}\BibitemShut {NoStop}%
\bibitem [{\citenamefont {Baer}\ \emph {et~al.}(1998)\citenamefont {Baer},
  \citenamefont {Head-Gordon},\ and\ \citenamefont {Neuhauser}}]{Baer:1998}%
  \BibitemOpen
  \bibfield  {author} {\bibinfo {author} {\bibfnamefont {R.}~\bibnamefont
  {Baer}}, \bibinfo {author} {\bibfnamefont {M.}~\bibnamefont {Head-Gordon}}, \
  and\ \bibinfo {author} {\bibfnamefont {D.}~\bibnamefont {Neuhauser}},\
  }\href@noop {} {\bibfield  {journal} {\bibinfo  {journal} {The Journal of
  chemical physics}\ }\textbf {\bibinfo {volume} {109}},\ \bibinfo {pages}
  {6219} (\bibinfo {year} {1998})}\BibitemShut {NoStop}%
\bibitem [{\citenamefont {Baer}\ and\ \citenamefont
  {Neuhauser}(2000)}]{Baer:2000}%
  \BibitemOpen
  \bibfield  {author} {\bibinfo {author} {\bibfnamefont {R.}~\bibnamefont
  {Baer}}\ and\ \bibinfo {author} {\bibfnamefont {D.}~\bibnamefont
  {Neuhauser}},\ }\href@noop {} {\bibfield  {journal} {\bibinfo  {journal} {The
  Journal of Chemical Physics}\ }\textbf {\bibinfo {volume} {112}},\ \bibinfo
  {pages} {1679} (\bibinfo {year} {2000})}\BibitemShut {NoStop}%
\bibitem [{\citenamefont {Baer}(2000)}]{Baer:2000b}%
  \BibitemOpen
  \bibfield  {author} {\bibinfo {author} {\bibfnamefont {R.}~\bibnamefont
  {Baer}},\ }\href@noop {} {\bibfield  {journal} {\bibinfo  {journal} {The
  Journal of Chemical Physics}\ }\textbf {\bibinfo {volume} {113}},\ \bibinfo
  {pages} {473} (\bibinfo {year} {2000})}\BibitemShut {NoStop}%
\bibitem [{\citenamefont {Alexandru}\ \emph {et~al.}(2022)\citenamefont
  {Alexandru}, \citenamefont {Basar}, \citenamefont {Bedaque},\ and\
  \citenamefont {Warrington}}]{Alexandru:2020wrj}%
  \BibitemOpen
  \bibfield  {author} {\bibinfo {author} {\bibfnamefont {A.}~\bibnamefont
  {Alexandru}}, \bibinfo {author} {\bibfnamefont {G.}~\bibnamefont {Basar}},
  \bibinfo {author} {\bibfnamefont {P.~F.}\ \bibnamefont {Bedaque}}, \ and\
  \bibinfo {author} {\bibfnamefont {N.~C.}\ \bibnamefont {Warrington}},\ }\href
  {\doibase 10.1103/RevModPhys.94.015006} {\bibfield  {journal} {\bibinfo
  {journal} {Rev. Mod. Phys.}\ }\textbf {\bibinfo {volume} {94}},\ \bibinfo
  {pages} {015006} (\bibinfo {year} {2022})},\ \Eprint
  {http://arxiv.org/abs/2007.05436} {arXiv:2007.05436 [hep-lat]} \BibitemShut
  {NoStop}%
\bibitem [{\citenamefont {Witten}(2011)}]{Witten:2010cx}%
  \BibitemOpen
  \bibfield  {author} {\bibinfo {author} {\bibfnamefont {E.}~\bibnamefont
  {Witten}},\ }\href@noop {} {\bibfield  {journal} {\bibinfo  {journal} {AMS/IP
  Stud. Adv. Math.}\ }\textbf {\bibinfo {volume} {50}},\ \bibinfo {pages} {347}
  (\bibinfo {year} {2011})},\ \Eprint {http://arxiv.org/abs/1001.2933}
  {arXiv:1001.2933 [hep-th]} \BibitemShut {NoStop}%
\bibitem [{\citenamefont {Witten}(2010)}]{Witten:2010zr}%
  \BibitemOpen
  \bibfield  {author} {\bibinfo {author} {\bibfnamefont {E.}~\bibnamefont
  {Witten}},\ }\href@noop {} {\  (\bibinfo {year} {2010})},\ \Eprint
  {http://arxiv.org/abs/1009.6032} {arXiv:1009.6032 [hep-th]} \BibitemShut
  {NoStop}%
\bibitem [{\citenamefont {Epelbaum}\ \emph {et~al.}(2009)\citenamefont
  {Epelbaum}, \citenamefont {Hammer},\ and\ \citenamefont
  {Meissner}}]{Epelbaum:2008ga}%
  \BibitemOpen
  \bibfield  {author} {\bibinfo {author} {\bibfnamefont {E.}~\bibnamefont
  {Epelbaum}}, \bibinfo {author} {\bibfnamefont {H.-W.}\ \bibnamefont
  {Hammer}}, \ and\ \bibinfo {author} {\bibfnamefont {U.-G.}\ \bibnamefont
  {Meissner}},\ }\href {\doibase 10.1103/RevModPhys.81.1773} {\bibfield
  {journal} {\bibinfo  {journal} {Rev. Mod. Phys.}\ }\textbf {\bibinfo {volume}
  {81}},\ \bibinfo {pages} {1773} (\bibinfo {year} {2009})},\ \Eprint
  {http://arxiv.org/abs/0811.1338} {arXiv:0811.1338 [nucl-th]} \BibitemShut
  {NoStop}%
\bibitem [{\citenamefont {Navr\'atil}\ \emph {et~al.}(2016)\citenamefont
  {Navr\'atil}, \citenamefont {Quaglioni}, \citenamefont {Hupin}, \citenamefont
  {Romero-Redondo},\ and\ \citenamefont {Calci}}]{Navratil:2016ycn}%
  \BibitemOpen
  \bibfield  {author} {\bibinfo {author} {\bibfnamefont {P.}~\bibnamefont
  {Navr\'atil}}, \bibinfo {author} {\bibfnamefont {S.}~\bibnamefont
  {Quaglioni}}, \bibinfo {author} {\bibfnamefont {G.}~\bibnamefont {Hupin}},
  \bibinfo {author} {\bibfnamefont {C.}~\bibnamefont {Romero-Redondo}}, \ and\
  \bibinfo {author} {\bibfnamefont {A.}~\bibnamefont {Calci}},\ }\href
  {\doibase 10.1088/0031-8949/91/5/053002} {\bibfield  {journal} {\bibinfo
  {journal} {Phys. Scripta}\ }\textbf {\bibinfo {volume} {91}},\ \bibinfo
  {pages} {053002} (\bibinfo {year} {2016})},\ \Eprint
  {http://arxiv.org/abs/1601.03765} {arXiv:1601.03765 [nucl-th]} \BibitemShut
  {NoStop}%
\bibitem [{\citenamefont {Tews}\ \emph {et~al.}(2020)\citenamefont {Tews},
  \citenamefont {Davoudi}, \citenamefont {Ekstr\"om}, \citenamefont {Holt},\
  and\ \citenamefont {Lynn}}]{Tews:2020hgp}%
  \BibitemOpen
  \bibfield  {author} {\bibinfo {author} {\bibfnamefont {I.}~\bibnamefont
  {Tews}}, \bibinfo {author} {\bibfnamefont {Z.}~\bibnamefont {Davoudi}},
  \bibinfo {author} {\bibfnamefont {A.}~\bibnamefont {Ekstr\"om}}, \bibinfo
  {author} {\bibfnamefont {J.~D.}\ \bibnamefont {Holt}}, \ and\ \bibinfo
  {author} {\bibfnamefont {J.~E.}\ \bibnamefont {Lynn}},\ }\href {\doibase
  10.1088/1361-6471/ab9079} {\bibfield  {journal} {\bibinfo  {journal} {J.
  Phys. G}\ }\textbf {\bibinfo {volume} {47}},\ \bibinfo {pages} {103001}
  (\bibinfo {year} {2020})},\ \Eprint {http://arxiv.org/abs/2001.03334}
  {arXiv:2001.03334 [nucl-th]} \BibitemShut {NoStop}%
\bibitem [{\citenamefont {van Kolck}(2020)}]{vanKolck:2020llt}%
  \BibitemOpen
  \bibfield  {author} {\bibinfo {author} {\bibfnamefont {U.}~\bibnamefont {van
  Kolck}},\ }\href {\doibase 10.3389/fphy.2020.00079} {\bibfield  {journal}
  {\bibinfo  {journal} {Front. in Phys.}\ }\textbf {\bibinfo {volume} {8}},\
  \bibinfo {pages} {79} (\bibinfo {year} {2020})},\ \Eprint
  {http://arxiv.org/abs/2003.06721} {arXiv:2003.06721 [nucl-th]} \BibitemShut
  {NoStop}%
\bibitem [{\citenamefont {Epelbaum}\ \emph {et~al.}(2022)\citenamefont
  {Epelbaum}, \citenamefont {Krebs},\ and\ \citenamefont
  {Reinert}}]{Epelbaum:2022cyo}%
  \BibitemOpen
  \bibfield  {author} {\bibinfo {author} {\bibfnamefont {E.}~\bibnamefont
  {Epelbaum}}, \bibinfo {author} {\bibfnamefont {H.}~\bibnamefont {Krebs}}, \
  and\ \bibinfo {author} {\bibfnamefont {P.}~\bibnamefont {Reinert}},\
  }\href@noop {} {\  (\bibinfo {year} {2022})},\ \Eprint
  {http://arxiv.org/abs/2206.07072} {arXiv:2206.07072 [nucl-th]} \BibitemShut
  {NoStop}%
\bibitem [{\citenamefont {Wiringa}\ \emph {et~al.}(1995)\citenamefont
  {Wiringa}, \citenamefont {Stoks},\ and\ \citenamefont
  {Schiavilla}}]{Wiringa:1994wb}%
  \BibitemOpen
  \bibfield  {author} {\bibinfo {author} {\bibfnamefont {R.~B.}\ \bibnamefont
  {Wiringa}}, \bibinfo {author} {\bibfnamefont {V.~G.~J.}\ \bibnamefont
  {Stoks}}, \ and\ \bibinfo {author} {\bibfnamefont {R.}~\bibnamefont
  {Schiavilla}},\ }\href {\doibase 10.1103/PhysRevC.51.38} {\bibfield
  {journal} {\bibinfo  {journal} {Phys. Rev. C}\ }\textbf {\bibinfo {volume}
  {51}},\ \bibinfo {pages} {38} (\bibinfo {year} {1995})},\ \Eprint
  {http://arxiv.org/abs/nucl-th/9408016} {arXiv:nucl-th/9408016} \BibitemShut
  {NoStop}%
\bibitem [{\citenamefont {Lovato}\ \emph {et~al.}(2022)\citenamefont {Lovato},
  \citenamefont {Bombaci}, \citenamefont {Logoteta}, \citenamefont {Piarulli},\
  and\ \citenamefont {Wiringa}}]{Lovato:2022apd}%
  \BibitemOpen
  \bibfield  {author} {\bibinfo {author} {\bibfnamefont {A.}~\bibnamefont
  {Lovato}}, \bibinfo {author} {\bibfnamefont {I.}~\bibnamefont {Bombaci}},
  \bibinfo {author} {\bibfnamefont {D.}~\bibnamefont {Logoteta}}, \bibinfo
  {author} {\bibfnamefont {M.}~\bibnamefont {Piarulli}}, \ and\ \bibinfo
  {author} {\bibfnamefont {R.~B.}\ \bibnamefont {Wiringa}},\ }\href {\doibase
  10.1103/PhysRevC.105.055808} {\bibfield  {journal} {\bibinfo  {journal}
  {Phys. Rev. C}\ }\textbf {\bibinfo {volume} {105}},\ \bibinfo {pages}
  {055808} (\bibinfo {year} {2022})},\ \Eprint
  {http://arxiv.org/abs/2202.10293} {arXiv:2202.10293 [nucl-th]} \BibitemShut
  {NoStop}%
\bibitem [{\citenamefont {Wiringa}\ and\ \citenamefont
  {Pieper}(2002)}]{Wiringa:2002ja}%
  \BibitemOpen
  \bibfield  {author} {\bibinfo {author} {\bibfnamefont {R.~B.}\ \bibnamefont
  {Wiringa}}\ and\ \bibinfo {author} {\bibfnamefont {S.~C.}\ \bibnamefont
  {Pieper}},\ }\href {\doibase 10.1103/PhysRevLett.89.182501} {\bibfield
  {journal} {\bibinfo  {journal} {Phys. Rev. Lett.}\ }\textbf {\bibinfo
  {volume} {89}},\ \bibinfo {pages} {182501} (\bibinfo {year} {2002})},\
  \Eprint {http://arxiv.org/abs/nucl-th/0207050} {arXiv:nucl-th/0207050}
  \BibitemShut {NoStop}%
\bibitem [{\citenamefont {Chen}\ and\ \citenamefont
  {Schmidt}(2022)}]{Chen:2022ndh}%
  \BibitemOpen
  \bibfield  {author} {\bibinfo {author} {\bibfnamefont {R.}~\bibnamefont
  {Chen}}\ and\ \bibinfo {author} {\bibfnamefont {K.~E.}\ \bibnamefont
  {Schmidt}},\ }\href {\doibase 10.1103/PhysRevC.106.044327} {\bibfield
  {journal} {\bibinfo  {journal} {Phys. Rev. C}\ }\textbf {\bibinfo {volume}
  {106}},\ \bibinfo {pages} {044327} (\bibinfo {year} {2022})},\ \Eprint
  {http://arxiv.org/abs/2204.10458} {arXiv:2204.10458 [nucl-th]} \BibitemShut
  {NoStop}%
\bibitem [{\citenamefont {{Foulkes}}\ \emph {et~al.}(2001)\citenamefont
  {{Foulkes}}, \citenamefont {{Mitas}}, \citenamefont {{Needs}},\ and\
  \citenamefont {{Rajagopal}}}]{Foulkes:2001}%
  \BibitemOpen
  \bibfield  {author} {\bibinfo {author} {\bibfnamefont {W.~M.}\ \bibnamefont
  {{Foulkes}}}, \bibinfo {author} {\bibfnamefont {L.}~\bibnamefont {{Mitas}}},
  \bibinfo {author} {\bibfnamefont {R.~J.}\ \bibnamefont {{Needs}}}, \ and\
  \bibinfo {author} {\bibfnamefont {G.}~\bibnamefont {{Rajagopal}}},\ }\href
  {\doibase 10.1103/RevModPhys.73.33} {\bibfield  {journal} {\bibinfo
  {journal} {Reviews of Modern Physics}\ }\textbf {\bibinfo {volume} {73}},\
  \bibinfo {pages} {33} (\bibinfo {year} {2001})}\BibitemShut {NoStop}%
\bibitem [{\citenamefont {Gezerlis}\ \emph {et~al.}(2013)\citenamefont
  {Gezerlis}, \citenamefont {Tews}, \citenamefont {Epelbaum}, \citenamefont
  {Gandolfi}, \citenamefont {Hebeler}, \citenamefont {Nogga},\ and\
  \citenamefont {Schwenk}}]{Gezerlis:2013ipa}%
  \BibitemOpen
  \bibfield  {author} {\bibinfo {author} {\bibfnamefont {A.}~\bibnamefont
  {Gezerlis}}, \bibinfo {author} {\bibfnamefont {I.}~\bibnamefont {Tews}},
  \bibinfo {author} {\bibfnamefont {E.}~\bibnamefont {Epelbaum}}, \bibinfo
  {author} {\bibfnamefont {S.}~\bibnamefont {Gandolfi}}, \bibinfo {author}
  {\bibfnamefont {K.}~\bibnamefont {Hebeler}}, \bibinfo {author} {\bibfnamefont
  {A.}~\bibnamefont {Nogga}}, \ and\ \bibinfo {author} {\bibfnamefont
  {A.}~\bibnamefont {Schwenk}},\ }\href {\doibase
  10.1103/PhysRevLett.111.032501} {\bibfield  {journal} {\bibinfo  {journal}
  {Phys. Rev. Lett.}\ }\textbf {\bibinfo {volume} {111}},\ \bibinfo {pages}
  {032501} (\bibinfo {year} {2013})},\ \Eprint {http://arxiv.org/abs/1303.6243}
  {arXiv:1303.6243 [nucl-th]} \BibitemShut {NoStop}%
\bibitem [{\citenamefont {Lynn}\ \emph {et~al.}(2016)\citenamefont {Lynn},
  \citenamefont {Tews}, \citenamefont {Carlson}, \citenamefont {Gandolfi},
  \citenamefont {Gezerlis}, \citenamefont {Schmidt},\ and\ \citenamefont
  {Schwenk}}]{Lynn:2015jua}%
  \BibitemOpen
  \bibfield  {author} {\bibinfo {author} {\bibfnamefont {J.~E.}\ \bibnamefont
  {Lynn}}, \bibinfo {author} {\bibfnamefont {I.}~\bibnamefont {Tews}}, \bibinfo
  {author} {\bibfnamefont {J.}~\bibnamefont {Carlson}}, \bibinfo {author}
  {\bibfnamefont {S.}~\bibnamefont {Gandolfi}}, \bibinfo {author}
  {\bibfnamefont {A.}~\bibnamefont {Gezerlis}}, \bibinfo {author}
  {\bibfnamefont {K.~E.}\ \bibnamefont {Schmidt}}, \ and\ \bibinfo {author}
  {\bibfnamefont {A.}~\bibnamefont {Schwenk}},\ }\href {\doibase
  10.1103/PhysRevLett.116.062501} {\bibfield  {journal} {\bibinfo  {journal}
  {Phys. Rev. Lett.}\ }\textbf {\bibinfo {volume} {116}},\ \bibinfo {pages}
  {062501} (\bibinfo {year} {2016})},\ \Eprint
  {http://arxiv.org/abs/1509.03470} {arXiv:1509.03470 [nucl-th]} \BibitemShut
  {NoStop}%
\bibitem [{\citenamefont {Piarulli}\ \emph {et~al.}(2015)\citenamefont
  {Piarulli}, \citenamefont {Girlanda}, \citenamefont {Schiavilla},
  \citenamefont {Navarro~P\'erez}, \citenamefont {Amaro},\ and\ \citenamefont
  {Ruiz~Arriola}}]{Piarulli:2014bda}%
  \BibitemOpen
  \bibfield  {author} {\bibinfo {author} {\bibfnamefont {M.}~\bibnamefont
  {Piarulli}}, \bibinfo {author} {\bibfnamefont {L.}~\bibnamefont {Girlanda}},
  \bibinfo {author} {\bibfnamefont {R.}~\bibnamefont {Schiavilla}}, \bibinfo
  {author} {\bibfnamefont {R.}~\bibnamefont {Navarro~P\'erez}}, \bibinfo
  {author} {\bibfnamefont {J.~E.}\ \bibnamefont {Amaro}}, \ and\ \bibinfo
  {author} {\bibfnamefont {E.}~\bibnamefont {Ruiz~Arriola}},\ }\href {\doibase
  10.1103/PhysRevC.91.024003} {\bibfield  {journal} {\bibinfo  {journal} {Phys.
  Rev. C}\ }\textbf {\bibinfo {volume} {91}},\ \bibinfo {pages} {024003}
  (\bibinfo {year} {2015})},\ \Eprint {http://arxiv.org/abs/1412.6446}
  {arXiv:1412.6446 [nucl-th]} \BibitemShut {NoStop}%
\bibitem [{\citenamefont {Piarulli}\ \emph {et~al.}(2018)\citenamefont
  {Piarulli} \emph {et~al.}}]{Piarulli:2017dwd}%
  \BibitemOpen
  \bibfield  {author} {\bibinfo {author} {\bibfnamefont {M.}~\bibnamefont
  {Piarulli}} \emph {et~al.},\ }\href {\doibase 10.1103/PhysRevLett.120.052503}
  {\bibfield  {journal} {\bibinfo  {journal} {Phys. Rev. Lett.}\ }\textbf
  {\bibinfo {volume} {120}},\ \bibinfo {pages} {052503} (\bibinfo {year}
  {2018})},\ \Eprint {http://arxiv.org/abs/1707.02883} {arXiv:1707.02883
  [nucl-th]} \BibitemShut {NoStop}%
\bibitem [{\citenamefont {Shi}\ and\ \citenamefont
  {Zhang}(2016)}]{Shi:2015lyu}%
  \BibitemOpen
  \bibfield  {author} {\bibinfo {author} {\bibfnamefont {H.}~\bibnamefont
  {Shi}}\ and\ \bibinfo {author} {\bibfnamefont {S.}~\bibnamefont {Zhang}},\
  }\href {\doibase 10.1103/PhysRevE.93.033303} {\bibfield  {journal} {\bibinfo
  {journal} {Phys. Rev. E}\ }\textbf {\bibinfo {volume} {93}},\ \bibinfo
  {pages} {033303} (\bibinfo {year} {2016})},\ \Eprint
  {http://arxiv.org/abs/1511.04084} {arXiv:1511.04084 [physics.comp-ph]}
  \BibitemShut {NoStop}%
\bibitem [{\citenamefont {Kingma}\ and\ \citenamefont
  {Ba}(2014)}]{Kingma:2014}%
  \BibitemOpen
  \bibfield  {author} {\bibinfo {author} {\bibfnamefont {D.~P.}\ \bibnamefont
  {Kingma}}\ and\ \bibinfo {author} {\bibfnamefont {J.}~\bibnamefont {Ba}},\
  }\href@noop {} {\enquote {\bibinfo {title} {Adam: A method for stochastic
  optimization},}\ } (\bibinfo {year} {2014}),\ \Eprint
  {http://arxiv.org/abs/1412.6980} {arXiv:1412.6980 [cs.LG]} \BibitemShut
  {NoStop}%
\bibitem [{\citenamefont {Kingma}\ and\ \citenamefont
  {Welling}(2013)}]{Kingma:2013}%
  \BibitemOpen
  \bibfield  {author} {\bibinfo {author} {\bibfnamefont {D.~P.}\ \bibnamefont
  {Kingma}}\ and\ \bibinfo {author} {\bibfnamefont {M.}~\bibnamefont
  {Welling}},\ }\href@noop {} {\enquote {\bibinfo {title} {Auto-encoding
  variational bayes},}\ } (\bibinfo {year} {2013}),\ \Eprint
  {http://arxiv.org/abs/1312.6114} {arXiv:1312.6114 [stat.ML]} \BibitemShut
  {NoStop}%
\bibitem [{\citenamefont {Seng}\ \emph {et~al.}(2019)\citenamefont {Seng},
  \citenamefont {Gorchtein},\ and\ \citenamefont
  {Ramsey-Musolf}}]{Seng:2018qru}%
  \BibitemOpen
  \bibfield  {author} {\bibinfo {author} {\bibfnamefont {C.~Y.}\ \bibnamefont
  {Seng}}, \bibinfo {author} {\bibfnamefont {M.}~\bibnamefont {Gorchtein}}, \
  and\ \bibinfo {author} {\bibfnamefont {M.~J.}\ \bibnamefont
  {Ramsey-Musolf}},\ }\href {\doibase 10.1103/PhysRevD.100.013001} {\bibfield
  {journal} {\bibinfo  {journal} {Phys. Rev. D}\ }\textbf {\bibinfo {volume}
  {100}},\ \bibinfo {pages} {013001} (\bibinfo {year} {2019})},\ \Eprint
  {http://arxiv.org/abs/1812.03352} {arXiv:1812.03352 [nucl-th]} \BibitemShut
  {NoStop}%
\bibitem [{\citenamefont {Hardy}\ and\ \citenamefont
  {Towner}(2020)}]{Hardy:2020qwl}%
  \BibitemOpen
  \bibfield  {author} {\bibinfo {author} {\bibfnamefont {J.~C.}\ \bibnamefont
  {Hardy}}\ and\ \bibinfo {author} {\bibfnamefont {I.~S.}\ \bibnamefont
  {Towner}},\ }\href {\doibase 10.1103/PhysRevC.102.045501} {\bibfield
  {journal} {\bibinfo  {journal} {Phys. Rev. C}\ }\textbf {\bibinfo {volume}
  {102}},\ \bibinfo {pages} {045501} (\bibinfo {year} {2020})}\BibitemShut
  {NoStop}%
\end{thebibliography}%

\end{document}